\documentclass[prd,10pt,aps,superscriptaddress,twocolumn,amsmath,amssymb,nofootinbib,longbibliography]{revtex4-1}

\usepackage{natbib}
\usepackage[colorlinks]{hyperref}
\hypersetup{linkcolor=blue,citecolor=blue,filecolor=blue,urlcolor=blue}

\usepackage{afterpage}
\usepackage{float} 

\usepackage{lipsum}
\usepackage{graphicx}
\usepackage{tabularx}
\usepackage{multirow}
\usepackage{array}
\usepackage[dvipsnames]{xcolor}

\usepackage{dcolumn}
\usepackage{bm}

\definecolor{green}{rgb}{0.19,0.64,0.54}
\definecolor{blue}{rgb}{0,0,1}
\definecolor{reddish}{rgb}{0.65, 0.2, 0.2}
\definecolor{darkgreen}{rgb}{0.2,0.7,0.3}
\definecolor{darkblue}{rgb}{0.3,0.40,0.48}
\definecolor{gray}{rgb}{.8,.8,.8}

\def\dd{\text{d}}

\begin{document}
\title{Orbital motion of primordial black holes crossing Solar-type stars}

\author{Vitorio A. \surname{De Lorenci}}
\email{delorenci@unifei.edu.br}

\affiliation{Instituto de F\'{\i}sica e Qu\'{\i}mica, Universidade
  Federal de Itajub\'a, Itajub\'a, Minas Gerais 37500-903, Brazil}

\author{David I. \surname{Kaiser}}
\email{dikaiser@mit.edu}

\affiliation{Department of Physics, Massachusetts Institute of Technology, Cambridge, Massachusetts 02139, USA}

\author{Patrick \surname{Peter}}
\email{peter@iap.fr}
\affiliation{{${\cal G}\mathbb{R}\varepsilon\mathbb{C}{\cal
      O}$}---Institut d'Astrophysique de Paris, CNRS and Sorbonne
  Universit\'e, UMR 7095 98 bis Boulevard Arago, 75014 Paris, France}

\begin{abstract}
Primordial black holes (PBHs) are hypothetical objects that could have originated from density fluctuations in a very early phase of our Universe. Recent observations restrict the masses that such PBHs could have, if they are to constitute all of dark matter today: $10^{17} \, {\rm g} \leq m \leq 10^{23} \, {\rm g}$. With such low masses, general relativity predicts that the corresponding radii for the PBHs would be atomic or subatomic in size. When captured by a star, such a tiny PBH could exhibit an orbit completely or partially inside the body of the star, without significantly changing its mass for quite a long time.
Here we examine the possible trajectories of a PBH that is captured by a Sun-like star. When in motion in the interior of the star, the amount of stellar mass that effectively interacts with the PBH will be a function of its distance to the center of the star. As a consequence, a strong effect on the shape of the orbits emerges, leading to PBH trajectories that could be open or closed, and exhibiting a rich variety of patterns.
\end{abstract}

\maketitle

\section{Introduction}
\label{introduction}

The existence of primordial black holes in the mass range $10^{17} \,{\rm g} \leq m \leq 10^{23} \, {\rm g}$ could be a possible explanation for dark matter (DM) \cite{Khlopov:2008qy,PhysRevD.87.023507,Smyth:2019whb,2020PhRvD.101l3514L,Carr_2021,10.21468/SciPostPhysLectNotes.48,Green:2020jor,Villanueva-Domingo:2021spv,Escriva:2022duf,Carr:2023tpt,Gorton:2024cdm}. The associated radii of such black holes range from subatomic scales to the size of a single DNA helix. Assuming they are the sole constituent of DM, they should be very abundant within galaxies and should be present in every star system, suggesting that they might be detectable based on their impact on the motions of visible objects \cite{Dror:2019twh,Li:2022oqo,BrownHeUnwin,Tran:2023jci,Bertrand:2023zkl,Cuadrat-Grzybowski:2024uph}.

A common question is how PBHs interact with their environment. In particular: what happens when a PBH crosses the bulk of a star in its orbit? Such questions arose around the time when the Large Hadron Collider at CERN began its first run, given the suggestion that high-energy proton-proton collisions could generate miniature black holes \cite{PhysRevD.78.035009}. More recent estimates show that, for an atomic-size black hole orbiting the interior of a typical star like our Sun, it would take approximately the star's life-time for the black hole to cause any significant disruption to the star \cite{Oncins:2022djq,Bellinger_2023,Caplan:2023ddo}. 

The detailed study of the motion of a small PBH that crosses the interior of a typical star in its path has not yet been fully described in the literature. When this topic has been discussed, the focus has usually remained on other effects, such as the energy loss of the PBH by means of dynamical friction due to its motion inside the gaseous environment of the star, and the eventual capture of the PBH at the stellar core \cite{Oncins:2022djq,Bellinger_2023,Caplan:2023ddo}. Investigations concerning the capture of PBHs by much more dense astronomical objects, such as neutron stars, and the characteristic signal of gravitational waves emitted by such systems, have been recently discussed \cite{2024PhRvD.109f3004B,2024arXiv240201838B,2024arXiv240408735B,2024arXiv240408057C}. In particular, Ref.~\cite{2024arXiv240201838B} considers some consequences of a distance-dependence of the source mass on the orbit of a PBH inside a neutron star for certain configurations.

In this work, we examine 
the issue of the orbital motion of a test particle (in this case, a small PBH) that can cross the interior of a Solar-type star in its trajectory. For the motion of DM particles near our own Sun, there is a remarkable congruence: the standard DM halo models, applied to the Milky Way, yield a typical velocity of the DM particles in the vicinity of the Sun of $v \sim 220 - 235 \, {\rm km \, s^{-1}}$ \cite{BinneyWong2017,PostiHelmi2018,Evans:2018bqy}. Meanwhile, the escape velocity for small-mass objects at a distance ${\cal R}$ from our Sun scales as $v_{\rm esc} \sim 600\, (R_\odot / {\cal R})^{1/2} \, {\rm km \, s^{-1}}$. Although arising from fundamentally different systems---the mass and spatial extent of our galaxy versus those of our Sun---together these estimates suggest that PBHs with $v < v_{\rm esc}$ could have been gravitationally captured by our Sun, for example, during its formation \cite{Capela:2012jz,Capela:2013yf,Esser:2022owk}. This congruence in typical speeds motivates a closer study of resulting PBH orbits around a Solar-like star within a typical galaxy.

Outside the star, the trajectory of the particle is described by the classical orbit equation, for which the total mass of the star is the source of the gravitational interaction. However, when crossing the star's bulk, the trajectory of the particle is influenced by changes in the solar mass distribution, which acts as the gravitational source of the interaction. Effectively, the source of interaction will consist of the spherical mass distribution whose radius is given by the particle's position. This relationship induces a strong precession effect on the particle's orbit. As discussed below, similar phenomena occur in the orbits of stars within a galaxy \cite{BinneyTremaine}, though in such cases the dynamics are governed by qualitatively different density profiles than the one considered here.

We consider the PBH motion over relatively short time-scales---exponentially shorter than the lifetime of the host star---for which we may consider the mass of both the PBH and of the star to remain fixed. Mass accretion by the PBH scales with the PBH cross section $\sim r_\textsc{bh}^2$, and hence remains highly suppressed over the time-scales considered here. Likewise, given the PBHs' small cross sections and the short time-scales of interest, we further neglect gravitational radiation \cite{2024arXiv240201838B,2024arXiv240408735B}, dynamical friction \cite{Ostriker_1999}, and hydrodynamic dragging effects \cite{Bellinger_2023}.

In Sec.~\ref{sec:eom}, we derive the equation governing the motion of a test particle of mass $m$ in a system whose source of gravity has a mass that depends on the particle's position, $M(r)$. A workable model for the mass density profile of the star is developed in Sec.~\ref{workable}, and is used to study the dynamics of a PBH around the center of a star in Sec.~\ref{sec_orbit}. Within the time-span considered in the numerical simulations, orbits can be generally classified as open when only one full rotation ($2\pi$ radians) is completed. However, if a larger angular variation is allowed, the results suggest that closed orbits exist, depending on the assumed initial conditions. Solutions for orbits completely or partially inside the star's mass distribution are illustrated. Final remarks are presented in Sec.~\ref{final}. Finally, a short discussion regarding the shape of the orbits, including the conditions for closed solutions to occur, is presented in the Appendix. Additionally, some suggestive numerical solutions for possible orbits (both open and closed) of a PBH crossing the star or residing completely within its interior are presented for illustrative purposes.

\section{Equations of motion}
\label{sec:eom}
In order to set the main notation used throughout the paper, we start by studying the Keplerian problem of the motion of a test particle of mass $m$ around a body of constant mass $M_{\odot}$, with $M_{\odot} \gg m$. We first consider the Newtonian dynamics for such a system, before bounding the magnitude of relativistic corrections for the scenarios of interest.

For an isolated two-body system, the total energy $E_\textsc{t}$ is conserved, and so is the total angular momentum ${\bm L}$, since no external torque acts on the particle. Therefore the motion of the test particle will lie in a plane, which we take to be the $xy$-plane. Expressing the velocity of the test particle in terms of the radial and angular coordinates $r$ and $\varphi$, it follows that $v^2 = {\dot r}^2 +r^2 {\dot \varphi}^2$, where a dot over a quantity denotes its time derivative. Then the energy conservation law reads 
\begin{align}
    \frac{1}{2}m{\dot r}^2 + \frac{1}{2}m r^2{\dot \varphi}^2 + U(r)  = 
    E_\textsc{t},
    \label{energy}
\end{align}
where $U(r)$ is the potential energy of the particle, which, for the particular case of the Keplerian problem, reads $U(r)=- GM_{\odot}m / r$. We define the specific angular momentum $\ell \doteq  L/m = r^2\dot\varphi$, introduce the convenient variable $u = 1/ r$, and express time derivatives in terms of the angular variable $\varphi$, such that $\dot r = -(1/u^2)u'\dot\varphi = -\ell u'$, where a prime denotes a derivative with respect to $\varphi$. Then taking the derivative of Eq.~(\ref{energy}) with respect to $\varphi$ yields
\begin{align}
    u'' + u = \frac{GM_{\odot}}{\ell^2}.
    \label{main}
\end{align}
The classical problem for constants $m$ and $M_{\odot}$ is thus solved by integrating Eq.~(\ref{main}) over time, resulting in the well-known conic sections, described by $r(\varphi)(1+ e \cos\varphi) = \ell^2/(2GM_{\odot})$. Here $e$ is the eccentricity parameter: $e=0$ corresponds to the circular solution, while $0 < e< 1$ leads to elliptic orbits and $e>1$ leads to open trajectories (parabolic and hyperbolic ones).

We are interested in the motion of a PBH of mass $m$ in orbit around a star. Given the small masses $m$ of interest, a PBH would be able to travel into the bulk of the star without short-time consequences. Over time-scales of interest, accretion onto the PBH remains negligible, so we consider both $m$ and $M_\odot$ to remain constant.

Consider the motion of a PBH of mass $m$ in orbit around the center of a star of total mass $M_{\odot}$ and radius $R$ in such a way that part or all of the PBH trajectory can be inside the body  of the star ($r<R$). Assuming the star is described by a spherical distribution of mass, the gravitational potential relevant for the particle's motion will be related to the amount of the star's mass between its center and the PBH's radial position. In this case, when the PBH is at a distance $r$ from the origin, the enclosed star mass that effectively interacts with it is given by
\begin{align}
    \label{Mdef}
    M(r) = \left\{ 
            \begin{array}{ll}
                4\pi \displaystyle\int_{0}^{r} \rho(r') {r'}^2\dd r', & r < R ,\\ \\
                M(R) \doteq M_{\odot}, & r \ge R .
            \end{array}
           \right.
\end{align}
Thus, at any point of its trajectory, the total energy of the particle will be given by Eq.~(\ref{energy}), but with $U(r)$ produced by the mass distribution $M(r)$ (see the Appendix for further details).

The angular momentum of the PBH remains conserved. The equation governing the PBH motion can be easily obtained from Newton's second law, which gives 
\begin{align}
    {\ddot r} - \frac{\ell^2}{r^3}  =  - \frac{GM(r)}{r^2}.
    \label{2ndlaw}
\end{align}
Using Eq.~(\ref{Mdef}) and rewriting Eq.~(\ref{2ndlaw}) in terms of the dimensionless variable $\eta(\varphi) \doteq R/r(\varphi)$, it follows that
\begin{align}
    \eta'' + \eta = \frac{GRM_\odot}{\ell^2}\frac{M(\eta)}{M_\odot}.
    \label{main2}
\end{align}
Once $\rho(r)$ is known, and thus $M(r)$, Eq.~(\ref{main2}) can be numerically integrated to yield the PBH trajectories for given initial conditions.

The leading correction from general relativity would modify Eq.~(\ref{main2}), contributing $3 GM \eta^2 / (Rc^2)$ to the right-hand side \cite{WaldBook}. This modification induces precession of planetary orbits---famously yielding the 43 arc-seconds of precession per century for the orbit of Mercury (the largest effect among the Solar System planets).

Once we consider motion of a test particle (such as a PBH) that traverses the bulk of the star, however, the significance of the leading-order relativistic correction diminishes. For such orbits, upon taking into account $M(r)$ as in Eq.~(\ref{Mdef}), the Newtonian dynamics already induce a strong precession. The ratio of the magnitude of the relativistic correction to the Newtonian contribution is given by $3(\ell/cr)^2$, where $r=r(\varphi)$. For the specific solutions studied in this work, this ratio is approximately $10^{-5}$. Consequently, when the particle's path intersects the star's bulk, the relativistic effect can be safely disregarded from our discussion, as it would only entail a minor adjustment to the precession effect already induced by $M(r)$. Meanwhile, over the short time-spans considered here, the relativistic precession for orbits exterior to the bulk of the star remains inconsequential. Hence we neglect such relativistic corrections and consider the Newtonian dynamics below.

\section{A workable model for the mass density profile}
\label{workable}
To investigate the motion a PBH around a typical star, we must first select a model that describes the mass-density profile of the star. An appropriate choice is to utilize a model based on the Sun \cite{2004PhRvL..92l1301B,2005ApJ...621L..85B,2010Ap&SS.328...13S,SolarModel}. Typically, the best fit to available data is represented by a fourth-degree polynomial, adjusted to prevent negative values for the mass-density near the boundary of the mass distribution (at $r=R$).

\begin{figure}[b]
    \includegraphics[scale=0.4]{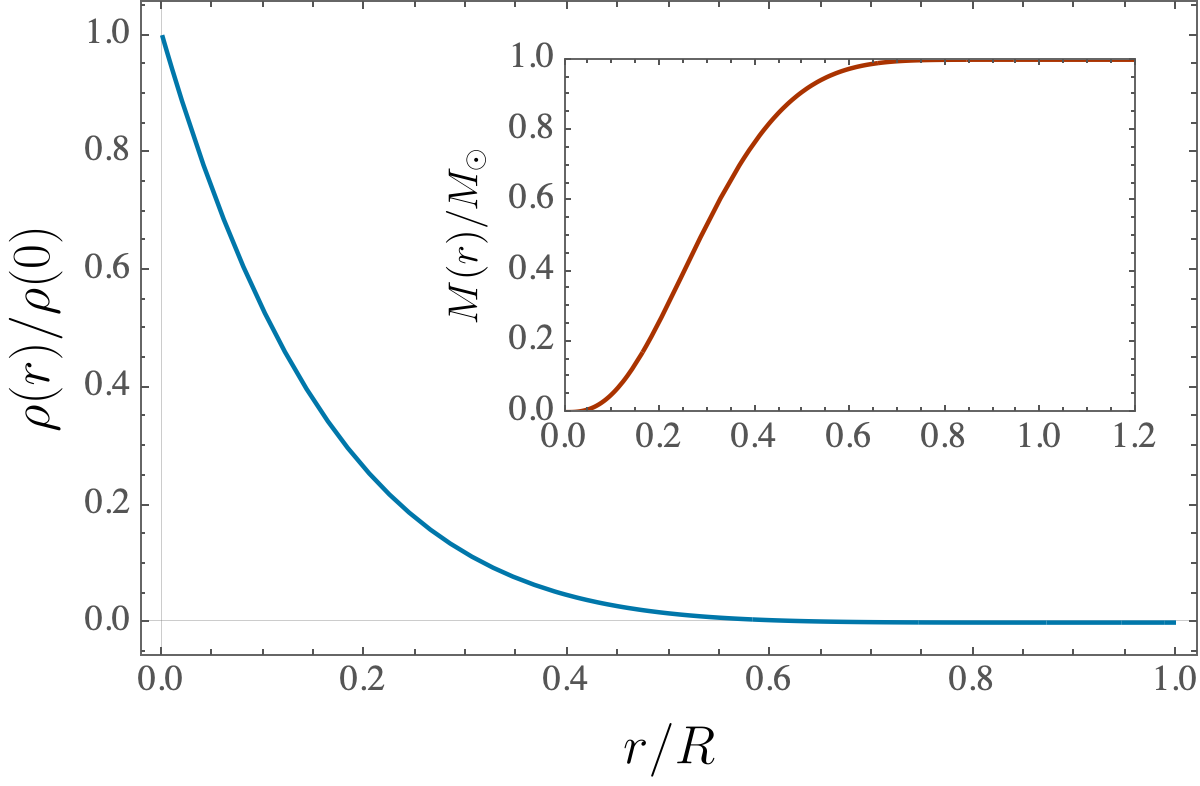}
    \caption{Toy model for the mass-density profile of the star. The inset shows the mass of the star effectively perceived by the PBH along its trajectory. Notice that the perceived mass diminishes as the PBH moves in the bulk of the star $r < R$, and equals $M_\odot$ outside the limits of the mass distribution $r\ge R$.}
    \label{fig0}
\end{figure}

For practical purposes, we instead adopt an alternative model that fits the data to a positive function of the form $\rho(r) = \rho(0)(1-r/R)^n$, where $\rho(0)$ and $n$ are parameters to be determined through fitting analysis. Using the dataset available in Ref.~\cite{SolarModel}, we obtain the following toy model to describe the mass-density function of a Solar-type star:
\begin{align}
    \rho(r) = \rho(0) \left(1-
    \frac{r}{R}\right)^6 \Theta \left(1 - \frac{r}{R}\right),
    \label{rho}
\end{align}
where $R$ is the radius of the star, $\rho(0) = 1.184 \times 10^5\, {\rm kg\, m^{-3}}$, and $\Theta(x)$ is the Heaviside step function, defined as $\Theta(x) = 1$ when $x>0$, $\Theta(0) = \frac12$ and $\Theta(x) = 0$ when $x<0$. The behaviour of $\rho(r)$ is depicted in Fig. \ref{fig0}.

From the model of $\rho (r)$ in Eq.~(\ref{rho}), the solar mass perceived by the PBH as it moves around or inside the star is given by
\begin{align}
    \frac{M(r)}{M_{\odot}} = &252 \left(\frac{r^9}{9 R^9}-\frac{3 r^8}{4 R^8}+\frac{15 r^7}{7 R^7}-\frac{10 r^6}{3 R^6}+\frac{3 r^5}{R^5} 
    \right. \nonumber \\
    &-\left. \frac{3 r^4}{2 R^4}+\frac{r^3}{3 R^3}\right) \Theta \left(1-\frac{r}{R}\right)+\Theta \left(\frac{r}{R}-1\right).
    \label{M(r)}
\end{align}
Notice that $M(r\ge R) =  M_{\odot}$, as needed. The behavior of this function is shown in the inset of Fig.~\ref{fig0}. As indicated in the figure, when the particle moves through the interior of the star, the mass that effectively contributes to the gravitational field at the particle's position is the mass that lies within the sphere centered at the star's center and whose radius is the radial position of the particle. On the other hand, when the particle travels outside the star, the total mass of the star, i.e. $M_\odot$, contributes to the gravitational field at the particle's position. 

The density profile for a star, such as in Eq.~(\ref{rho}), is qualitatively different from the typical density profiles used to model dynamics of individual stars within a large galaxy. In the case of galactic dynamics, the salient feature that density profiles aim to capture---even in the simple limit of a spherically symmetric, coarse-grained mass density---is a function $\rho (r)$ that is nearly flat at small radii (near the core of the galaxy), and which falls off as $r^{-1}$ at larger distances from the galactic core. An example is the classic Plummer model, for which $\rho (r) = \rho (0) [ 1 + (r / R)^2]^{-5/2}$ \cite{BinneyTremaine}. The resulting function of the enclosed mass for galactic dynamics, $M(r)$, then rises quasi-linearly with distance over the range $\frac{1}{2} R  \leq r \leq R$, unlike the steeper profile of $M(r)$ for the Solar model as shown in Fig.~\ref{fig0}.

In addition to these differences in the spatial profiles for $\rho(r)$ and $M(r)$ for typical stars compared to typical galaxies, we also consider a wider range of orbital motions in this work than are typical in galactic dynamics. In particular, as we discuss in the next section, for PBHs in orbit around a Sun-like star, the PBH trajectories may either be contained entirely within the star, with $r < R$, or may extend beyond the volume of the star, repeatedly crossing into the region $r > R$ on each orbit. In contrast, stars within a galaxy are typically assumed to remain within the effective radius $R$ of that galaxy, even as their orbits may describe complicated, non-Keplerian trajectories \cite{BinneyTremaine}.  

\section{Orbits}
\label{sec_orbit}
In this section, we examine numerical solutions of Eq.~(\ref{main2}) associated with some representative initial conditions for the motion of a PBH. In order to simplify the analysis, we set initial conditions at $\varphi =0$, such that $r'(\varphi) = 0$. In other words, we set initial conditions at a turning point of the orbit. With this choice, the parameter $\ell$ is conveniently given by $\ell = r(\varphi=0) v(\varphi=0)$. Additionally, in terms of the dimensionless  variable $\eta(r)$, the star disk has unit radius: $\eta(R) = 1$. The star appears as a solid disk (colored orange) in all figures hereafter. 

The fact that the source-mass $M(r)$ is a distance-dependent function in the equation of motion of the test particle, while it is travelling within the interior of the star, leads to a multitude of possible trajectories. These trajectories can be open or closed depending on the initial conditions. 

As noted above, the typical velocity of dark matter particles---including PBHs---that are gravitationally bound within a large halo can be estimated from the standard galactic DM halo models \cite{BinneyWong2017,PostiHelmi2018,Evans:2018bqy}. When applied to dark matter in the vicinity of our Solar System, this yields $v \sim 220 - 235\, {\rm km \, s^{-1}}$. Hence in our simulations we consider initial velocities $v(0) \sim {\cal O} (10^2) \, {\rm km \, s^{-1}}$.

\subsection{Initial conditions such that $r(\varphi=0) < R$}
\label{sec-interior}
\begin{figure}[h]
    \includegraphics[scale=0.3]{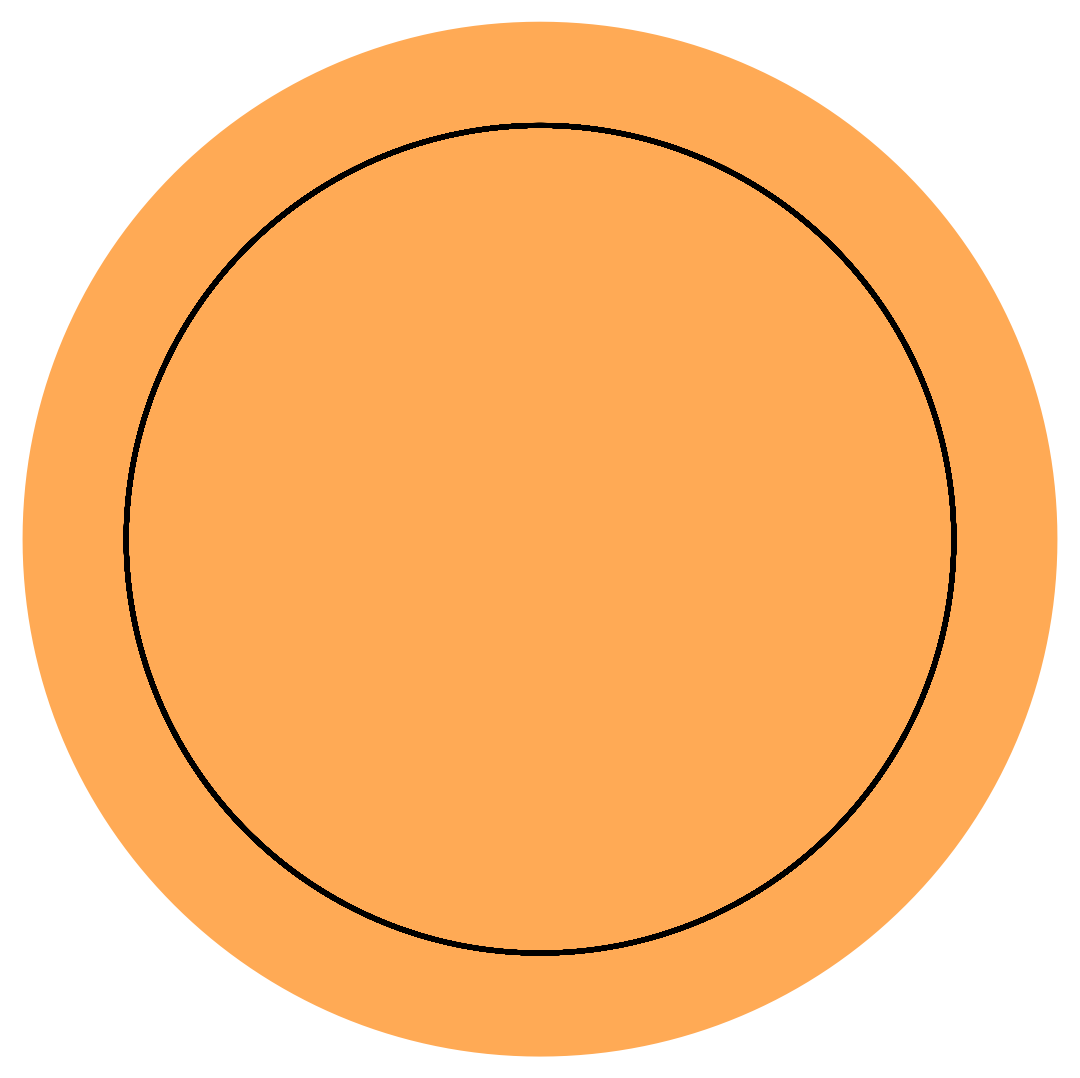}
    \caption{Orbit of the PBH in the bulk of the star. The plot shows a set of 100 complete revolutions ($\Delta \varphi = 200 \pi$) of the PBH inside the star, depicted by the closed solid line inside the orange disk, which represents the star.}
    \label{fig_1_48p81}
\end{figure}

\begin{figure}[h]
    \includegraphics[scale=0.22]{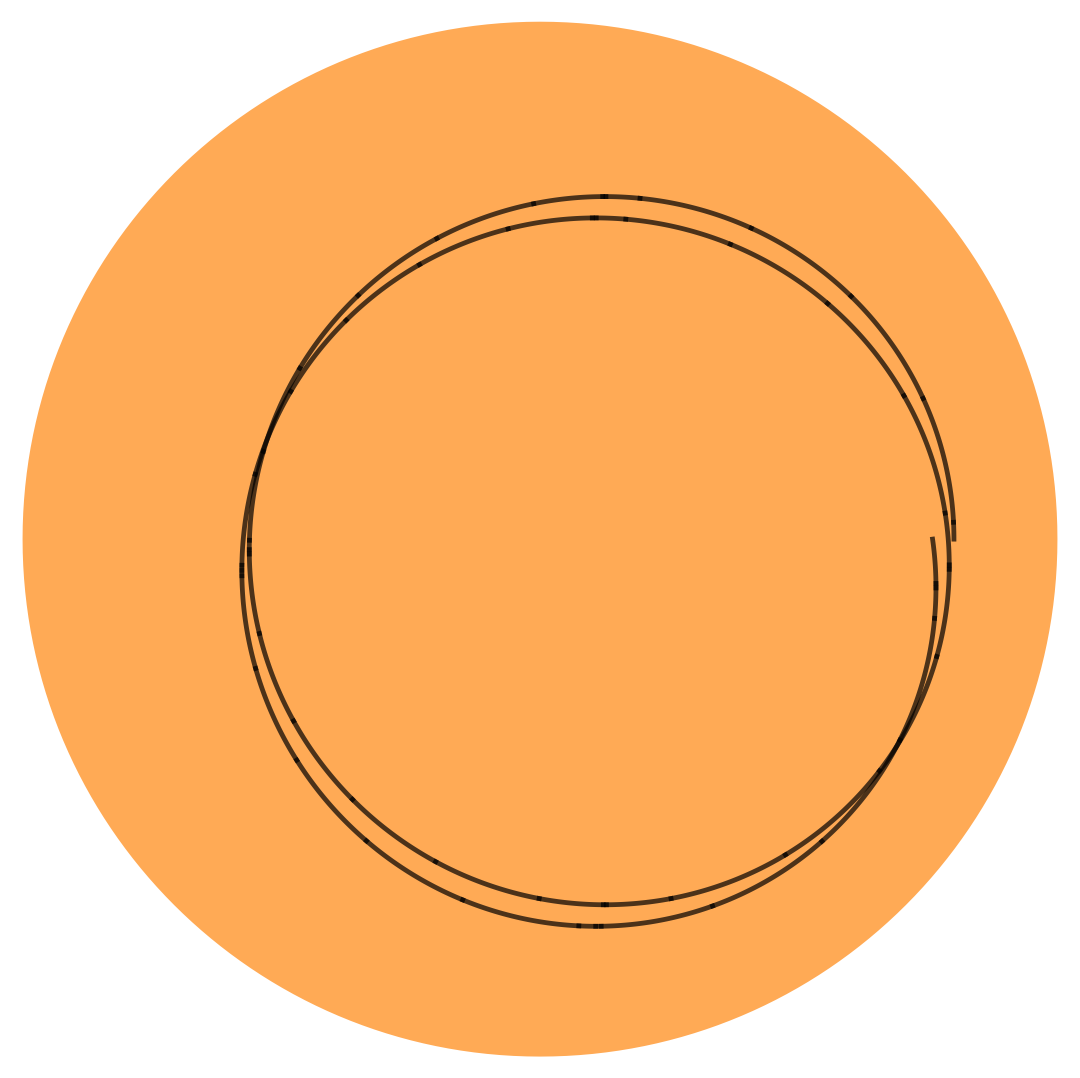}
     \includegraphics[scale=0.22]{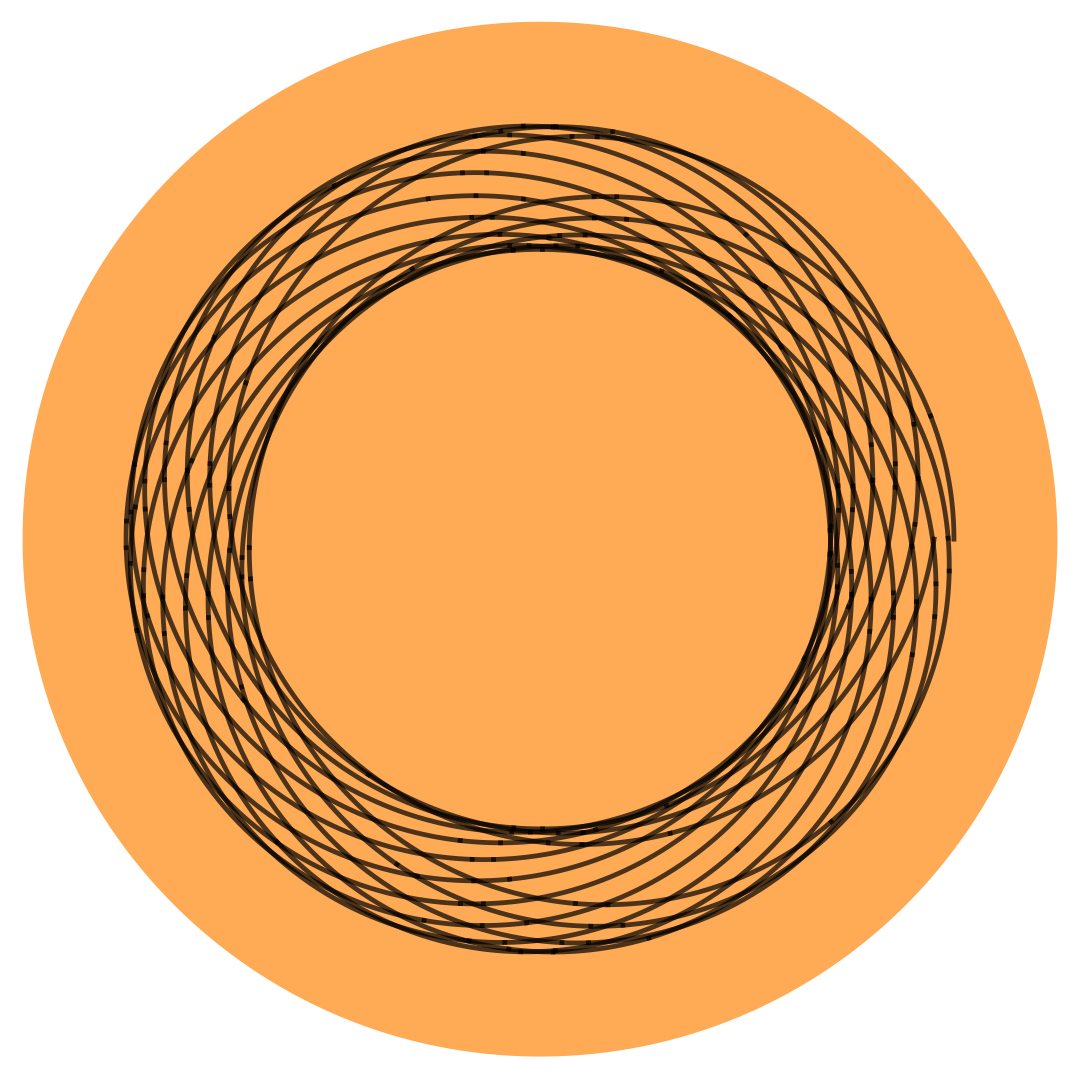}
    \caption{Open orbit of the PBH in the bulk of the star corresponding to initial conditions $r(0) = 4R/5$ and $v(0) = 440\, {\rm km\, s^{-1}}$. The left plot depicts two complete revolutions ($\Delta\varphi = 4\pi$) while the right plot illustrates a series of fifteen complete revolutions of the PBH inside the star.}
    \label{fig_1_44}
\end{figure}

We start by analyzing some solutions in the interior of the star. As shown in Fig.~\ref{fig_1_48p81}, initial conditions such that $r(0)=\frac45 R$ and $v(0) = 488.18\, {\rm km\, s^{-1}}$ lead to a stable, closed trajectory. 
This figure shows one hundred complete revolutions of the PBH around the center of the star. If we slightly modify these initial conditions, the outcome will be an open trajectory. For instance, when we keep the same value for the initial distance but decrease the initial velocity to $v(0) = 440\, {\rm km\, s^{-1}}$, we obtain an open trajectory that evolves in time as illustrated in Fig.~\ref{fig_1_44}, where the integration covers two (the left panel) and fifteen (right panel) complete revolutions of the open orbit.

\begin{figure}[h]
    \includegraphics[scale=0.3]{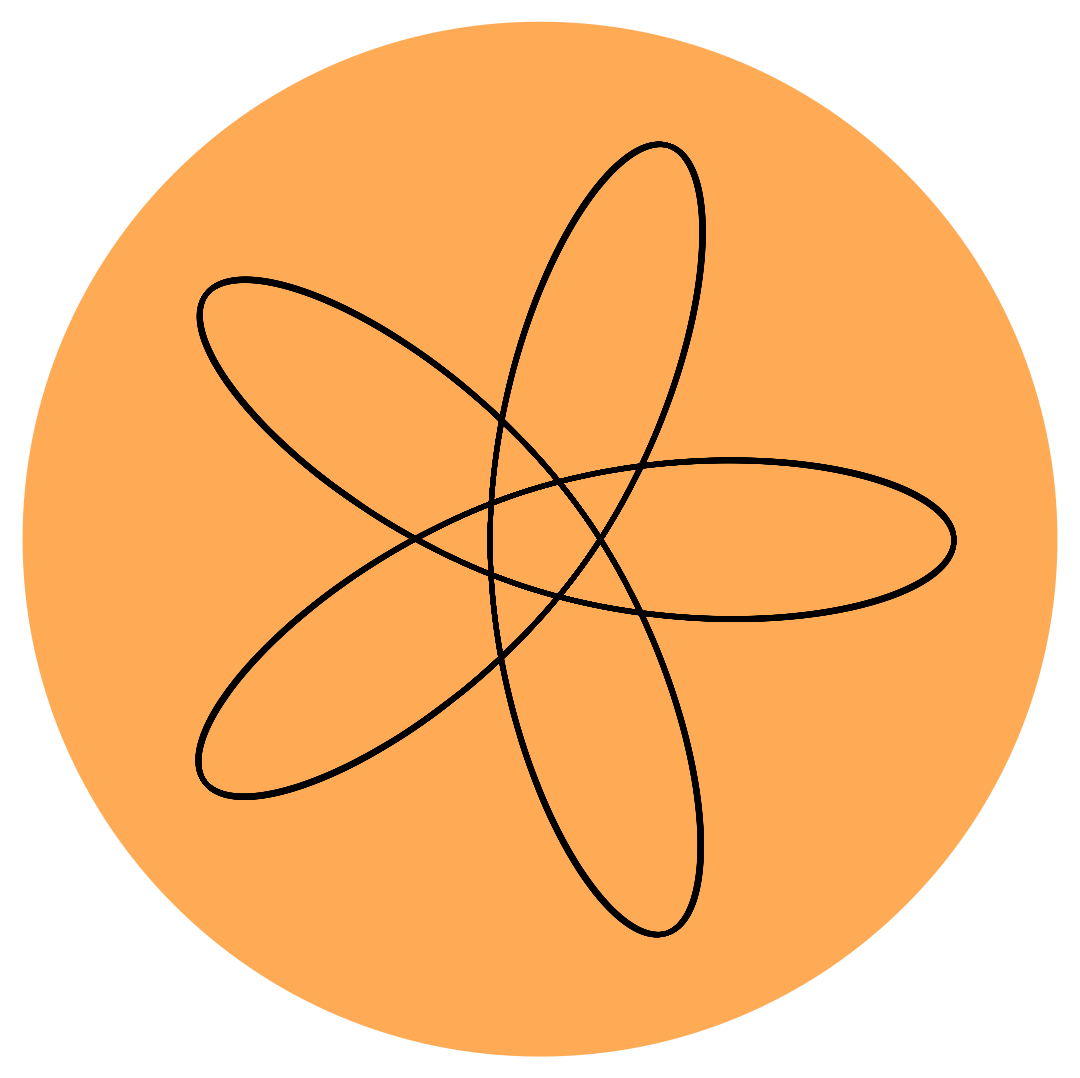}
    \caption{Closed orbit of the PBH in the bulk of the star corresponding to initial conditions $r(0) = 4R/5$ and $v(0) = 129\, {\rm km\, s^{-1}}$. The plot shows a set of 100 revolutions of the orbit, i.e. $\Delta \varphi = 200 \pi$.}
    \label{fig_5_12p9}
\end{figure}

Upon further modifying the value of the initial velocity, many other closed trajectories are possible. As another example, when we choose $v(0) = 129\, {\rm km\, s^{-1}}$, which yields a closed trajectory in which five symmetric laces (5 petals) appear, as depicted in Fig.~\ref{fig_5_12p9}. Once the symmetric five-petals closed-trajectory is formed, the complete orbit repeats itself each $\Delta\varphi = 6\pi$, without further changes.

\begin{figure}[h]
    \includegraphics[scale=0.22]{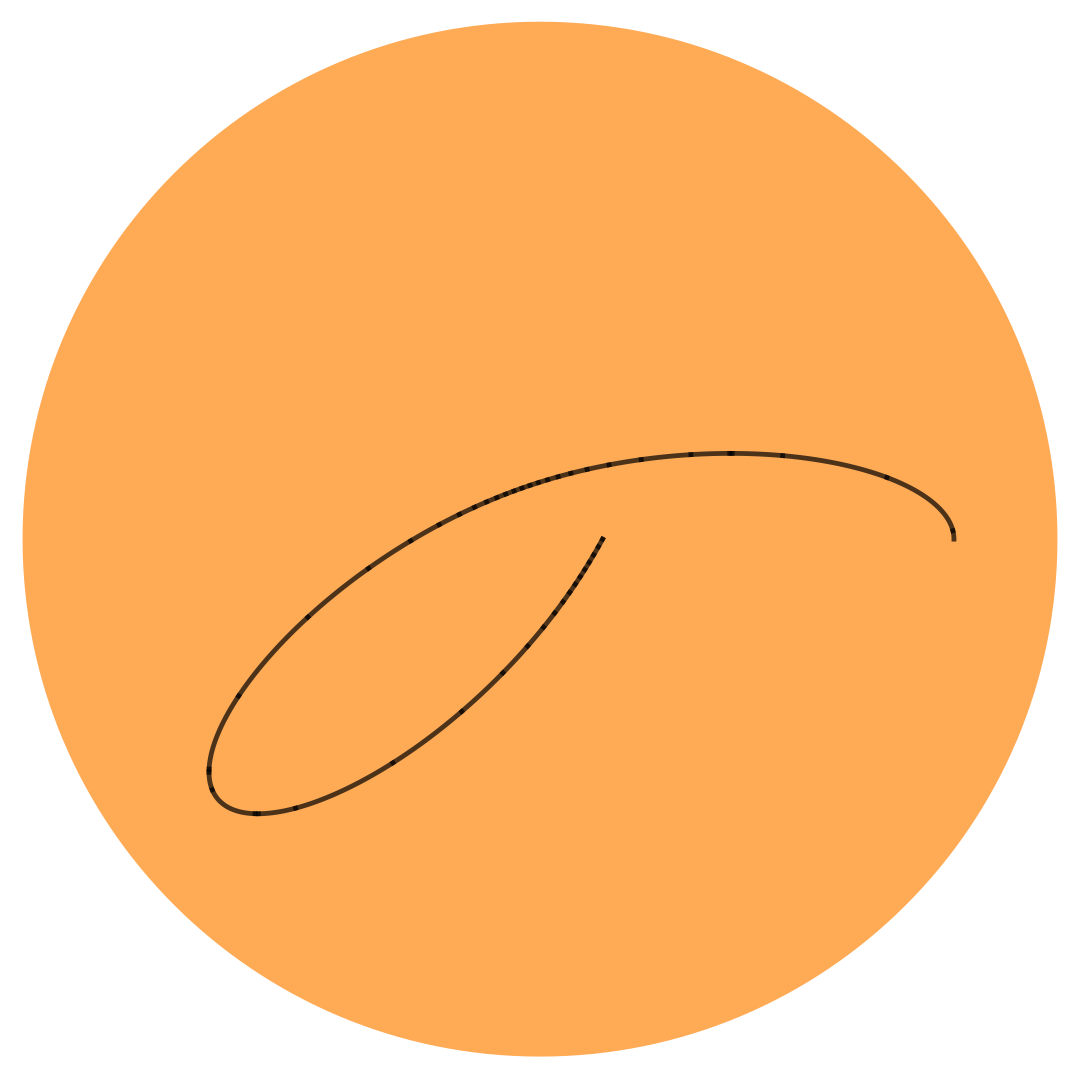}
    \includegraphics[scale=0.22]{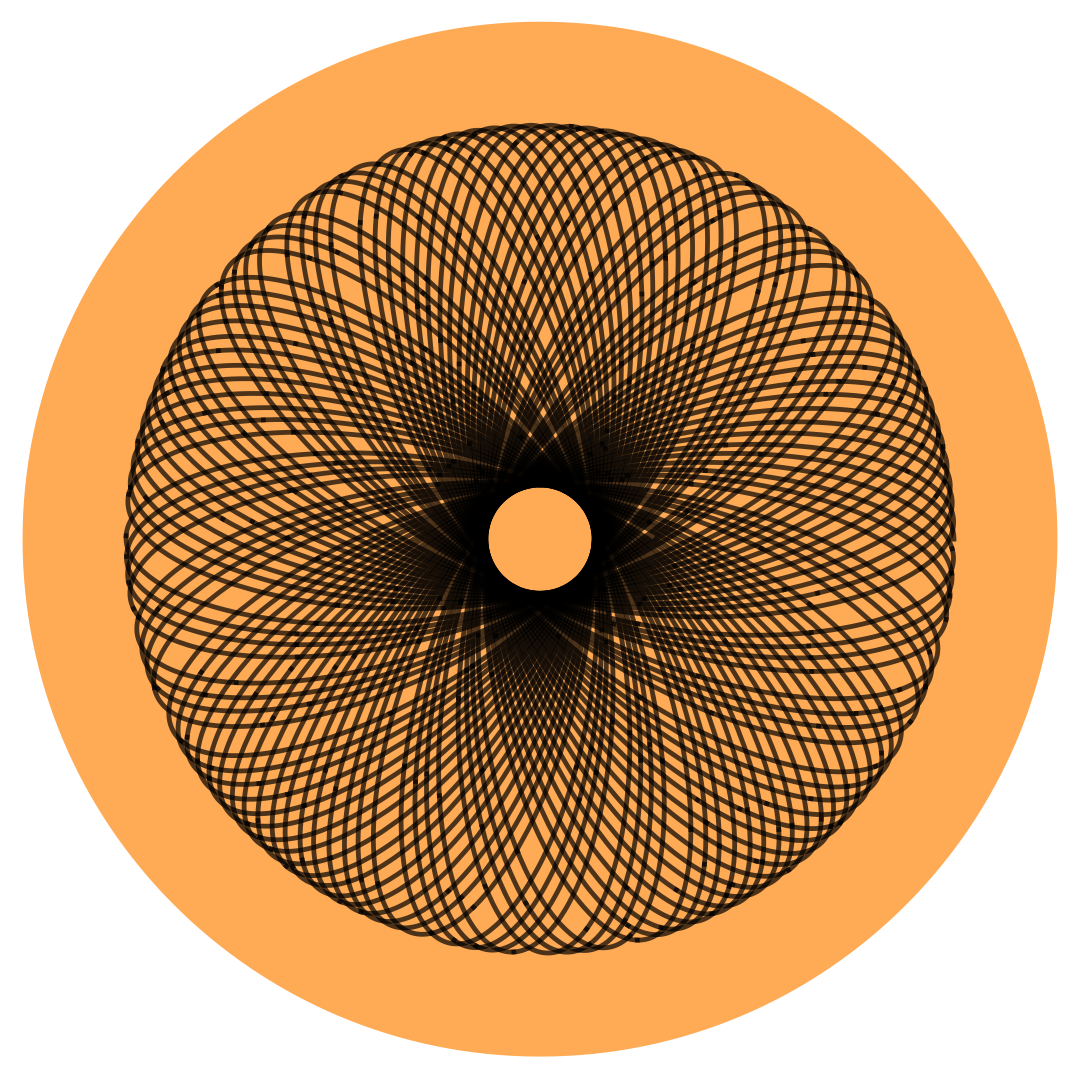}
    \caption{Open orbit of the PBH in the bulk of the star corresponding to initial conditions $r(0) = 4R/5$ and $v(0) = 139\, {\rm km\, s^{-1}}$. The left plot depicts only one revolution of the orbit ($\Delta \varphi = 2 \pi$), while the right plot depicts a series of 70 revolutions, i.e. $\Delta \varphi = 140 \pi$.}
    \label{fig_5_13p9_140turns}
\end{figure}

Increasing the initial velocity to $v(0) = 139\, {\rm km\, s^{-1}}$, the resulting trajectory will be open, as depicted in Fig.~\ref{fig_5_13p9_140turns}, where a series of seventy revolutions are shown in the right-hand plot. 

Several other examples of interior orbits are exhibited in the Appendix. 
The box in Fig.~\ref{fig_interior} illustrates some of the possible closed orbits related to $r(0) = \frac45 R$, while Fig.~\ref{fig_interior_open} depicts open orbits.

\begin{figure}[h]
    \includegraphics[scale=0.3]{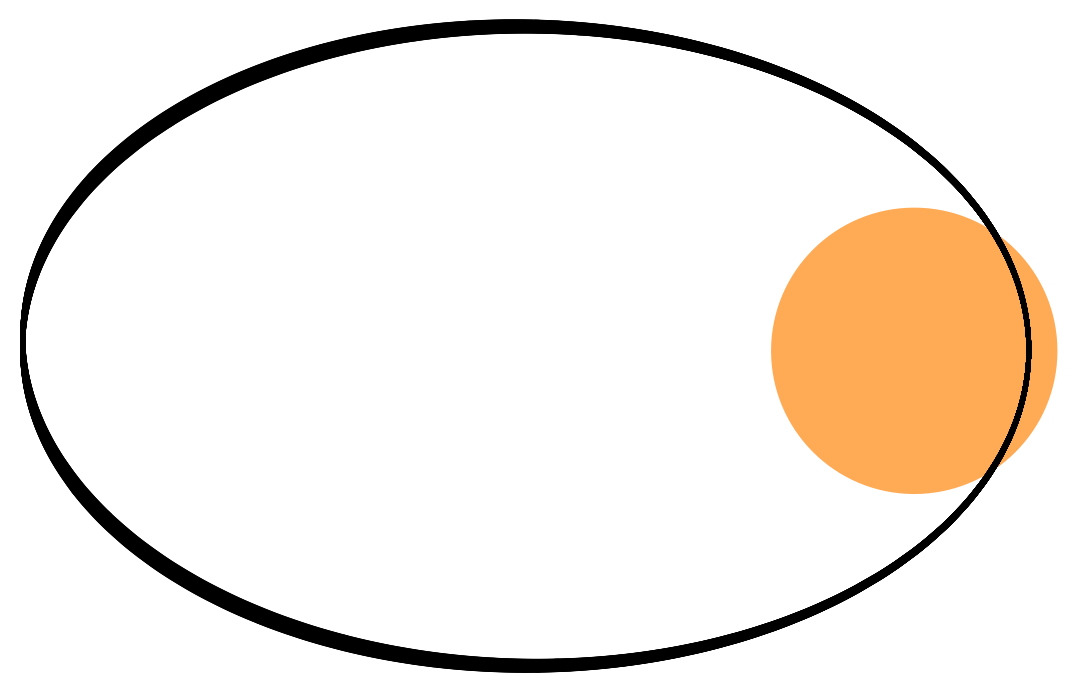}
    \caption{Open orbit of the PBH corresponding to initial conditions $r(0) = 4R/5$ and $v(0) = 650 \, {\rm km\, s^{-1}}$. For these initial conditions the orbit is semi-interior. In this example the plot includes a set of one hundred revolutions, i.e. $\Delta \varphi = 200 \pi$. The (Newtonian) perihelion precession effect can be seen by direct inspection.}
    \label{fig_out_65}
\end{figure}

If we keep increasing the initial velocity, the PBH will eventually evolve to an orbit that extends beyond the boundaries of the star, as shown in the example depicted in Fig.~\ref{fig_out_65}.
In this example we integrated through one hundred revolutions. The orbit is open and it is  possible to see the effect of perihelion evolution. This Newtonian effect is much larger than the one produced by relativistic corrections, which would not produce a visible shift within the time-span considered in these figures. The maximum distance that the particle achieves from the attraction center in this plot is about $6R$. Bounded trajectories with $r(0) = \frac45 R$ are only feasible for initial velocities smaller than $v(0) \approx 690 \, {\rm km\, s^{-1}}$; for larger initial velocities the PBH escapes to infinity. The same analysis can be employed for any other value for the internal initial position ($r<R$) of the particle. 

\subsection{Initial conditions such that $r(\varphi=0) > R$}
Let us now examine some solutions for a PBH with an initial position in the exterior region. For $r (\varphi = 0) > R$, there will exist two distinct cases: initial conditions such that the resulting orbits will be completely outside the star, and those leading to orbits that cross the star disk. The former case is not of our interest, since those orbits are the well-known Keplerian orbits. The latter is the subject of the subsequent analysis. 

\begin{figure}[h]
    \includegraphics[scale=0.3]{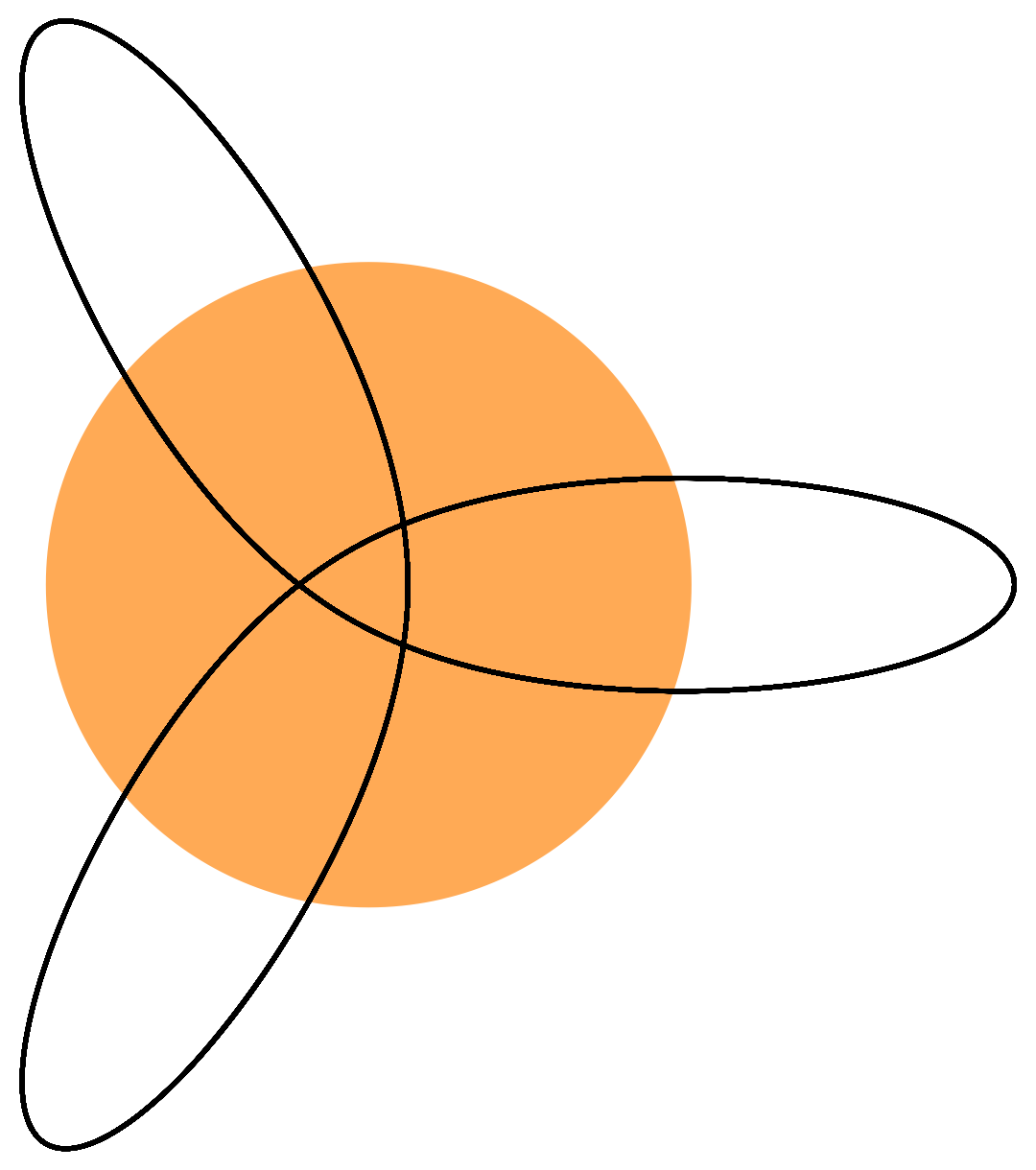}
    \caption{Closed orbit of the PBH with initial conditions $r(0)=2R$ and $v(0) = 71.115 \, {\rm km\, s^{-1}}$. The plot shows a set of two-hundred revolutions ($\Delta \varphi = 400 \pi$), which, in this case, corresponds to one hundred closed orbits.}
    \label{fig_3out_7p111}
\end{figure}

In order to investigate the motion of a test particle in this new case, we set its initial position at $r = 2R$ and adjust values for its initial velocity. If we choose $v(0) = 71.115 \, {\rm km\, s^{-1}}$, the resulting trajectory will be closed (within the time-span considered in the numerical integration, that is $\Delta\varphi = 400\pi$), as depicted in Fig.~\ref{fig_3out_7p111}. 
\begin{figure}[h]
    \includegraphics[scale=0.19]{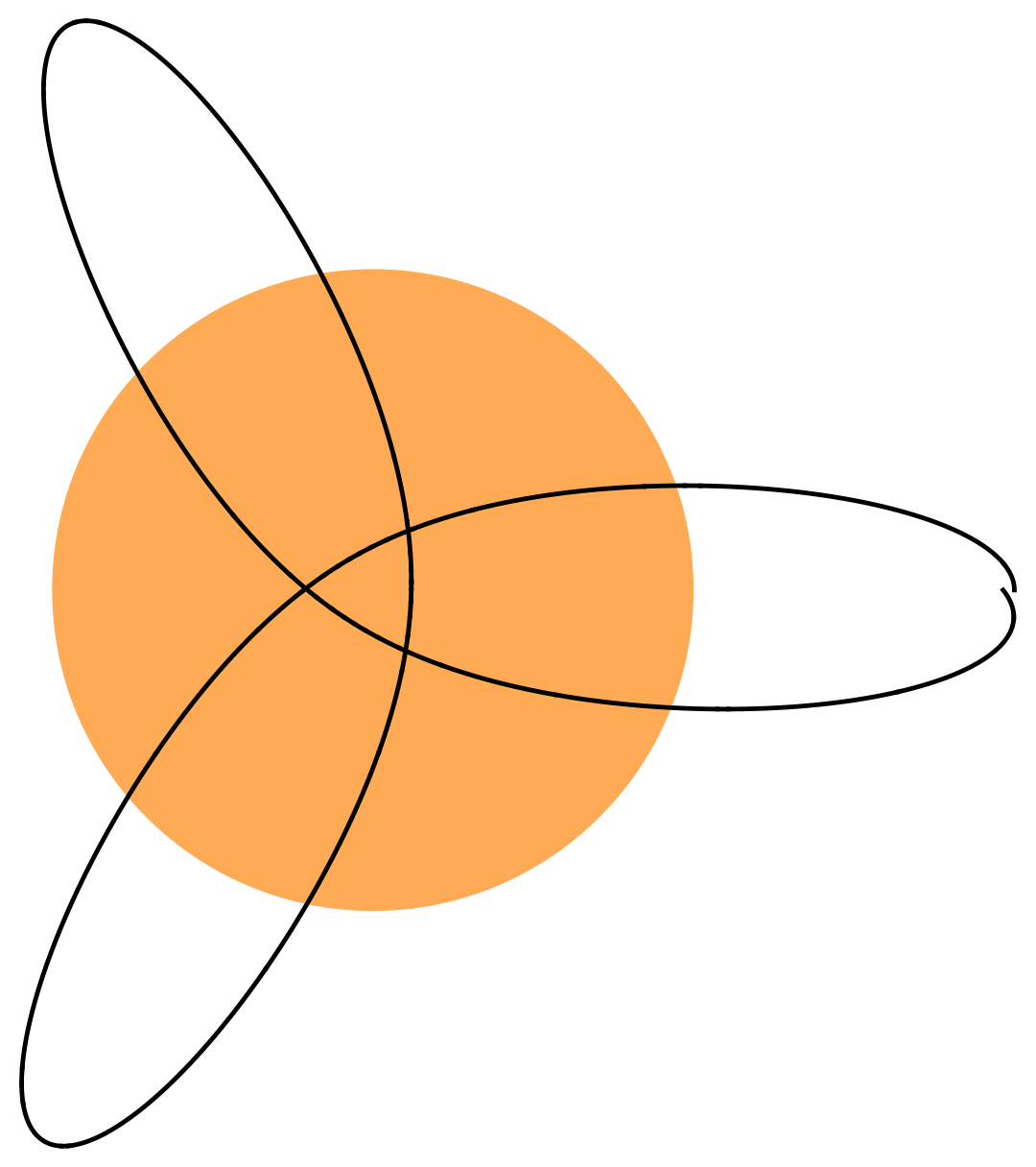}
     \includegraphics[scale=0.22]{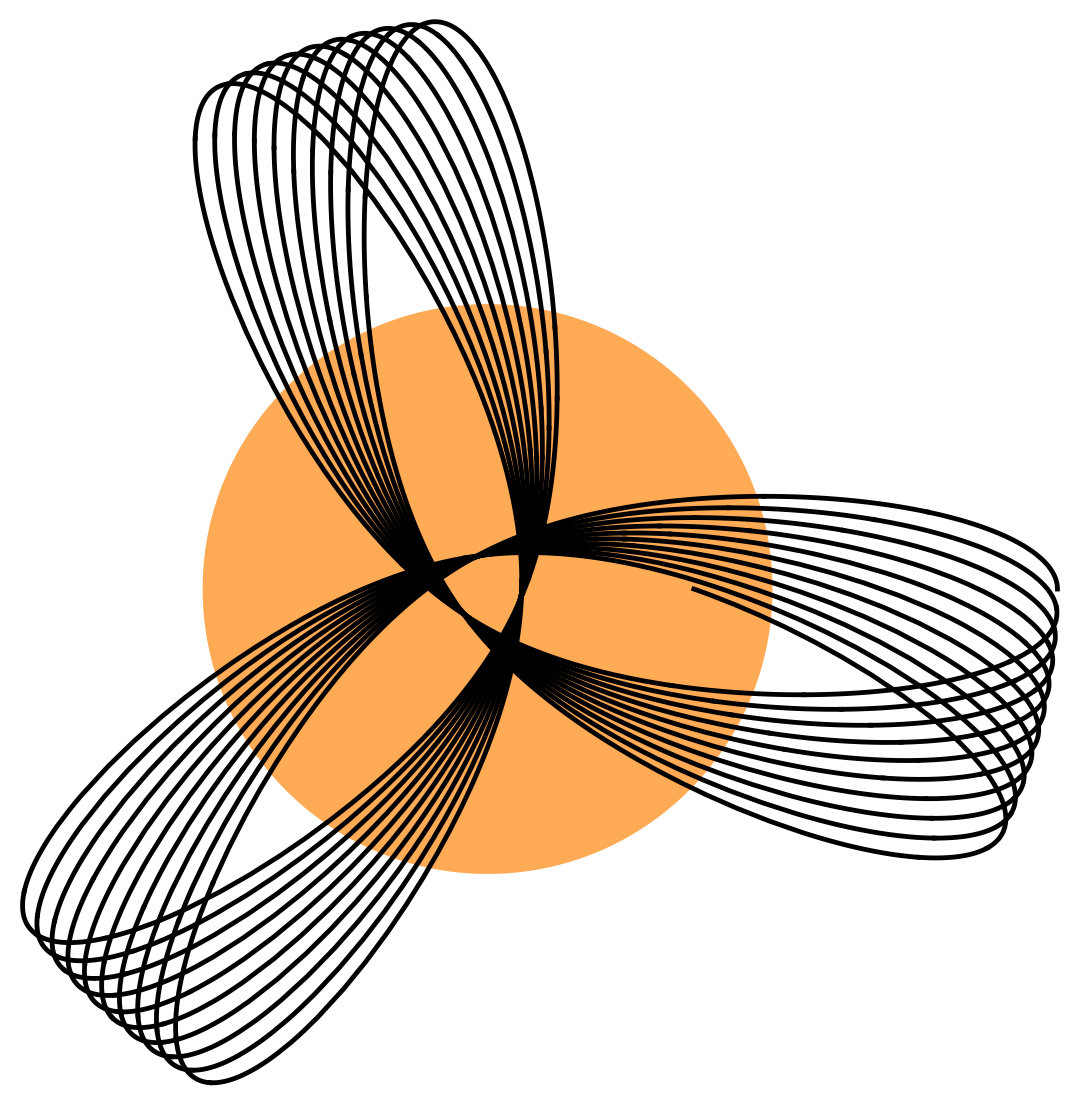}
    \caption{Open orbit of the PBH in the bulk of the star corresponding to initial conditions $r(0) = 2R$ and $v(0) = 70.115 \, {\rm km\, s^{-1}}$. The left plot depicts two complete revolutions ($\Delta\varphi = 4\pi$) while the right plot illustrates a series of twenty complete revolutions of the PBH around the center of  the star.}
    \label{fig_open_out_6p311}
\end{figure}
In this figure, the three-petals shape is completed at each $4\pi$ variation of $\varphi$. In a total of one hundred complete closed orbits ($\Delta\varphi = 400\pi$), no signal of further evolution of this shape is observed. (In fact, the value of the initial velocity can always be adjusted in order to keep the shape of the orbit closed when a longer time-span is allowed in the solution.) 
However, if we perturb the initial velocity, we obtain an open trajectory, as shown in Fig.~\ref{fig_open_out_6p311}. 

In the same way as discussed in Sec.~\ref{sec-interior}, there will be a certain number of solutions that yield closed orbits and many others describing open trajectories. The closed orbits form symmetric figures like the ones exhibited in Fig.~\ref{figs_out_closed} in the Appendix. 
A set of open trajectories obtained by changing the initial velocities implemented in the plots in Fig.~\ref{figs_out_closed} by $8 \, {\rm km\, s^{-1}}$ are presented in Fig.~\ref{figs_out_open}.

\begin{figure}[h]
    \includegraphics[scale=0.3]{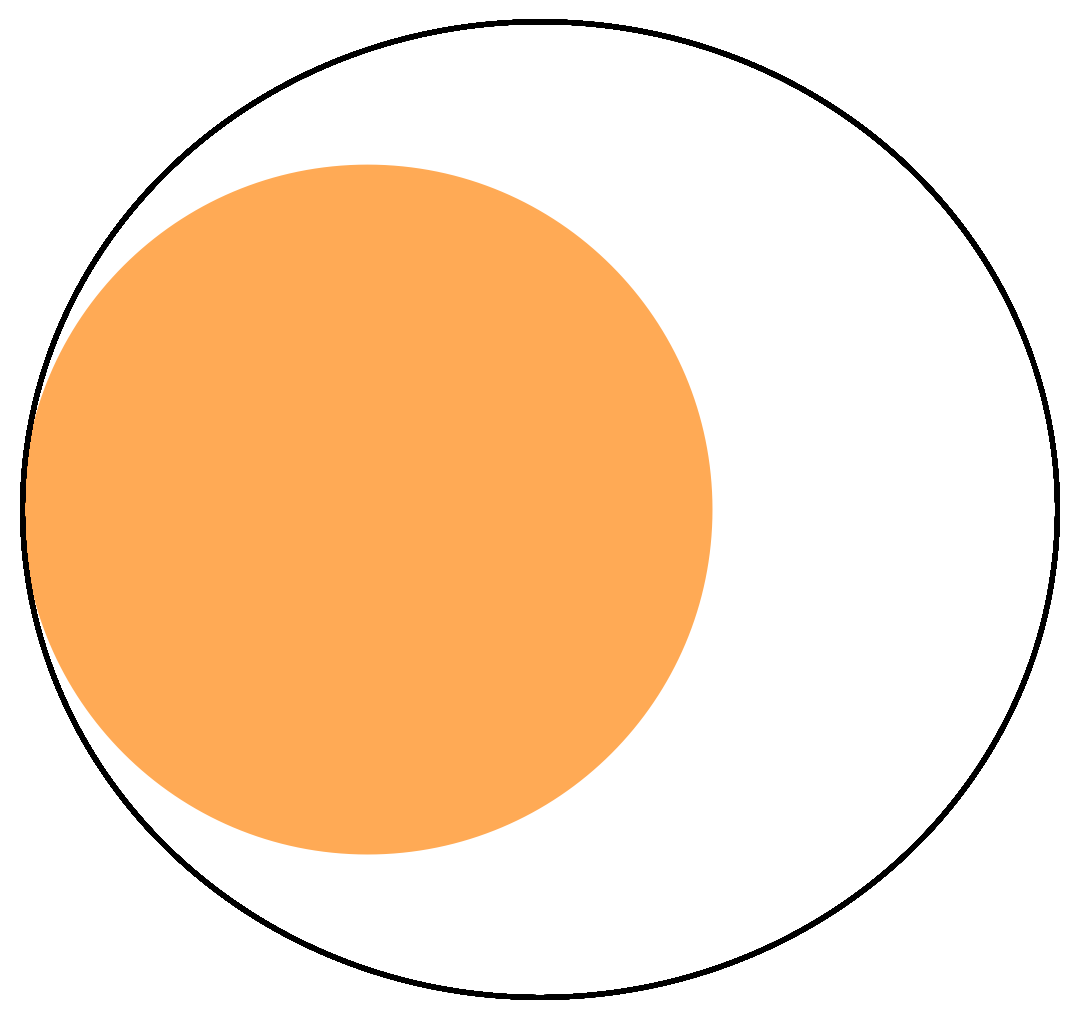}
    \caption{Closed orbit of the PBH with initial conditions $r(0)=2R$ and $v(0) = 252.14 \, {\rm km\, s^{-1}}$. The plot shows a set of one hundred complete revolutions ($\Delta \varphi = 200 \pi$).}
    \label{fig_1out_25p214}
\end{figure}

It is interesting to observe that the number of ``petals" in a closed orbit does not obey a direct proportion with the magnitude of the initial velocity. However, as we increase the initial velocity, there will be an extreme value for which the orbit will start to proceed completely outside the star, as 
illustrated in Fig.~\ref{fig_1out_25p214}. In this extreme case, the orbit touches the star disk at only one point, and for larger values of $v(0)$ the orbit will always be closed, as the Keplerian regime will be achieved (ignoring relativistic corrections).

\section{Discussion}
\label{final}
Primordial black holes (PBHs) are fascinating theoretical objects, which might help address the decades-long puzzle of dark matter \cite{Khlopov:2008qy,PhysRevD.87.023507,Smyth:2019whb,2020PhRvD.101l3514L,Carr_2021,10.21468/SciPostPhysLectNotes.48,Green:2020jor,Villanueva-Domingo:2021spv,Escriva:2022duf,Carr:2023tpt,Gorton:2024cdm}. If PBHs do exist and constitute a large fraction of the total dark-matter abundance, then they should be ubiquitous throughout the Universe. In particular, they should encounter stars with some regularity. Given such encounters, it is of interest to consider scenarios in which PBHs become gravitationally bound to a host star, in addition to considering how transient PBH flybys might affect the motion of visible objects \cite{Dror:2019twh,Li:2022oqo,BrownHeUnwin,Tran:2023jci,Bertrand:2023zkl,Cuadrat-Grzybowski:2024uph}.

In our analysis, we have identified both closed and open orbits for a small-mass PBH around the center of a Solar-type star. We have identified such orbits for initial conditions such that the PBH begins either inside or outside the star. By taking into account the nontrivial mass-density distribution within the star, $\rho (r)$, we have identified such effects as precession of the PBH perihelion, which arise strictly from the Newtonian dynamics. 

We have numerically evolved the PBH trajectories up to ${\cal O} (100)$ complete orbits for a range of initial conditions. For PBH orbits near a Solar-type star, with $M \sim M_\odot$, $R \sim 10^6 \, {\rm km}$, and $v(0) \sim {\cal O} (100 \, {\rm km \, s^{-1}})$, the time-scale of a typical orbit is $t_{\rm orbit} \sim R / v(0) \sim 10^4 \, {\rm s}$, so that our simulations correspond to $t_{\rm sim} \sim 100 \, t_{\rm orbit} \sim 10^6 \, {\rm s}$. This is a sufficiently short time-scale for which we may self-consistently neglect accretion onto the PBH as well as energy loss via dynamical friction \cite{Ostriker_1999} or hydrodynamic dragging effects \cite{Bellinger_2023}. Likewise, by considering PBH motion near a Solar-type star rather than near a more dense object such as a neutron star \cite{2024PhRvD.109f3004B,2024arXiv240201838B,2024arXiv240408735B,2024arXiv240408057C}, relativistic corrections to the Newtonian dynamics remain strongly suppressed.

Over exponentially longer time-scales than those considered here, small-mass PBHs can become captured by a host star and settle in the star's core \cite{Oncins:2022djq,Bellinger_2023,Caplan:2023ddo}. To estimate the time-scale on which dynamical friction becomes non-negligible, we may use the scaling in Ref.~\cite{2024arXiv240408057C}: $t_{\rm dyn} \sim 10^{-1} (M / m) \, t_{\rm orbit}$. For a PBH at the lower end of the allowed dark-matter mass range ($10^{17} \, {\rm g})$ encountering a Solar-type star along the trajectory depicted in Fig.~\ref{fig_open_out_6p311}, dynamical friction would become relevant on a time-scale $t_{\rm dyn} \sim 10^{12}$ years, longer than the current age of the universe. For a PBH at the upper end of the dark-matter mass range ($10^{23} \, {\rm g})$, we expect $t_{\rm dyn} \sim 10^6$ years, consistent with the simulations in Ref.~\cite{Bellinger_2023}.

A feature of the two-body system that could affect the PBH trajectories on the time-scales we consider here is rotation of the star. If the star rotates sufficiently rapidly, its shape will deviate from sphericity, such that $M \rightarrow M(r, \varphi)$. A non-spherical distribution of the star's mass, in turn, would affect PBH orbital trajectories, such as causing deviations from planar motion. (The material within stars typically undergoes {\it differential} rotation \cite{1998ApJ...505..390S}, which could drive even more complicated dynamics for an orbiting PBH.) We leave for further research the question of whether such non-spherical stellar mass distributions could eliminate the closed-orbit trajectories identified here.

In the absence of stellar rotation, we identified several sets of closed orbits. Naturally, the trajectory of a test particle (such as the small-mass PBHs we consider here) depends on the assumed initial conditions for its position and velocity, which generally lead to trajectories that never close. However, for some specific choices of initial conditions, closed shapes can be possible after a certain number of full rotations of the orbit. Some of these solutions are depicted in Figs.~\ref{fig_interior} and \ref{figs_out_closed}. 

\begin{figure}[!htb]
    \centering
        \includegraphics[width=0.322\linewidth]{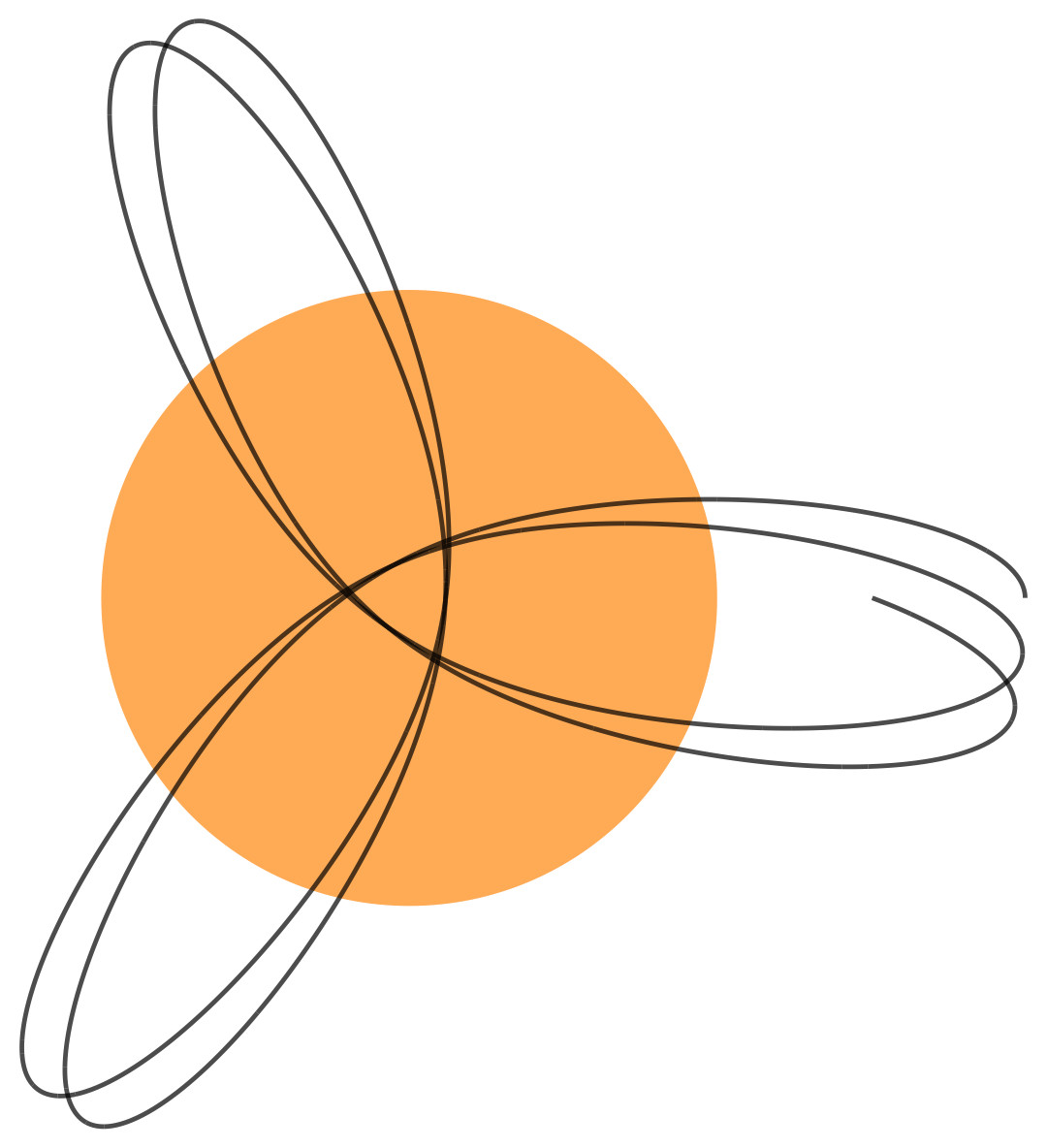}
        \includegraphics[width=0.322\linewidth]{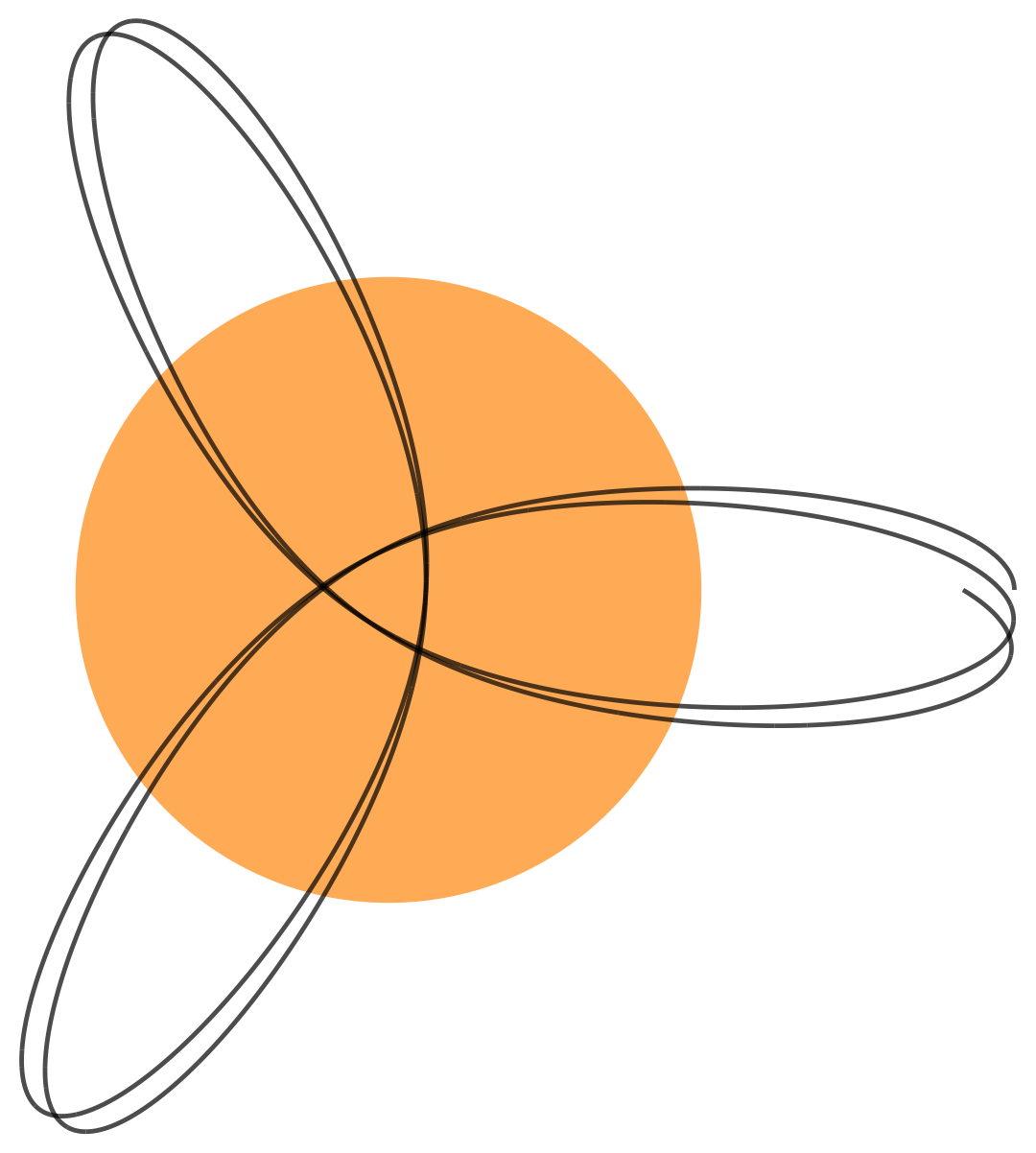}
        \includegraphics[width=0.322\linewidth]{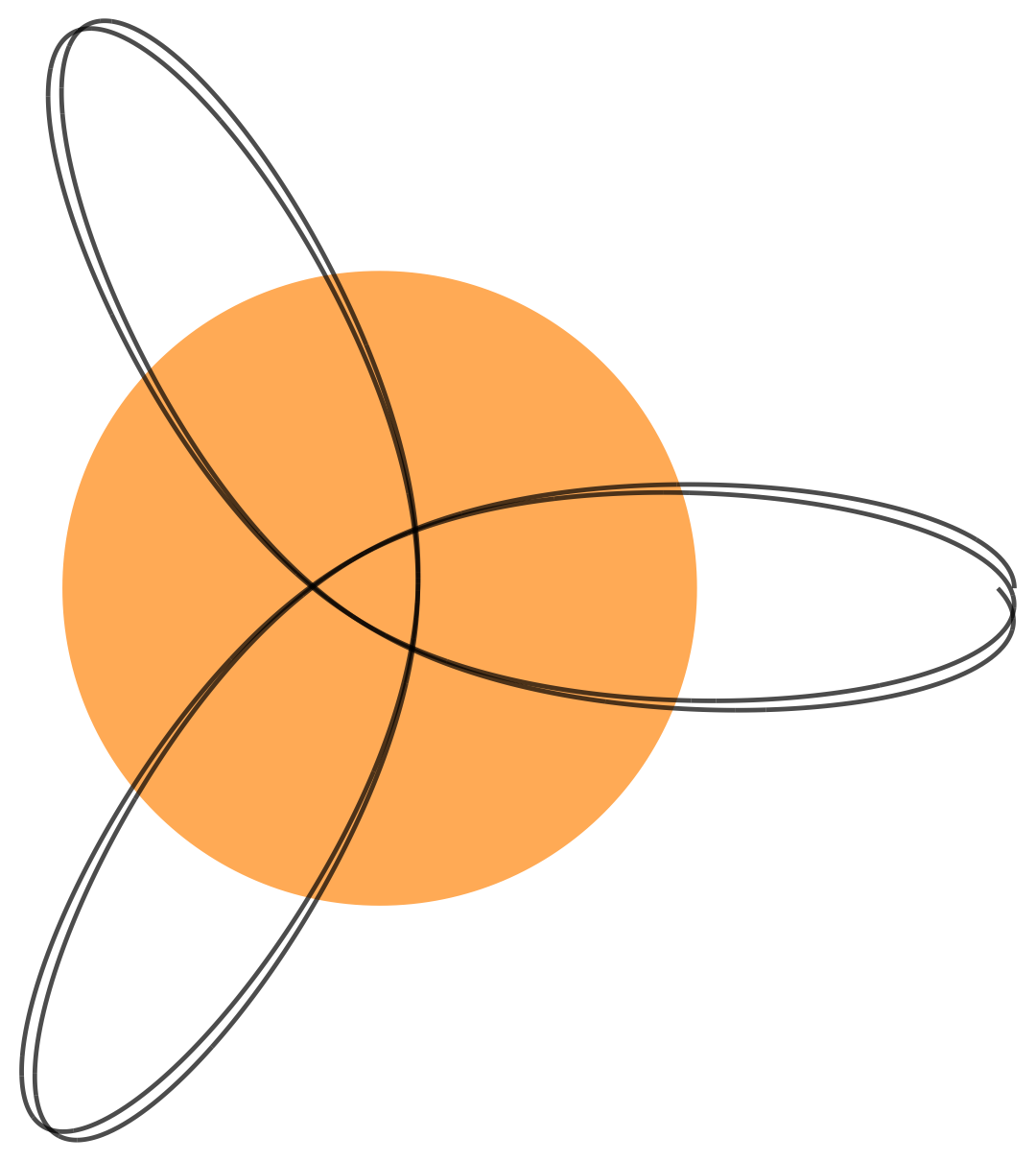}
        \\
        \includegraphics[width=0.322\linewidth]{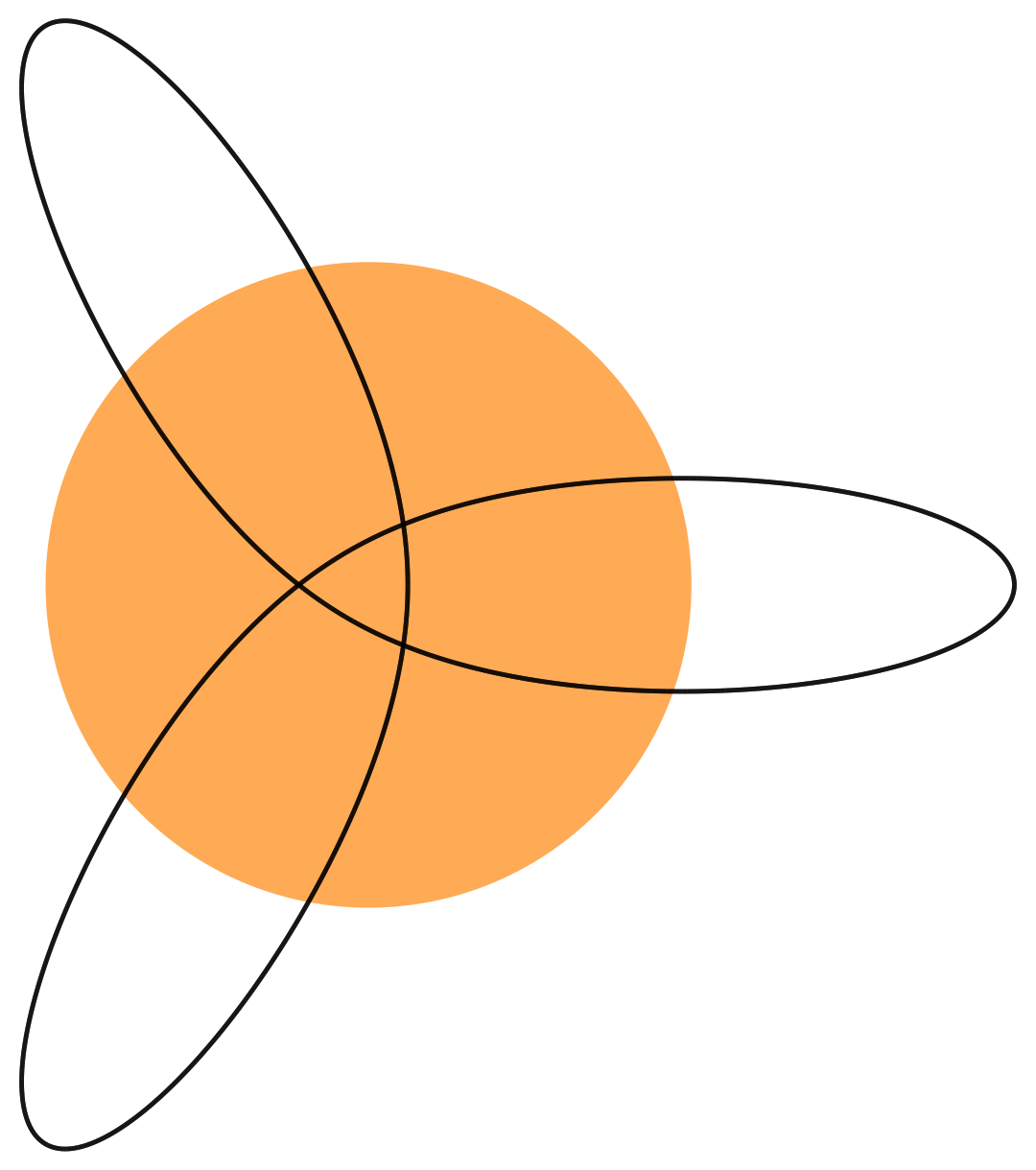}
        \includegraphics[width=0.322\linewidth]{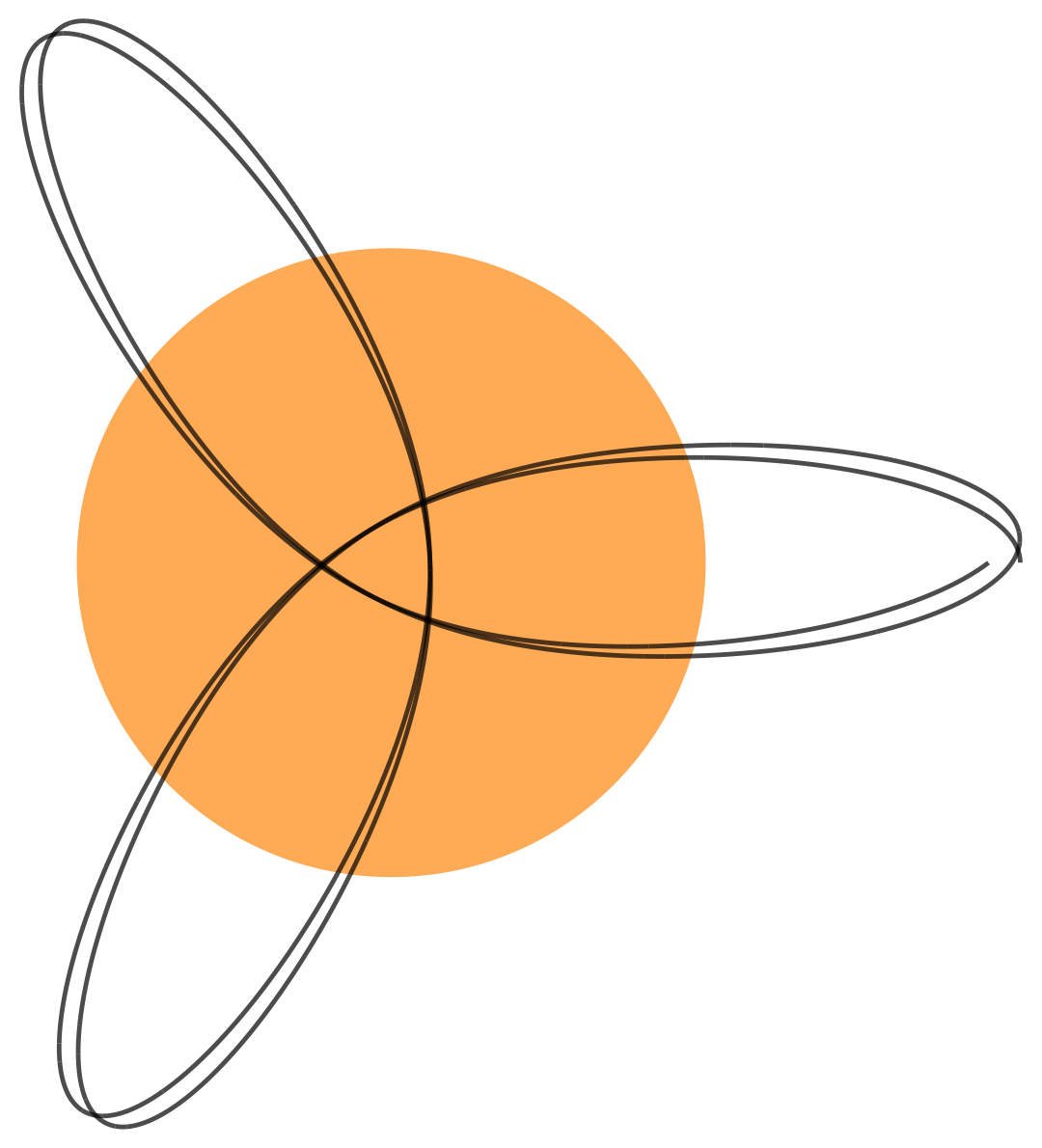}
        \includegraphics[width=0.322\linewidth]{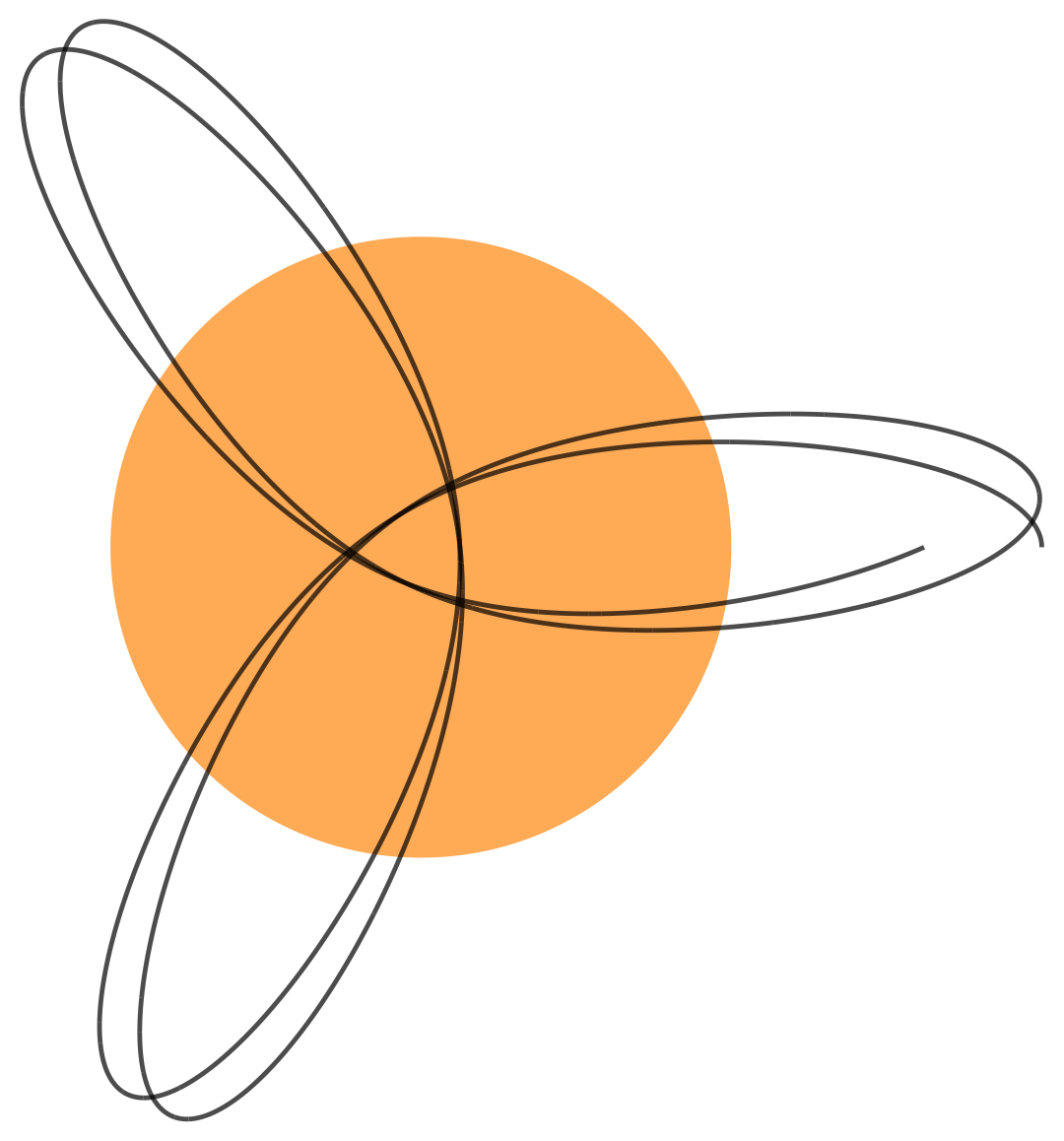}
    \caption{The occurrence of a closed trajectory is here illustrated as the initial condition is slightly changed. In this figure, only four full rotations are shown ($\Delta\varphi = 8\pi$).}
    \label{fig3p}
\end{figure}

Fig.~\ref{fig3p} illustrates how a closed orbital trajectory is achieved by smoothly varying the initial conditions. In this figure only six intermediate solutions were selected, among infinite possibilities. 
The initial position of the particle was kept fixed, $r(\varphi=0)=2R$, and the initial velocity $v(\varphi=0)$ was chosen to vary in the window (69 - 73)$\, {\rm km\,  s^{-1}}$. For the initial choice of $v(0)=69\, {\rm km \,  s^{-1}}$ (left-top panel in this figure) the trajectory is clearly open. However, as $v(0)$ increases, the trajectory evolves, eventually passing by one configuration exhibiting a closed trajectory: the left-bottom panel, for which $v(0)\approx 71.115\, {\rm km \,  s^{-1}}$. In this specific case the closed path is formed after two full rotations ($4\pi$ radians). 

Similar analysis can be repeated for all closed trajectories shown in the figures discussed throughout the paper: each closed solution is surrounded by an infinite set of open solutions. Thus the set of closed trajectories is discrete and of measure zero within the continuous set of open solutions. 

For the closed orbits identified here, if a perturbation is introduced, such as the leading-order relativistic correction, a closed trajectory will revert to an open solution, which would be visible with a sufficiently long integration time for the numerical simulations. However, by further adjusting the initial conditions, a closed solution can be restored even in such cases. In other words, if the leading relativistic correction is incorporated, one can still find a discrete set of closed solutions for the trajectory of the test particle. 

These results suggest that a rare but rich class of closed solutions for the orbits of test particles in systems consisting of a source mass that depends on the position of the test particle may exist in astrophysical environments.
\vspace{0.2in}

\begin{acknowledgments}
We thank Elba Alonso-Monsalve, Lucas Martins Barreto Alves, Thomas Baumgarte, Earl Bellinger, Matt Caplan, Lucas Ruiz, and Stuart Shapiro for helpful discussions.
V.~A.~D.~L. is supported in part by the Brazilian research agency CNPq
(Conselho Nacional de Desenvolvimento Cient\'{\i}fico e Tecnol\'ogico)
under Grant No. 302492/2022-4. Portions of this work were conducted in MIT's Center for Theoretical Physics and supported in part by the U.~S.~Department of Energy under Contract No.~DE-SC0012567.
\end{acknowledgments}

\appendix
\section*{Appendix: Remarks on the shape of the orbits}
\label{setoffigs}
The expression for the energy conservation law, as stated in Eq.~(\ref{energy}), can be conveniently expressed as $\dot r^2= (2/m)(E_\textsc{t} - V)$, where $V$ represents the effective potential energy $V(r) = m \Phi(r) + m\ell^2/(2r^2)$. The term proportional to $\ell$ accounts for the centrifugal contribution arising from the particle's angular motion and $\Phi(r)$ is the interaction potential. Given $\Phi(r)$, bounded solutions exist if $E_\textsc{t} - V > 0$, which determines a limited region where $r_{\rm in} \le r \le r_{\rm ext}$. Here $r_{\rm in}$ and $r_{\rm ext}$ are turning points and represent the internal and external radii of a ring-shaped region within which the paths evolve. Their values depend on the initial conditions governing the particle's motion. The angular distance $\delta \varphi$ between a maximum position $r=r_{\rm ext}$ (apocenter) and its nearest minimum $r=r_{\rm in}$ (pericenter) along the particle trajectory can be formally obtained as \cite{1978mmcm.book.....A}
$$
\delta \varphi = \int_{r_{\rm in}}^{r_{\rm ext}} \frac{\ell/r^2}{\sqrt{(2/m)(E_\textsc{t}-V})}\dd r.
$$
This is the angle between two adjacent apsidal vectors, which are the vectors that locate the turning points of an orbit. In fact, it can be shown that the orbit is invariant under reflection about the apsidal vectors \cite{2002clme.book.....G}. This expression can be used to classify the solutions: an orbit will be closed if $\delta\varphi = 2\pi (p/q)$, where $p$ and $q$ are integers, otherwise it will be open and dense in the ring-shaped region. 
There are only two potentials for which {\em all} bounded orbits are closed: the harmonic $\Phi(r) \sim r^2$ and the Newtonian $\Phi(r) \sim r^{-1}$ potentials. This result is guaranteed by the well-known Bertrand's theorem \cite{2002clme.book.....G}.

The gravitational potential related to the mass distribution examined in Sec.~\ref{workable} can be conveniently expressed in the form $\Phi(r) = -G\mu(r)/r$, where $\mu(r)$ is a mass function defined by
\begin{widetext}
\begin{align}
\frac{\mu(r)}{M} = \left(-\frac{7 r^9}{2 R^9}+\frac{27 r^8}{R^8}-\frac{90 r^7}{R^7}+\frac{168 r^6}{R^6}-\frac{189 r^5}{R^5}
+\frac{126 r^4}{R^4}-\frac{42 r^3}{R^3}+\frac{9 r}{2 R}\right) \Theta \left(1-\frac{r}{R}\right)+ \Theta \left(\frac{r}{R}-1\right).
\nonumber
\end{align}
\end{widetext}
%
%
Notice that $M(r) = -r^2 \dd [\mu(r)/r]/\dd r$ in both regions, $r<R$ and $r>R$, as expected. 

\begin{figure}[h]
    \includegraphics[scale=0.4]{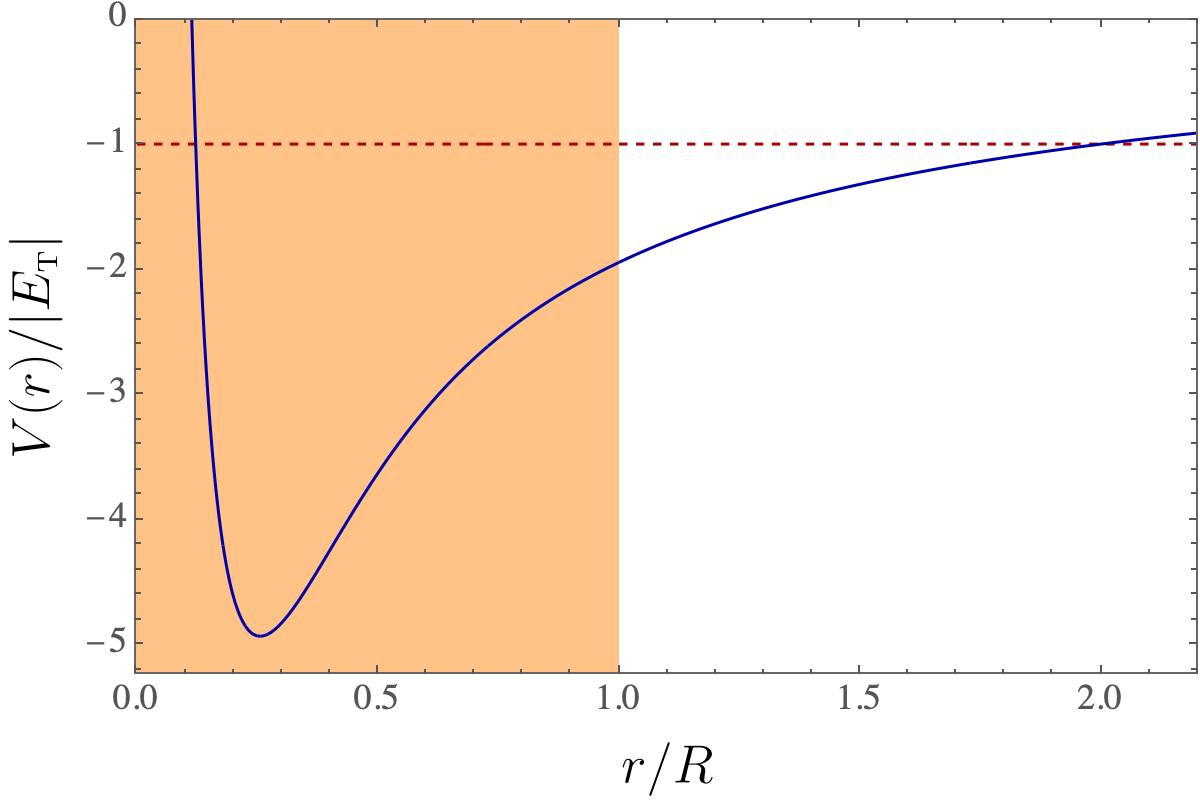}
    \caption{This figure illustrates the effective potential $V(r)$ obtained for the initial conditions used in Fig.~\ref{fig_3out_7p111}. The orange region represents the interior of the star of radius $R$. The turning points of the orbit are determined by the crossing of the curve with the 
    horizontal dashed line, which represents the total energy of the particle.}
    \label{effective}
\end{figure}

In order to show a specific example, the effective potential energy corresponding to the three petal orbit explored in Fig.~\ref{fig_3out_7p111} is depicted in Fig.~\ref{effective}. In such configuration $\delta\varphi = 2\pi / 3.$

With this prescription, closed orbits can be found for given initial conditions, as described through the text, and are further investigated in the examples below.

\begin{figure}[!htb]
 \centering
    \begin{minipage}{\textwidth}
    \begin{tabularx}{\textwidth}{>{\centering\arraybackslash}X|>{\centering\arraybackslash}X|>{\centering\arraybackslash}X|>{\centering\arraybackslash}X}
            \multicolumn{4}{c}{$r(0)=4R/5$} \\
            \hline
            \includegraphics[width=\linewidth]{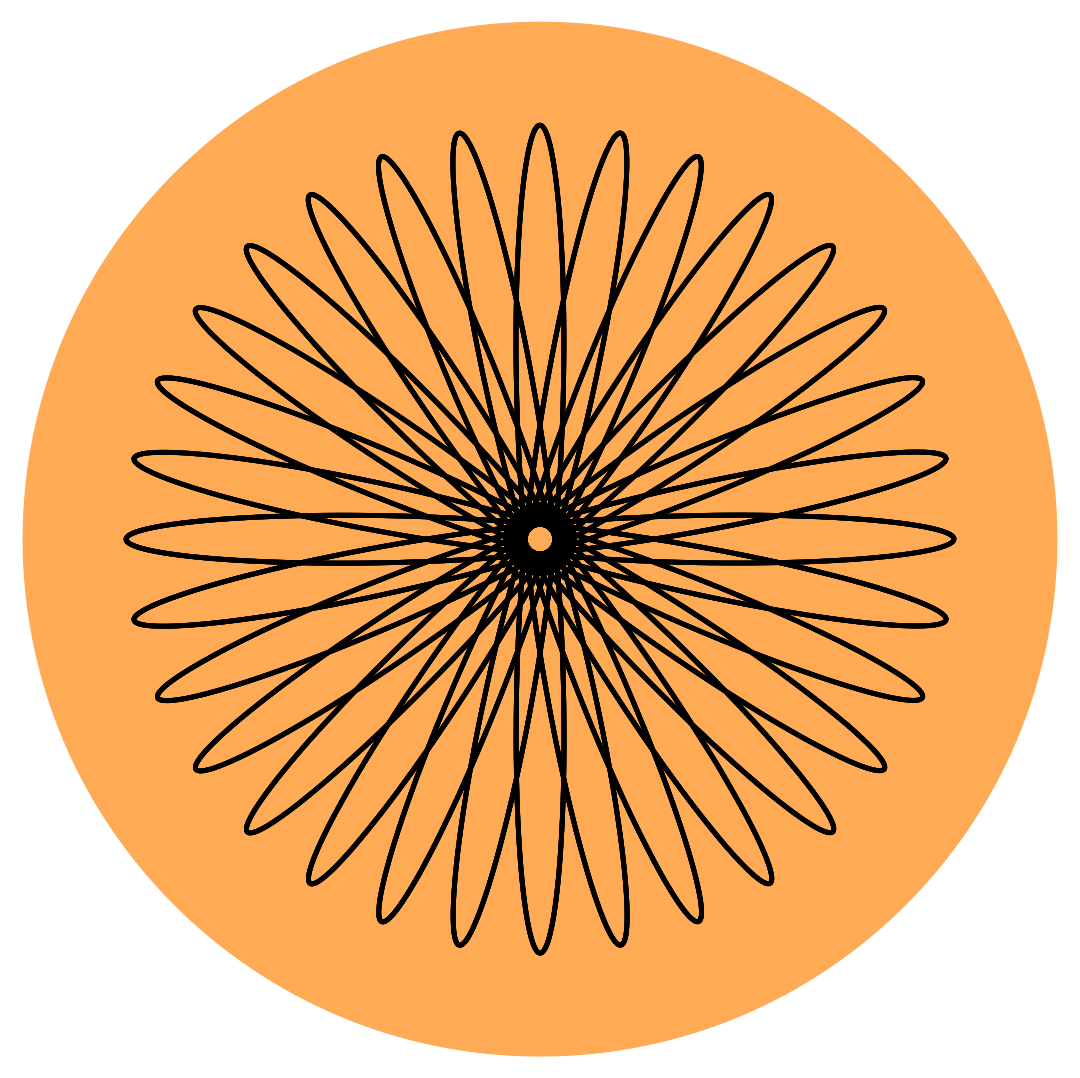} {\small $v(0)=39.534\, {\rm km\, s^{-1}}$}
            & \includegraphics[width=\linewidth]{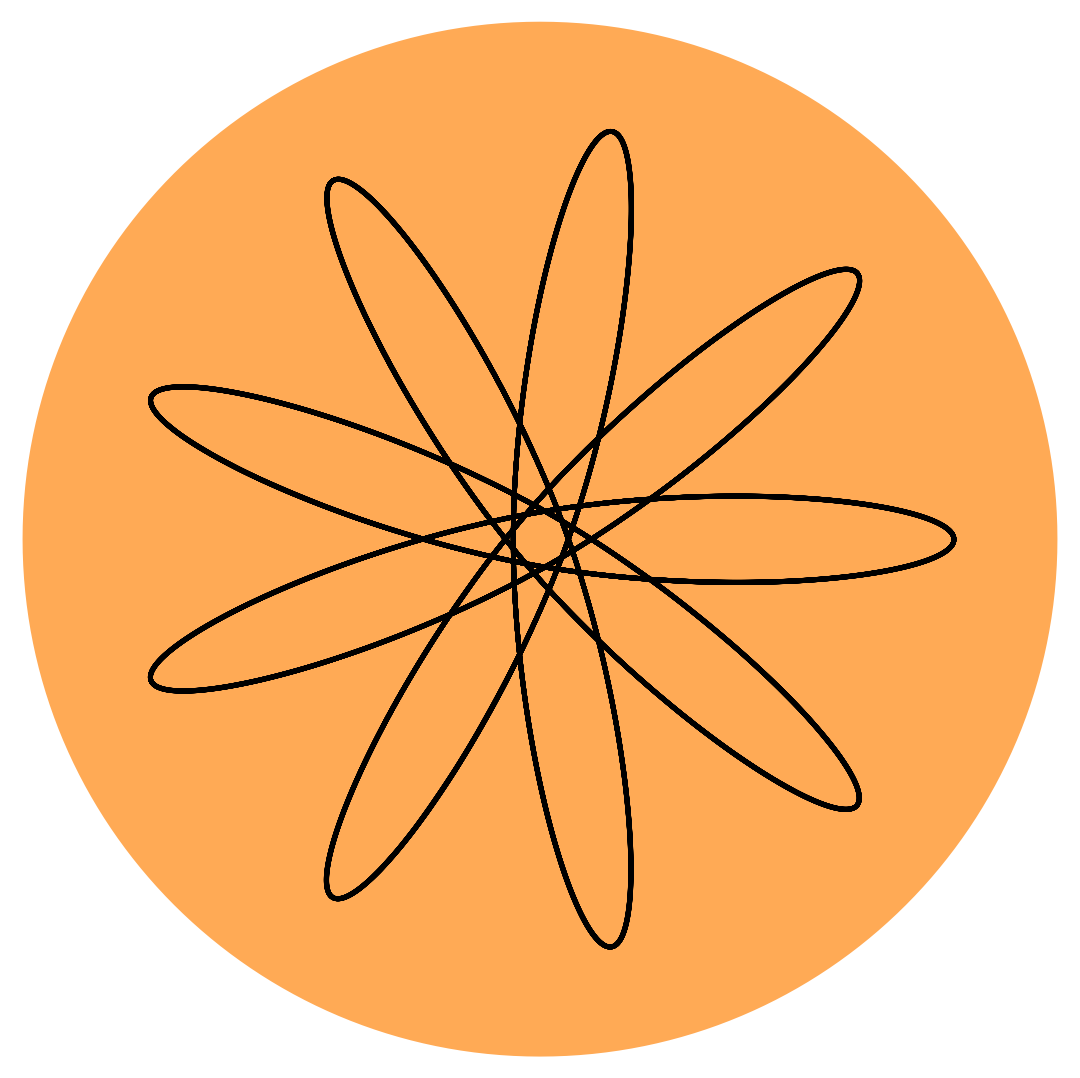} {\small $v(0)=70.925\, {\rm km\, s^{-1}}$}  
            & \includegraphics[width=\linewidth]{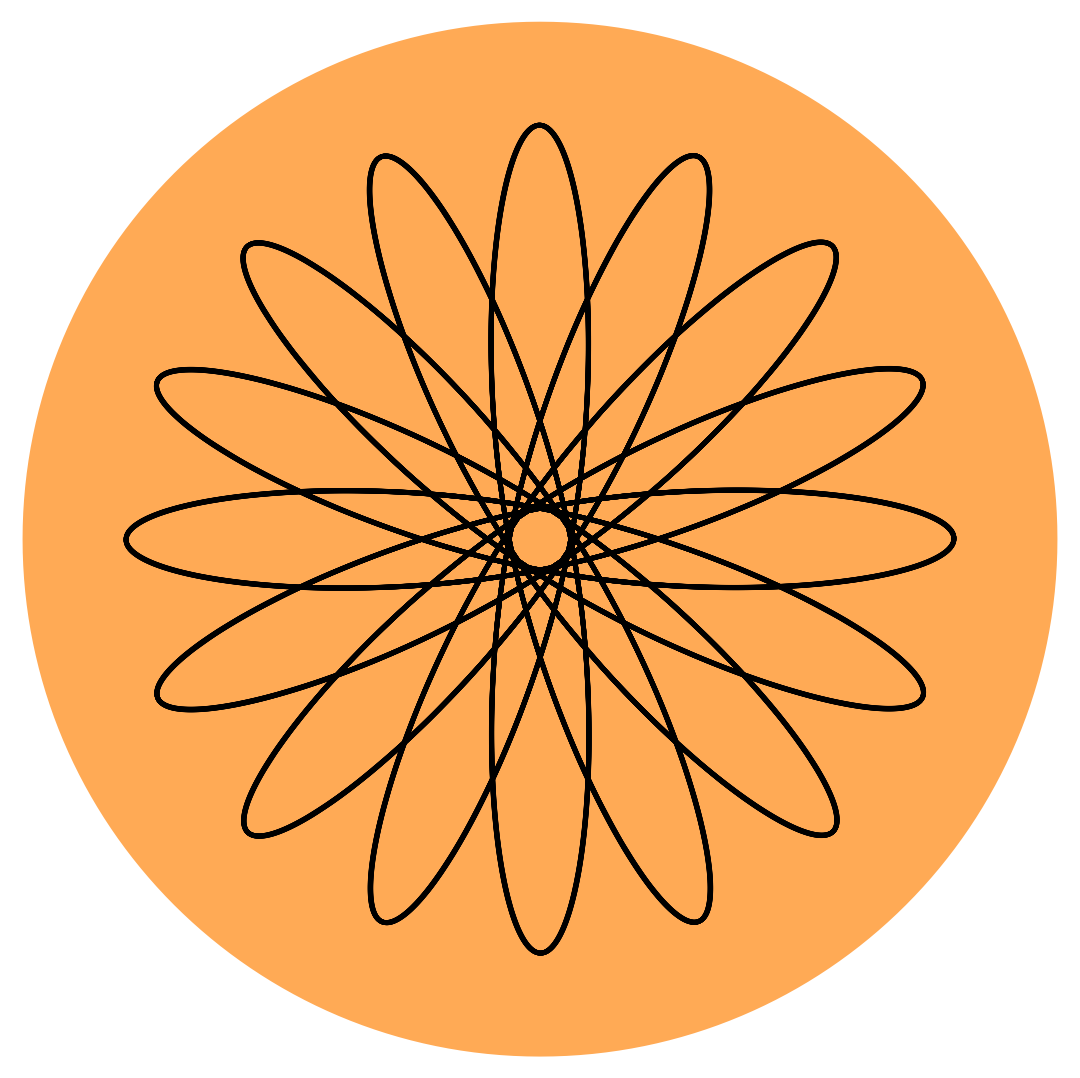} {\small $v(0)=79.880\, {\rm km\, s^{-1}}$} 
            & \includegraphics[width=\linewidth]{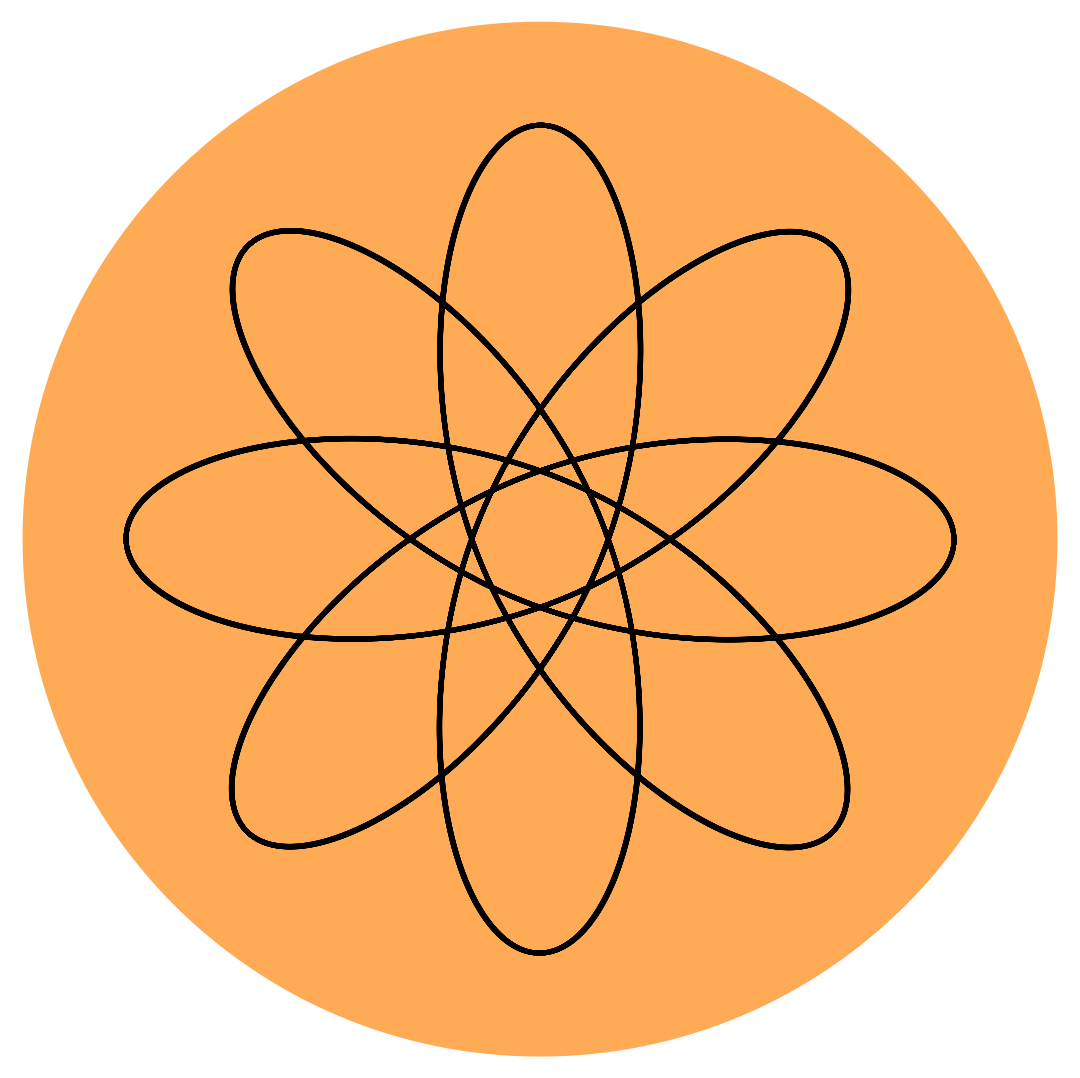} {\small $v(0)=161.30\, {\rm km\, s^{-1}}$} \\
            \hline
            \includegraphics[width=\linewidth]{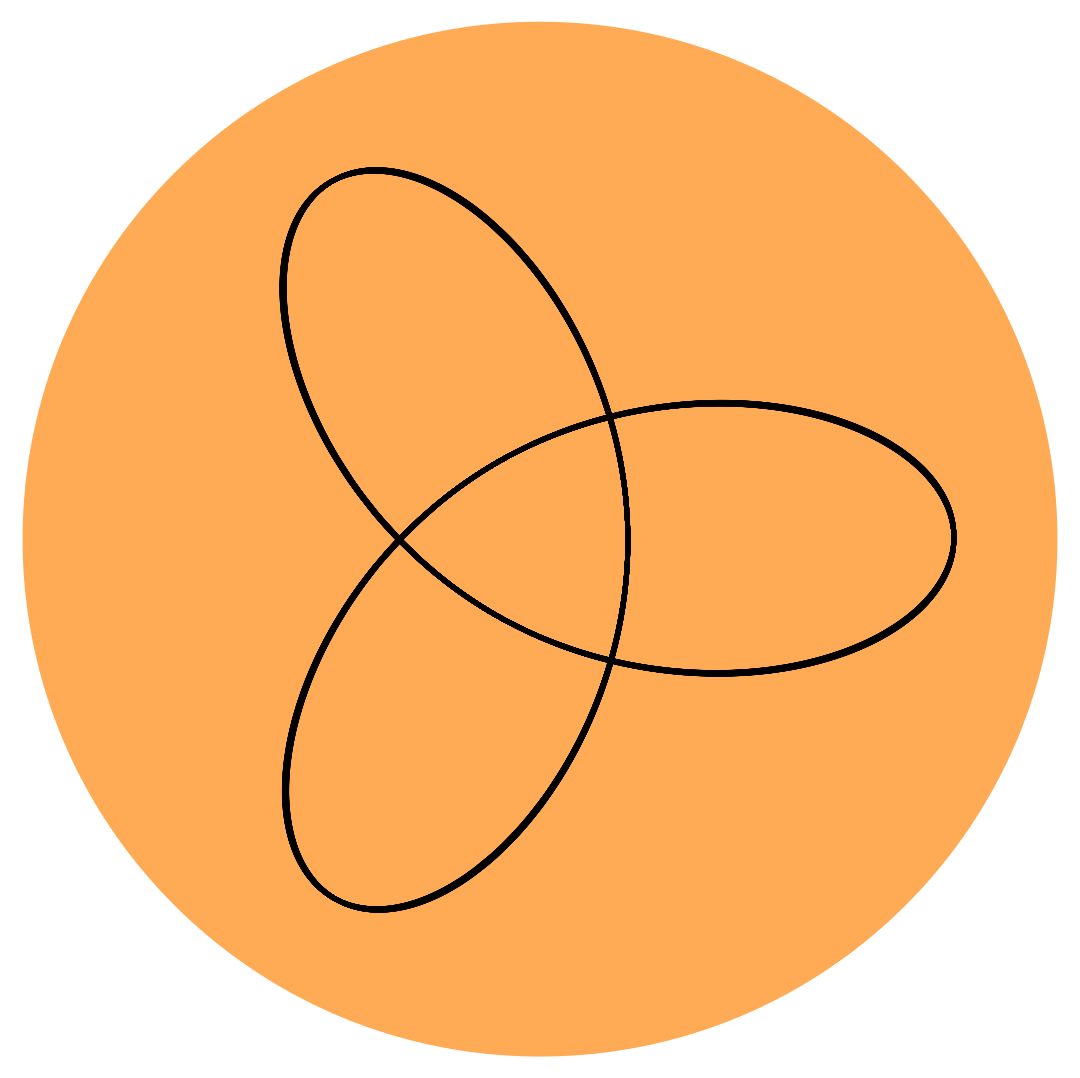} {\small $v(0)=212.99\, {\rm km\, s^{-1}}$} 
            & \includegraphics[width=\linewidth]{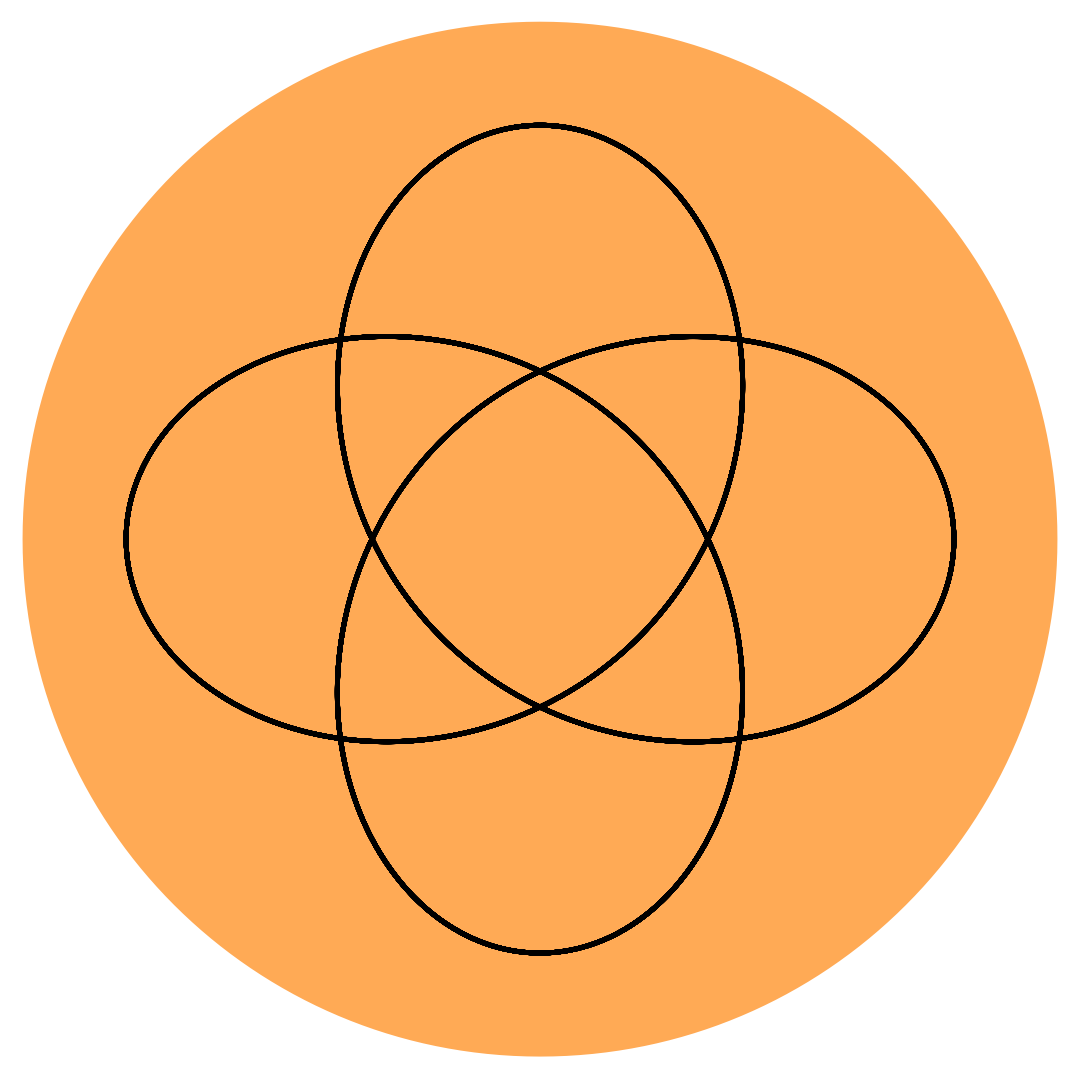} {\small $v(0)=302.52\, {\rm km\, s^{-1}}$} 
            & \includegraphics[width=\linewidth]{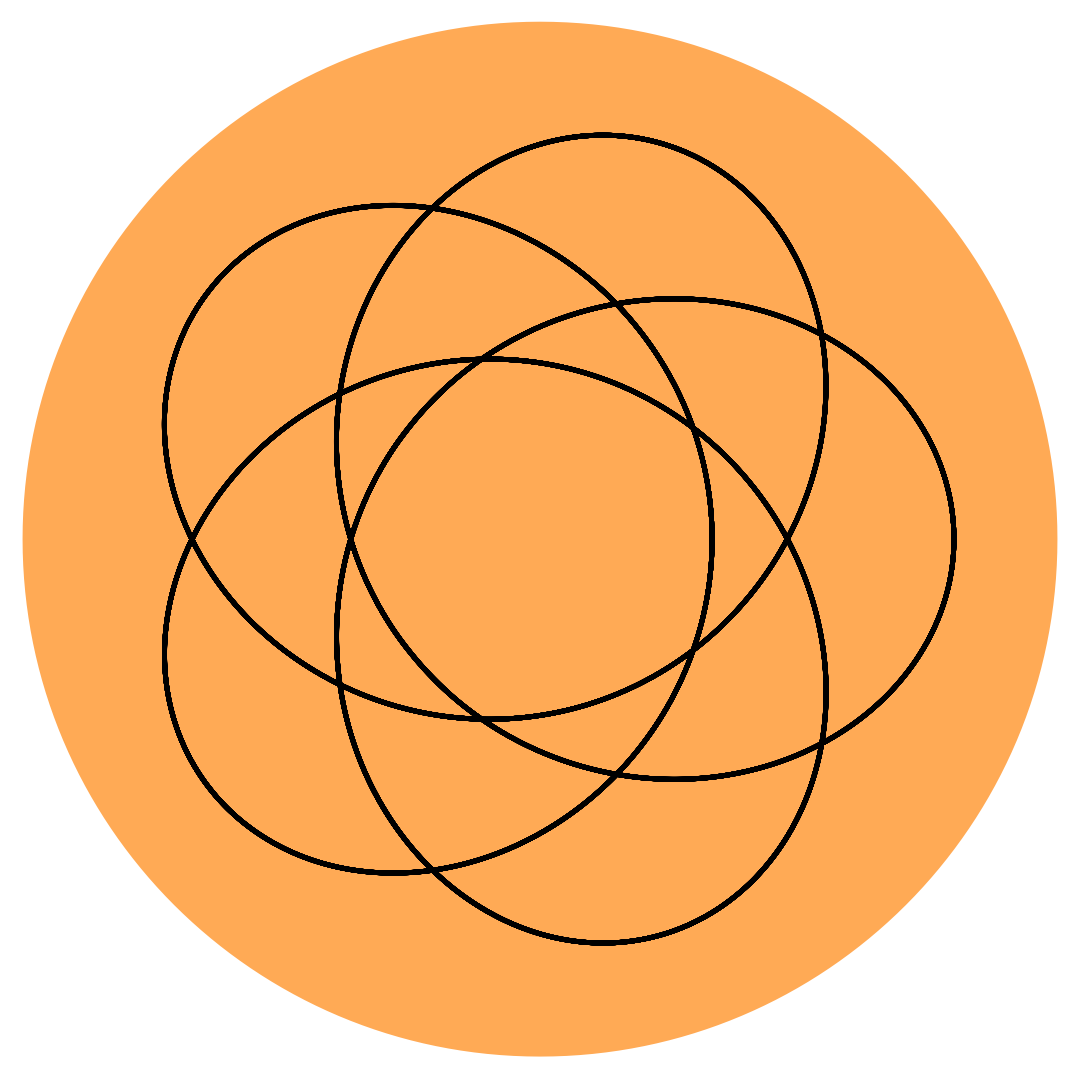} {\small $v(0)=345.85\, {\rm km\, s^{-1}}$} 
            & \includegraphics[width=\linewidth]{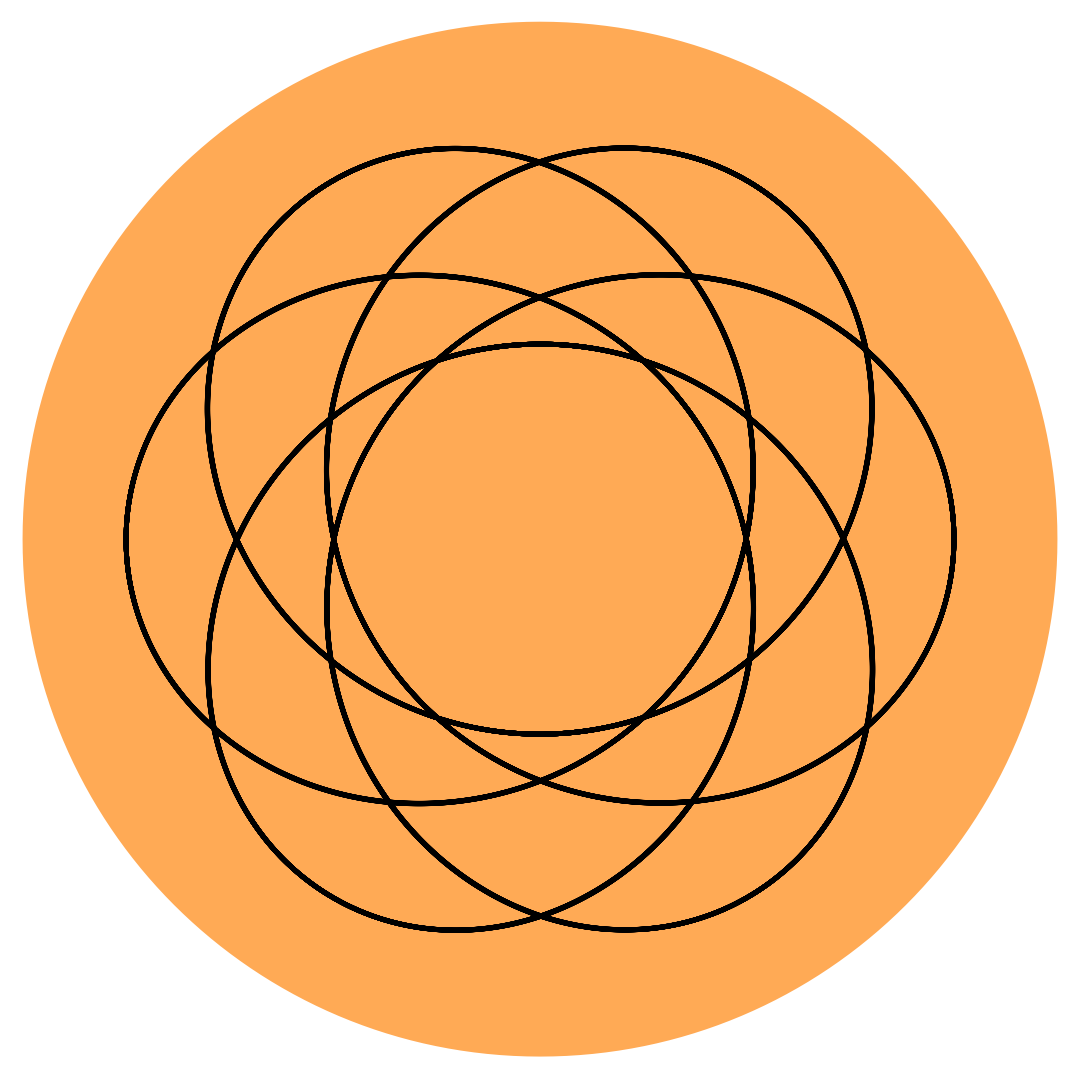}  {\small $v(0)=370.79\, {\rm km\, s^{-1}}$} \\
            \hline
        \end{tabularx}
        \caption{Closed orbits of a PBH in the interior of a star whose mass-density profile is described by the idealized model discussed in Sec.~\ref{workable}. The initial position was set to be $r(\varphi = 0) = \frac45 R$ for all cases, and the initial velocity at $\varphi=0$ is indicated at the bottom of each figure. The interval of integration in these figures is $200\pi$. The orange disk represents the star.}
        \label{fig_interior}
    \end{minipage}

    \vspace{1cm}

    \begin{minipage}{\textwidth}
    \begin{tabularx}{\textwidth}{>{\centering\arraybackslash}X|>{\centering\arraybackslash}X|>{\centering\arraybackslash}X|>{\centering\arraybackslash}X}
            \multicolumn{4}{c}{$r(0)=4R/5$} \\  
            \hline
            \includegraphics[width=\linewidth]{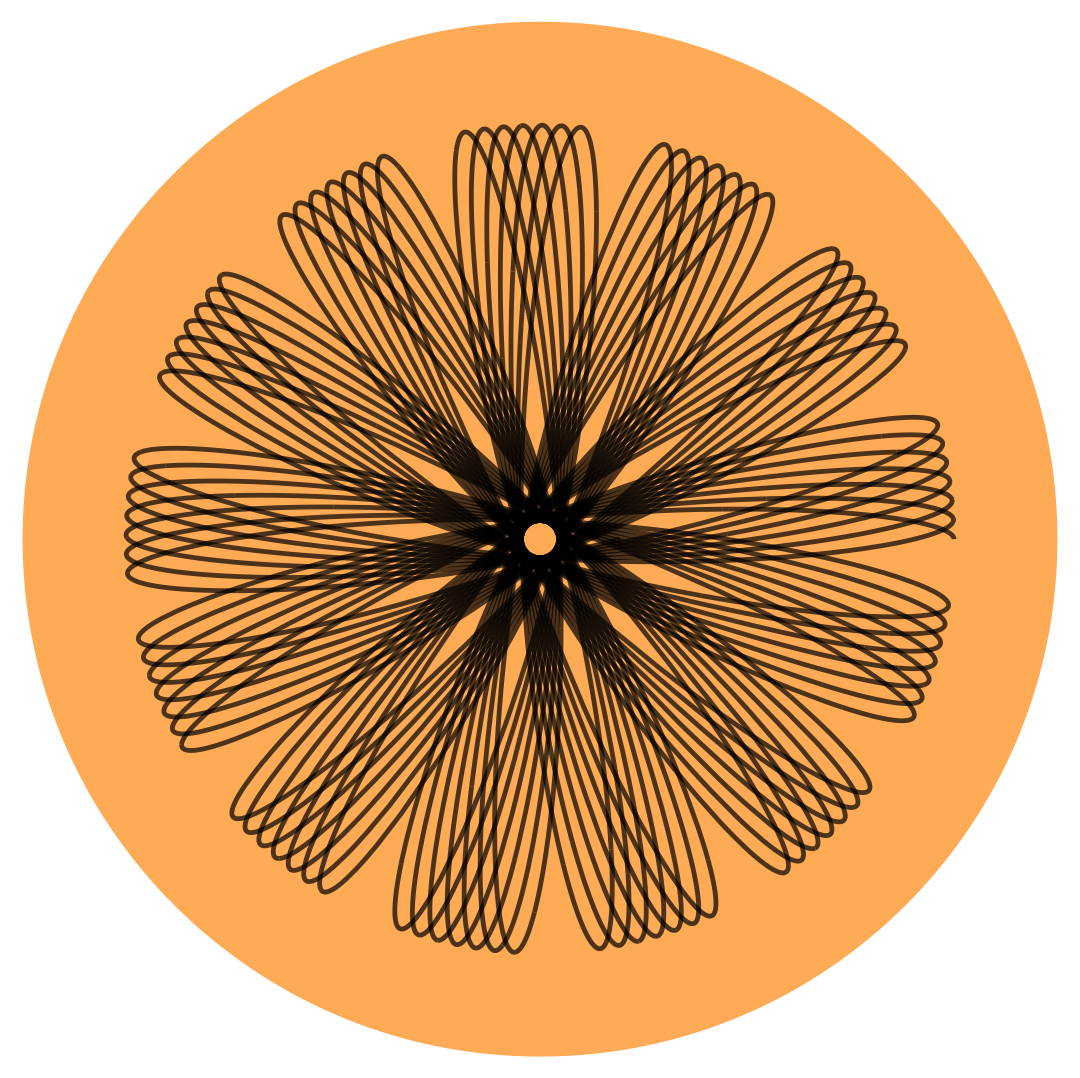} {\small $v(0)=49.534\, {\rm km\, s^{-1}}$}
            & \includegraphics[width=\linewidth]{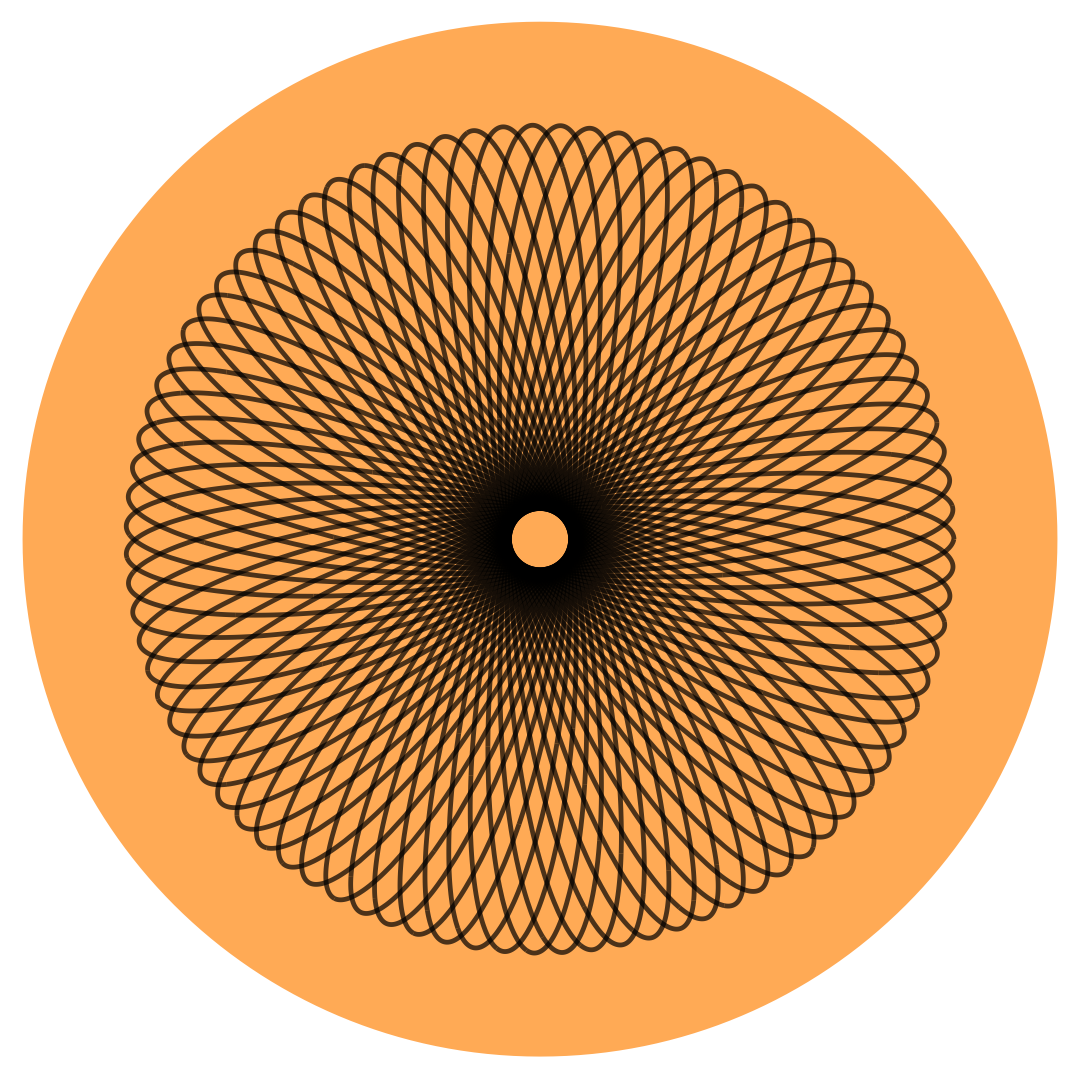} {\small $v(0)=80.925\, {\rm km\, s^{-1}}$}  
            & \includegraphics[width=\linewidth]{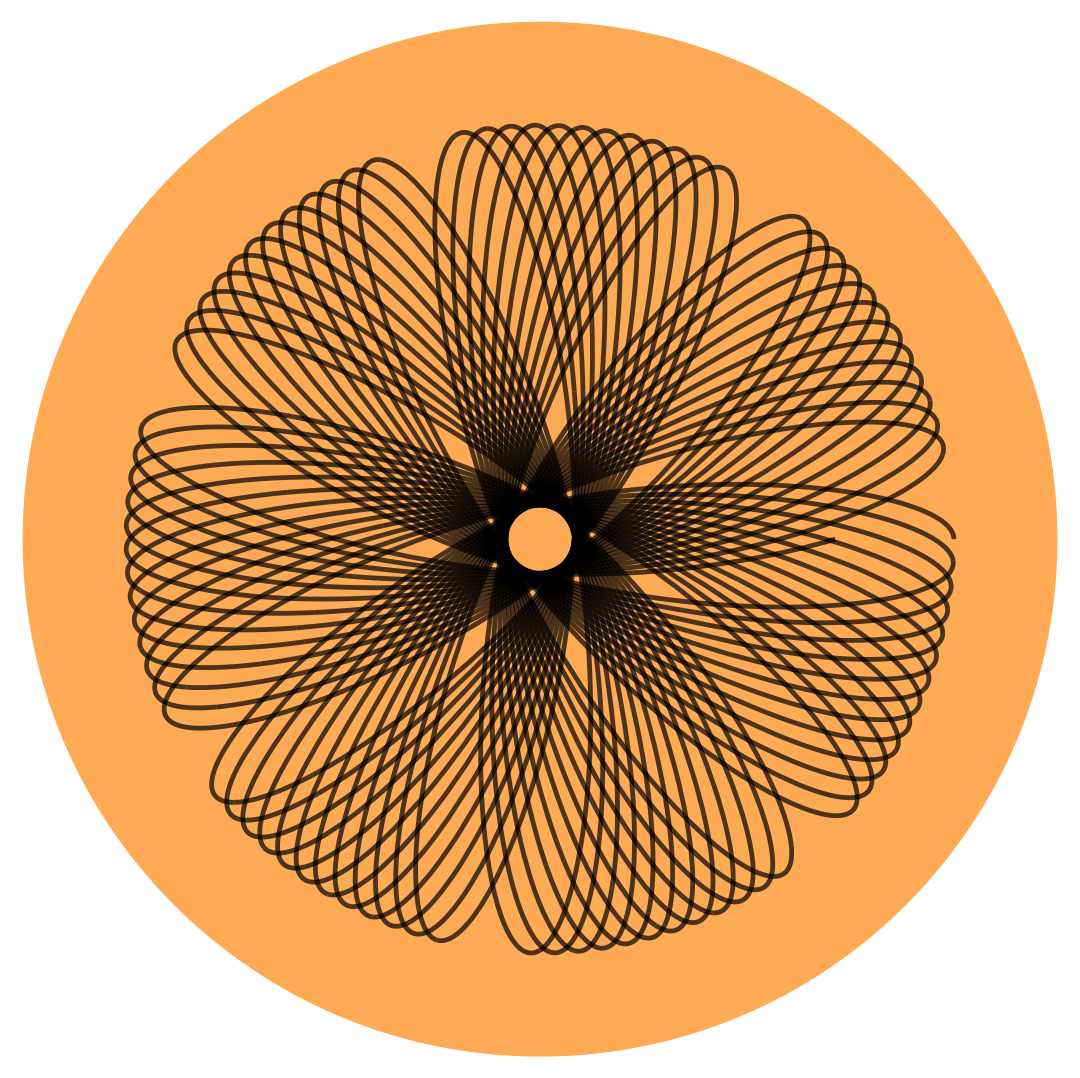} {\small $v(0)=89.880\, {\rm km\, s^{-1}}$} 
            & \includegraphics[width=\linewidth]{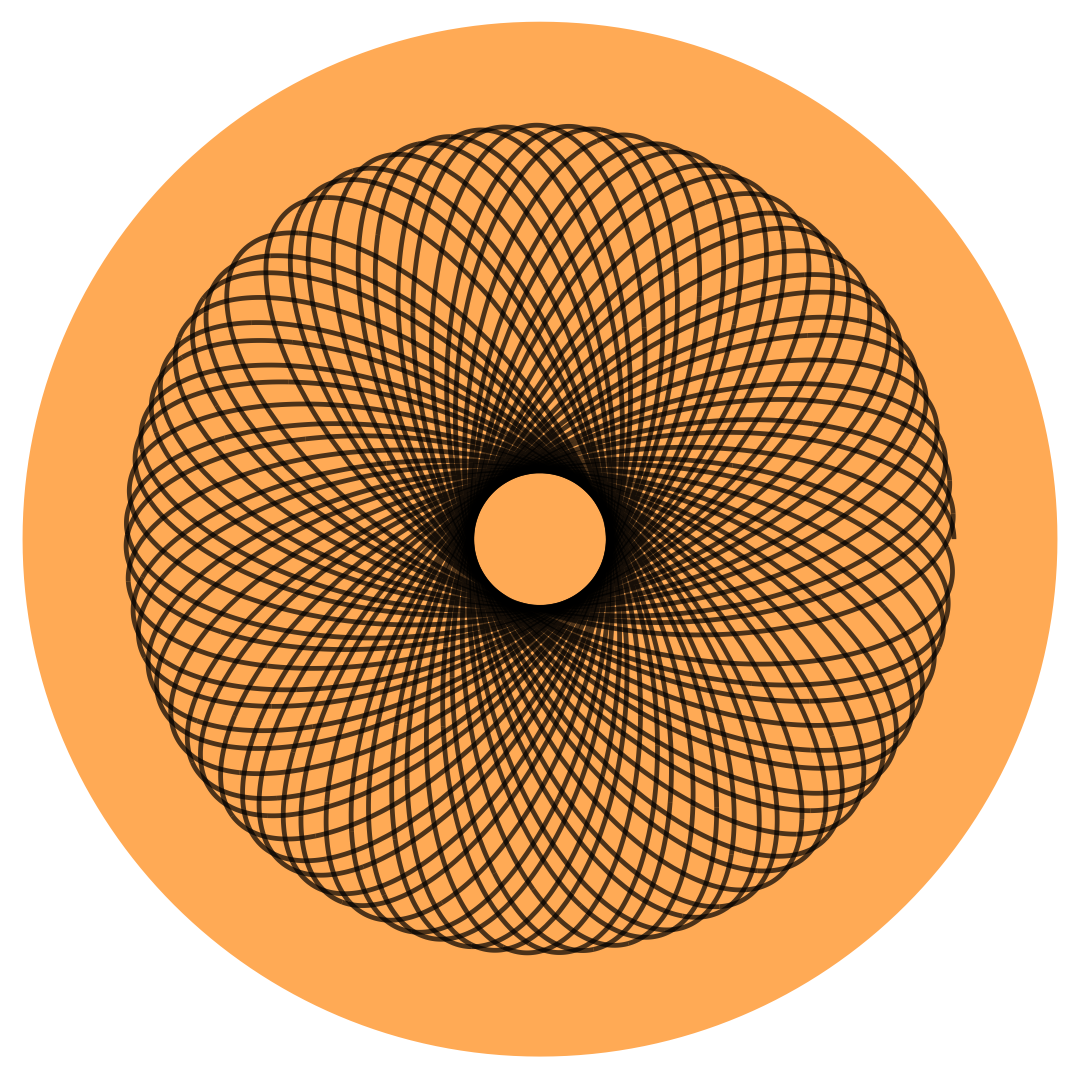} {\small $v(0)=171.30\, {\rm km\, s^{-1}}$} \\
            \hline
            \includegraphics[width=\linewidth]{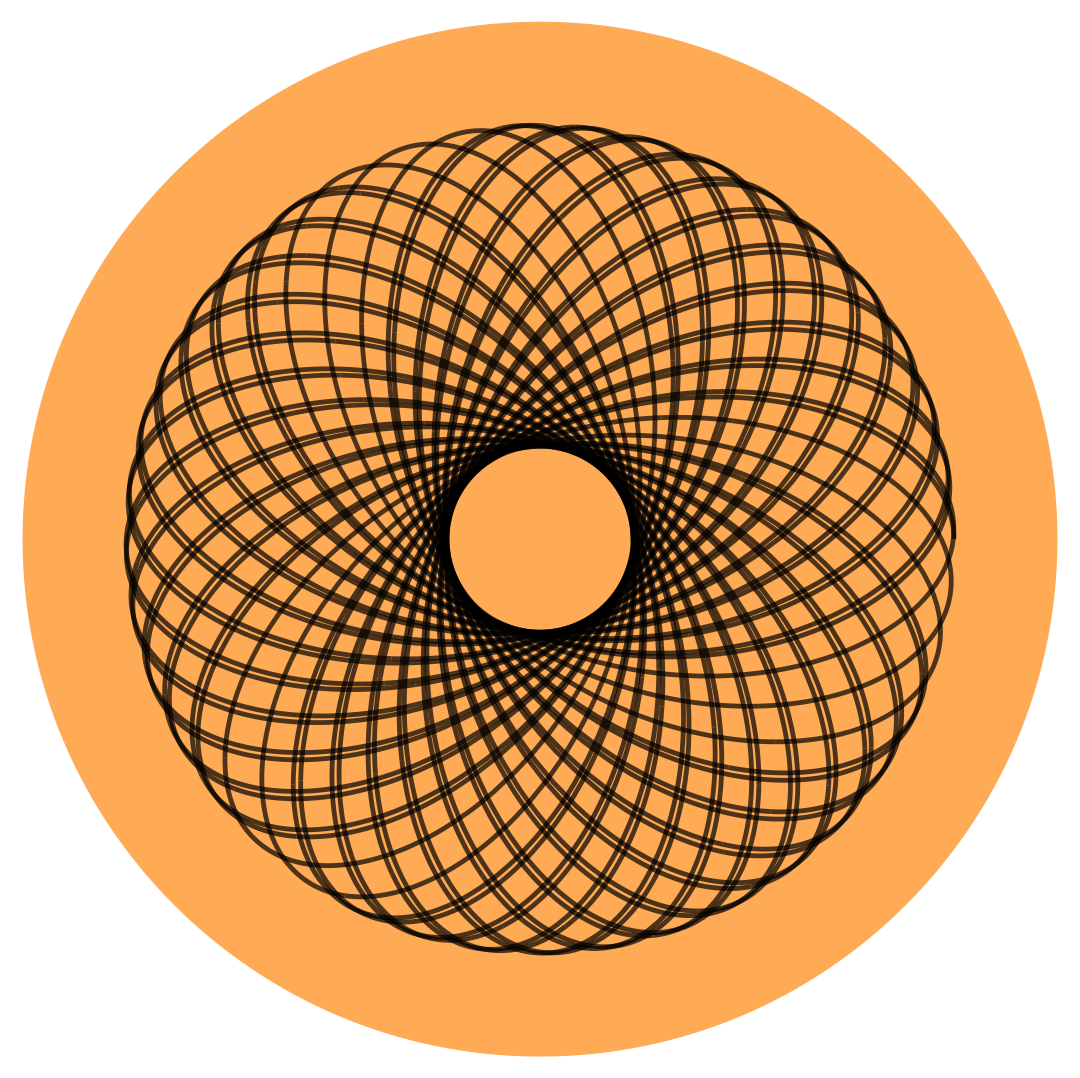} {\small $v(0)=222.99\, {\rm km\, s^{-1}}$} 
            & \includegraphics[width=\linewidth]{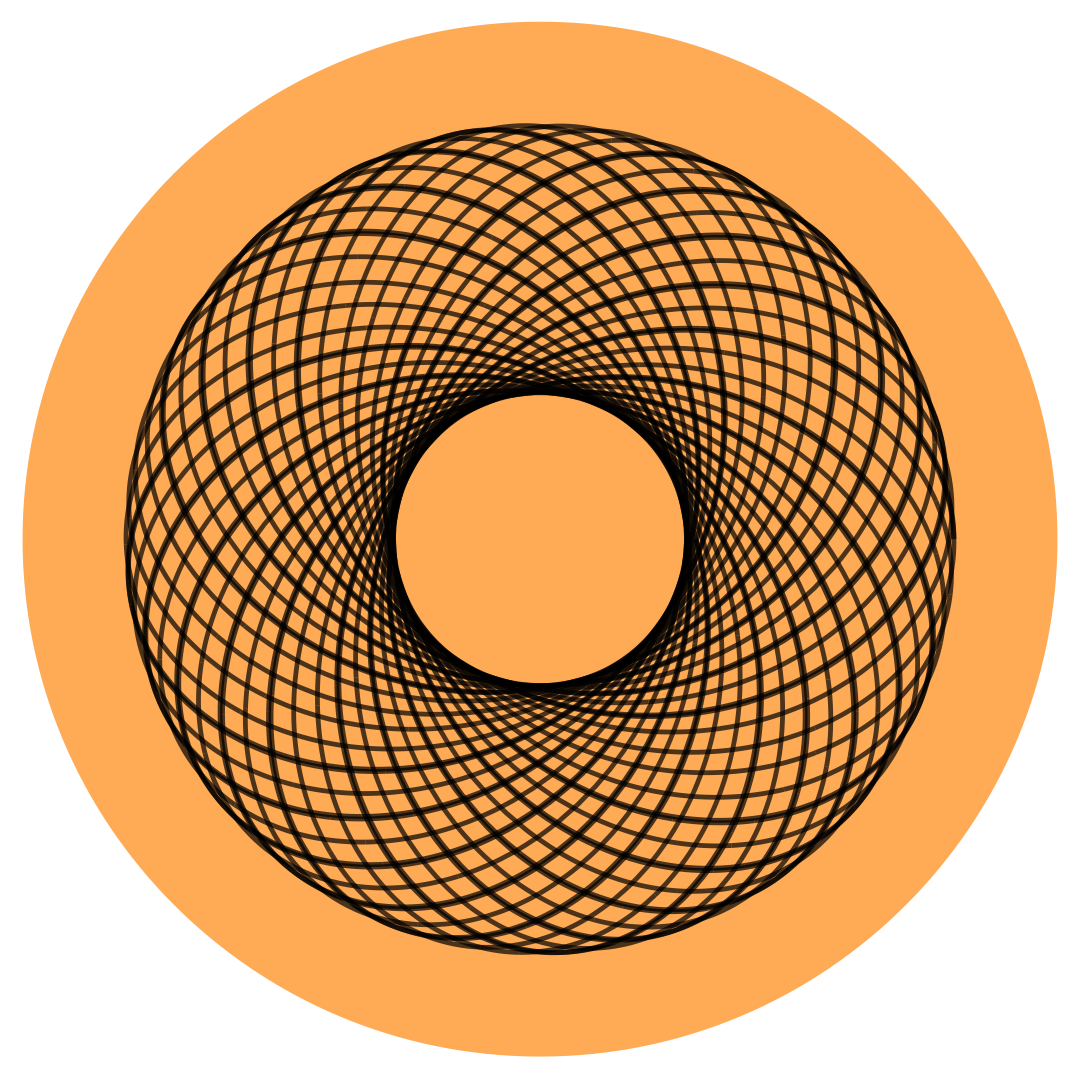} {\small $v(0)=312.55\, {\rm km\, s^{-1}}$} 
            & \includegraphics[width=\linewidth]{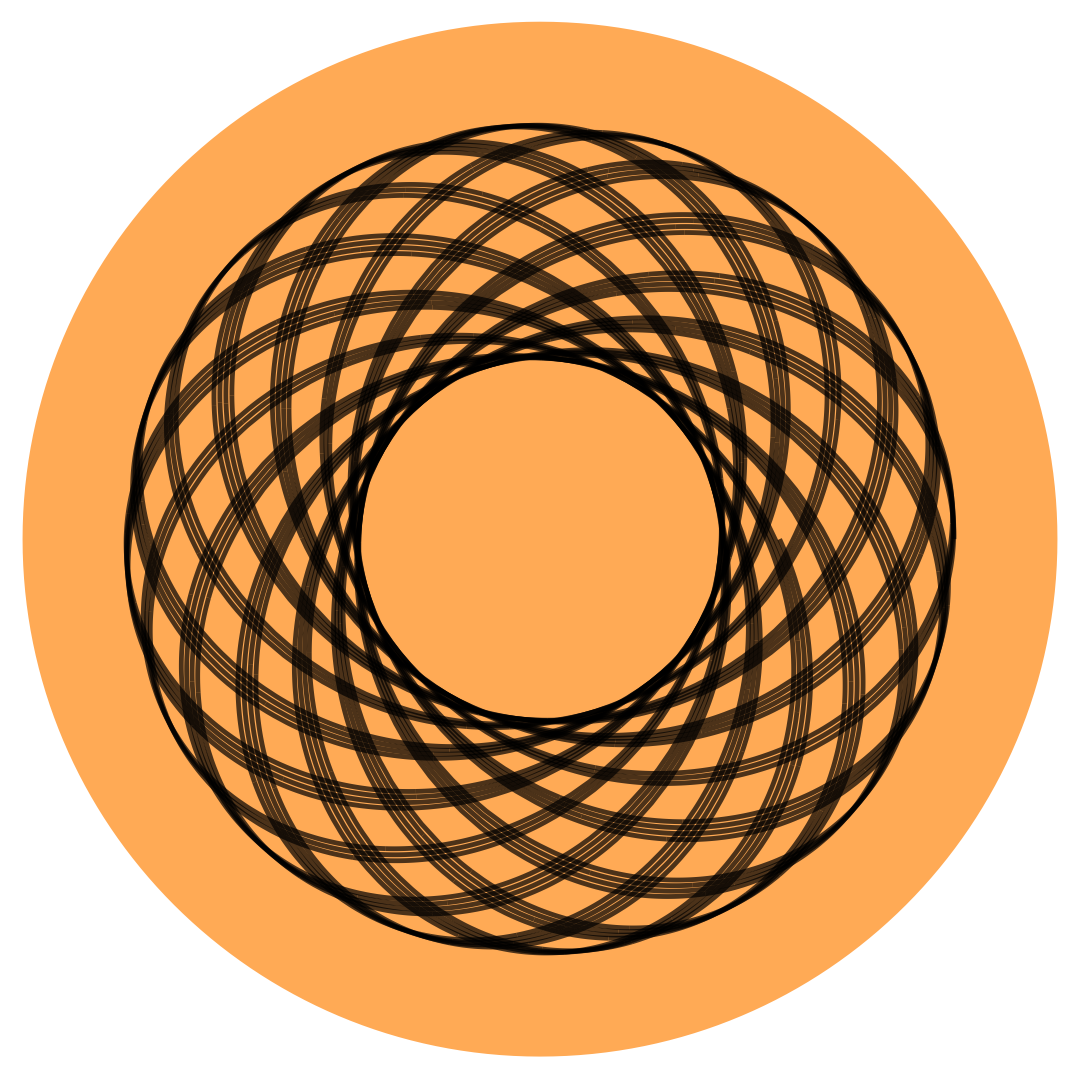} {\small $v(0)=355.85\, {\rm km\, s^{-1}}$} 
            & \includegraphics[width=\linewidth]{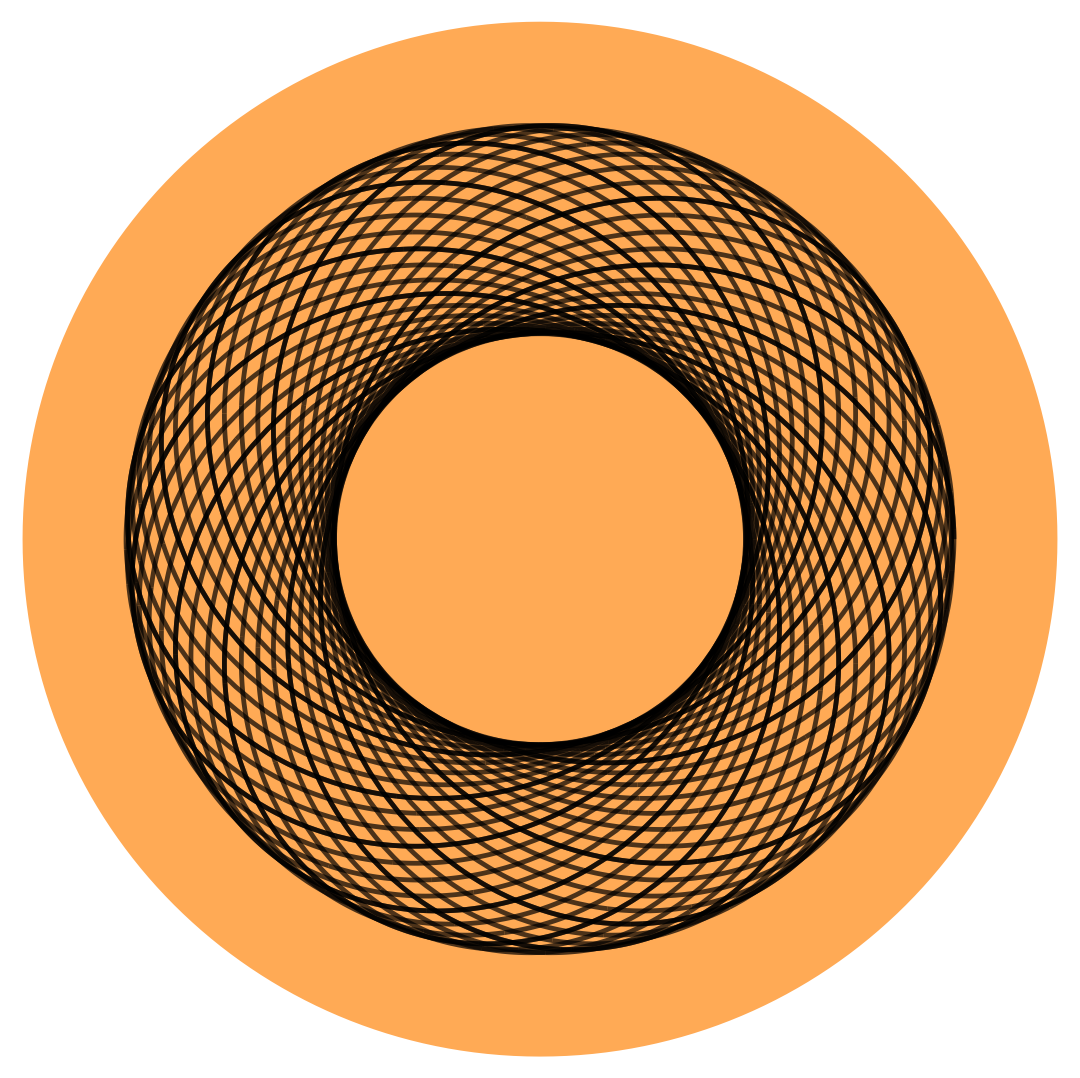} {\small $v(0)=380.79\, {\rm km\, s^{-1}}$} \\
            \hline
        \end{tabularx}
        \caption{Here the initial velocity at $\varphi=0$ is perturbed compared to the initial velocities for orbits depicted in Fig.~\ref{fig_interior}. The resulting orbits are now open. The plots illustrate a series of fifty revolutions, corresponding to an integration interval of $\Delta \varphi = 100\pi$.}
        \label{fig_interior_open}
    \end{minipage}
\end{figure}

In Fig.~\ref{fig_interior}, initial conditions such that the initial position of the PBH is inside the star were chosen in such a way that, within the precision set in the simulations, closed trajectories could be found (among several others that are not shown here). These trajectories always remain inside the star, and the interval of integration was limited to $200\pi$, which means that the circular orbit solution accumulates 100 complete revolutions around the center of the star. The other plots in this figure complete their closed shape over intervals larger than $2\pi$. For instance, for the initial condition $v(\varphi = 0) = 129.00 \, {\rm km \, s^{-1}}$ the complete five-petals orbit only appears after $\Delta \varphi = 6\pi$. 

In Fig.~\ref{fig_interior_open}, the initial conditions used in Fig.~\ref{fig_interior} were modified by adding $10 \, {\rm km \, s^{-1}}$ to the initial velocities. The resulting trajectories are all open, and the plots cover an interval of integration of $\Delta\varphi = 100\pi$. The patterns appear as the orbits evolve.

\begin{figure}
    \centering
    \begin{minipage}{\textwidth}
    \begin{tabularx}{\textwidth}{>{\centering\arraybackslash}X|>{\centering\arraybackslash}X|>{\centering\arraybackslash}X|>{\centering\arraybackslash}X}
        \multicolumn{4}{c}{$r(0)=2R$} \\
        \hline
          \includegraphics[width=\linewidth]{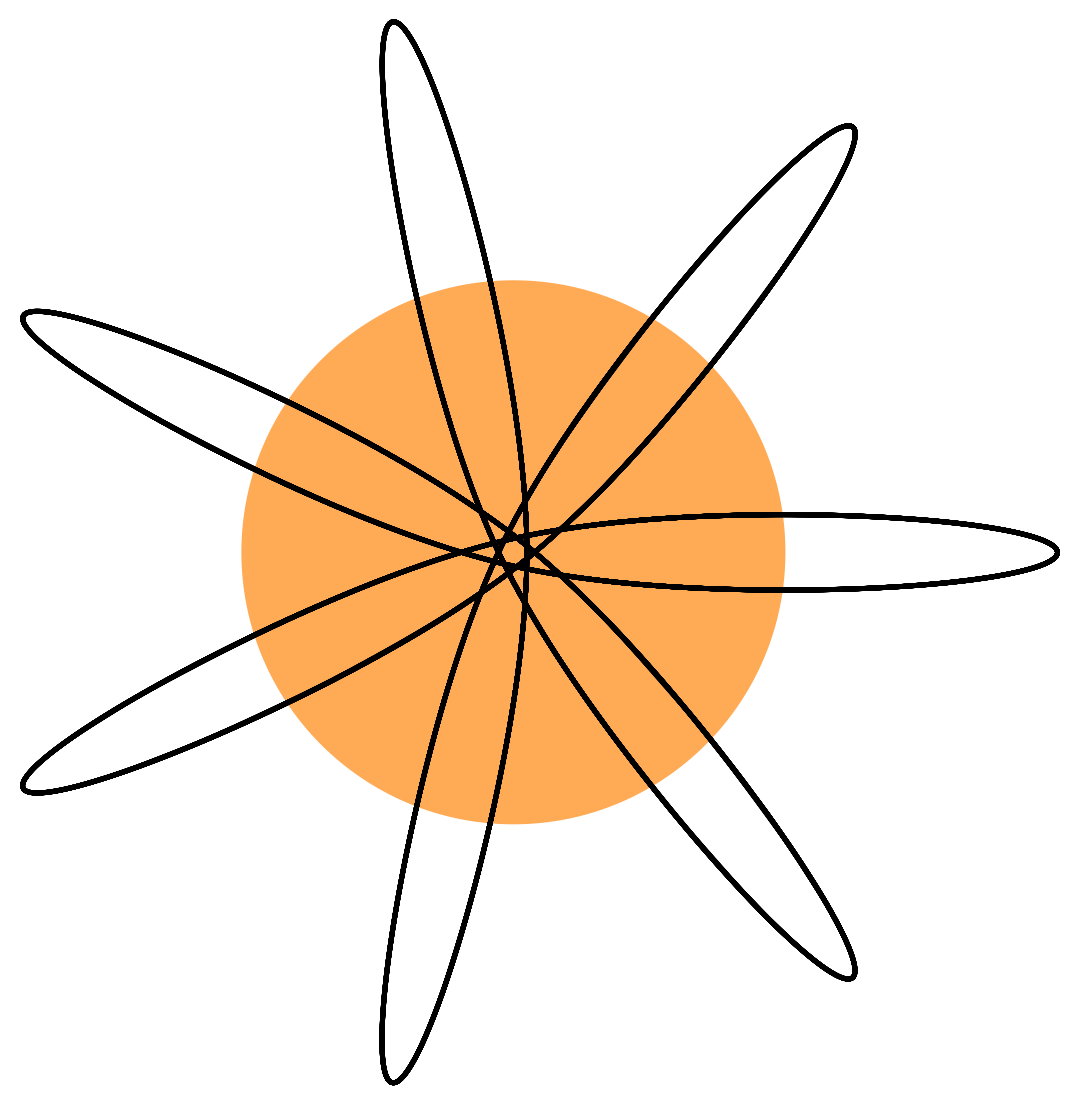} {\small $v(0)=30.196\, {\rm km\, s^{-1}}$}
         & \includegraphics[width=\linewidth]{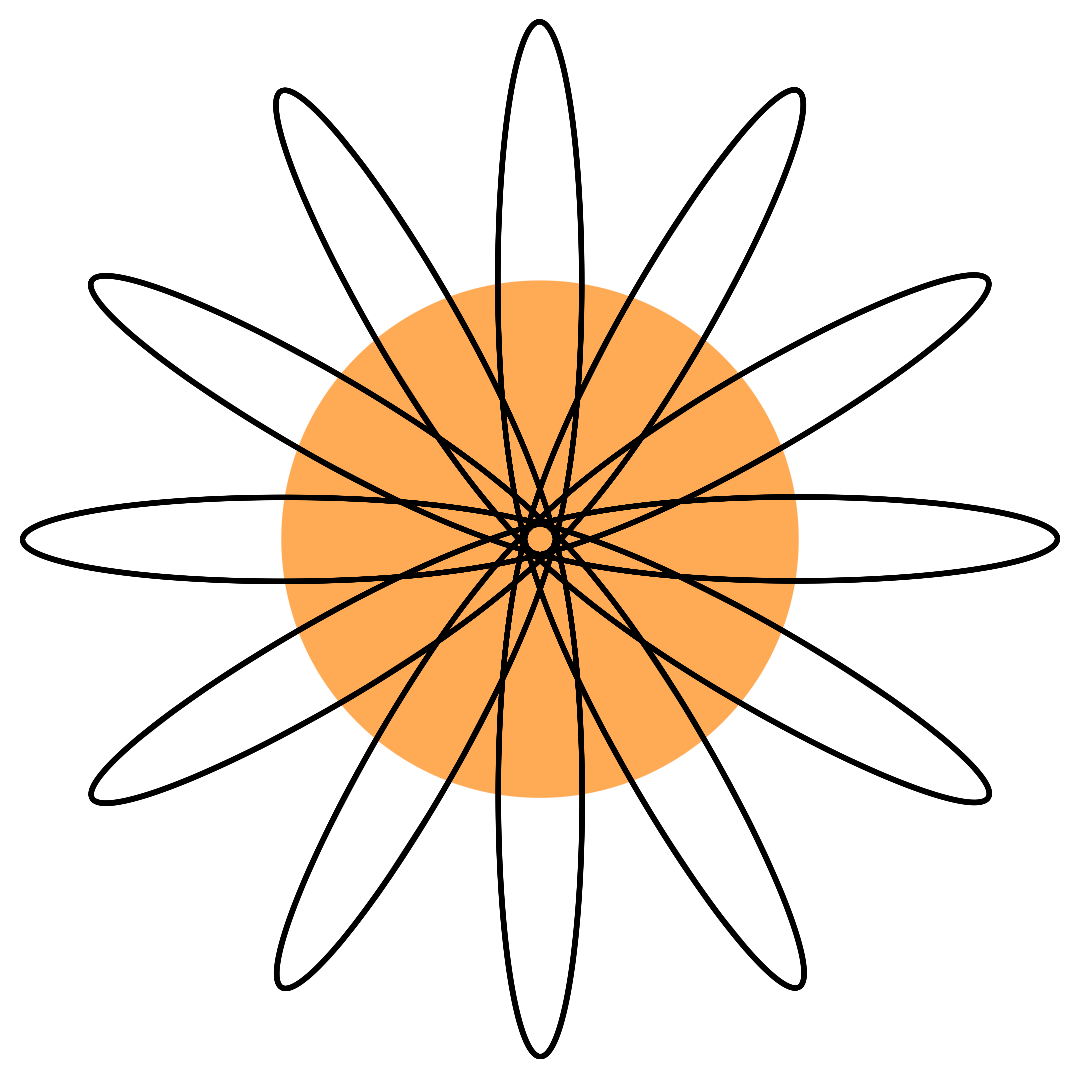} {\small $v(0)=35.3\, {\rm km\, s^{-1}}$}  
         & \includegraphics[width=\linewidth]{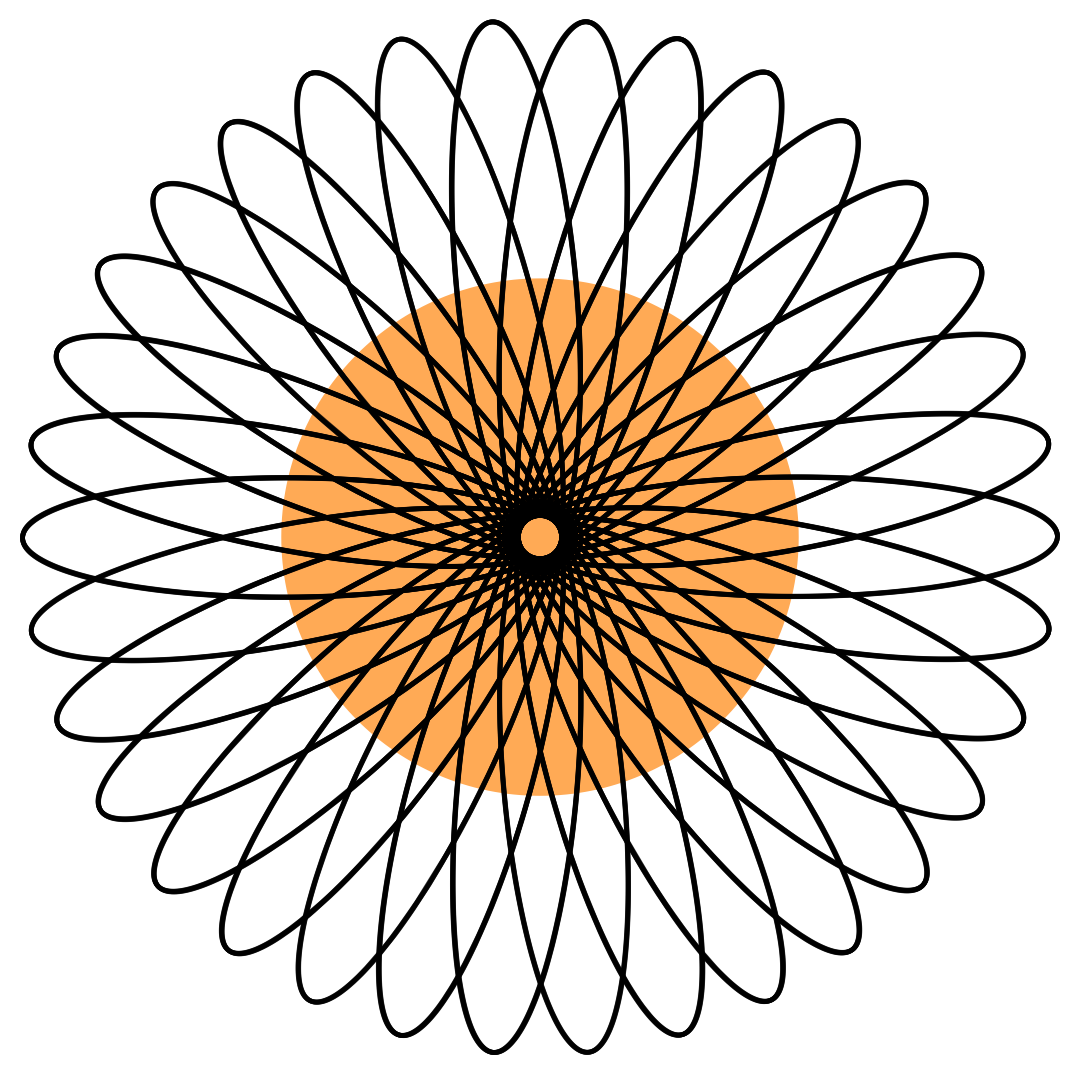} {\small $v(0)=50.065\, {\rm m s^{-1}}$} 
         & \includegraphics[width=\linewidth]{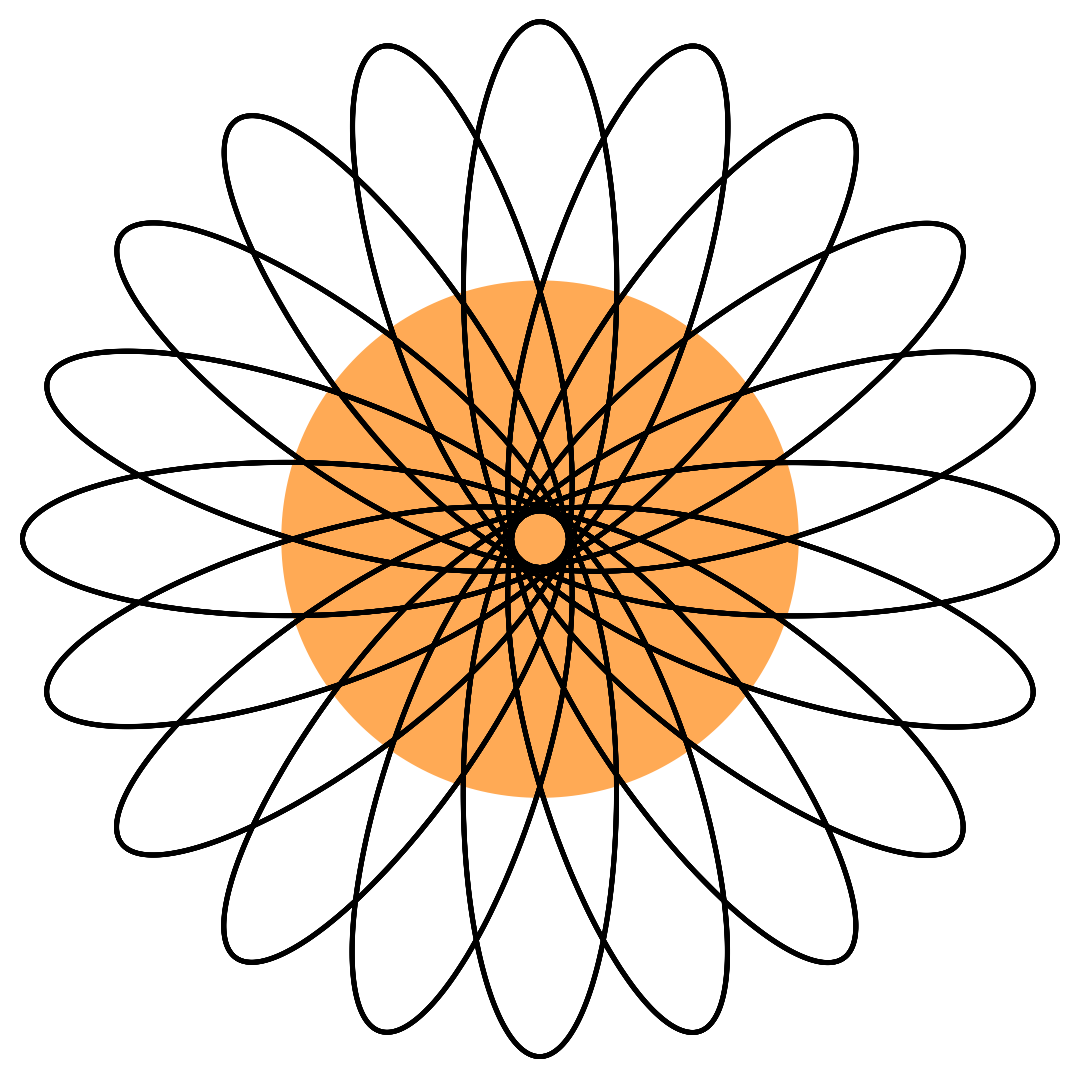} {\small $v(0)=63.98\, {\rm km\, s^{-1}}$} 
          \\ \hline
          \includegraphics[width=\linewidth]{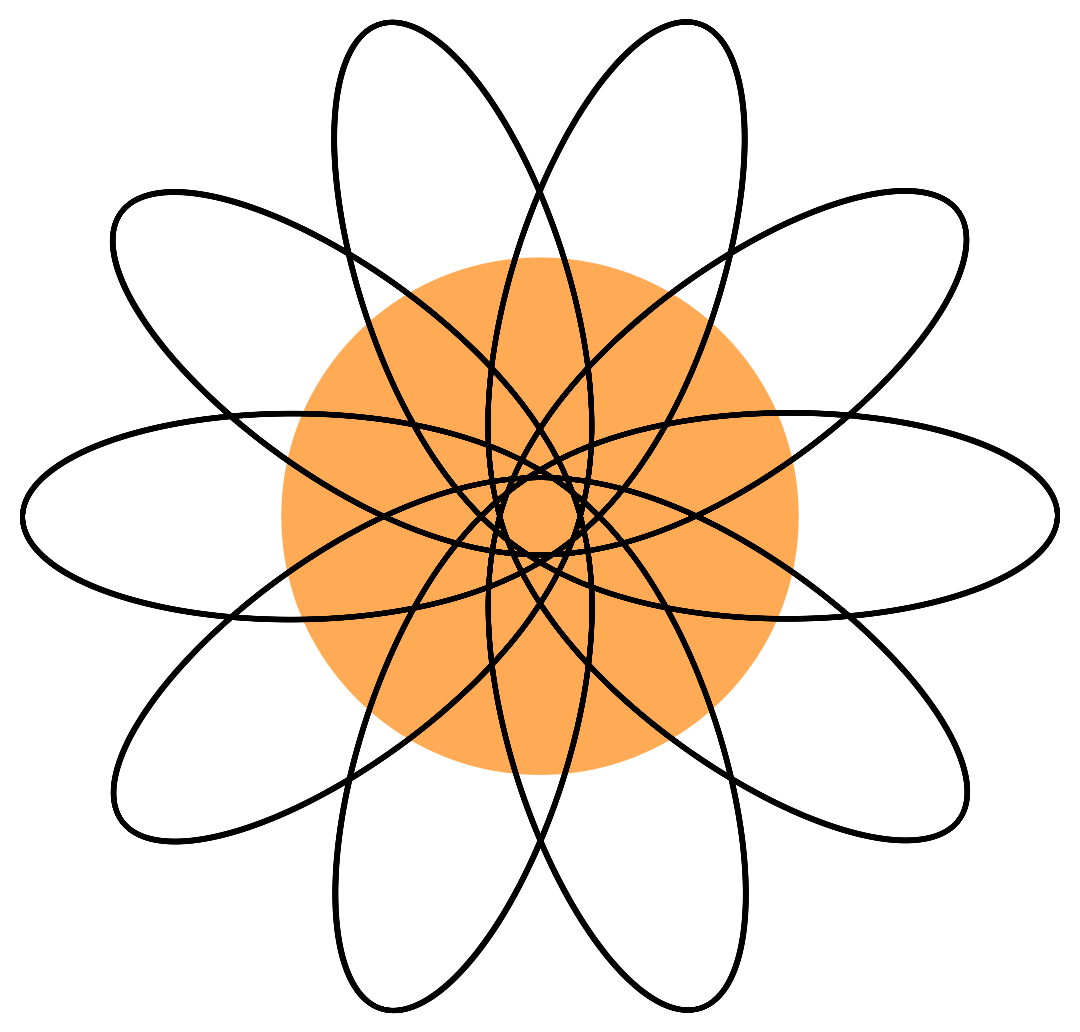} {\small $v(0)=85.25\, {\rm km\, s^{-1}}$} 
         & \includegraphics[width=\linewidth]{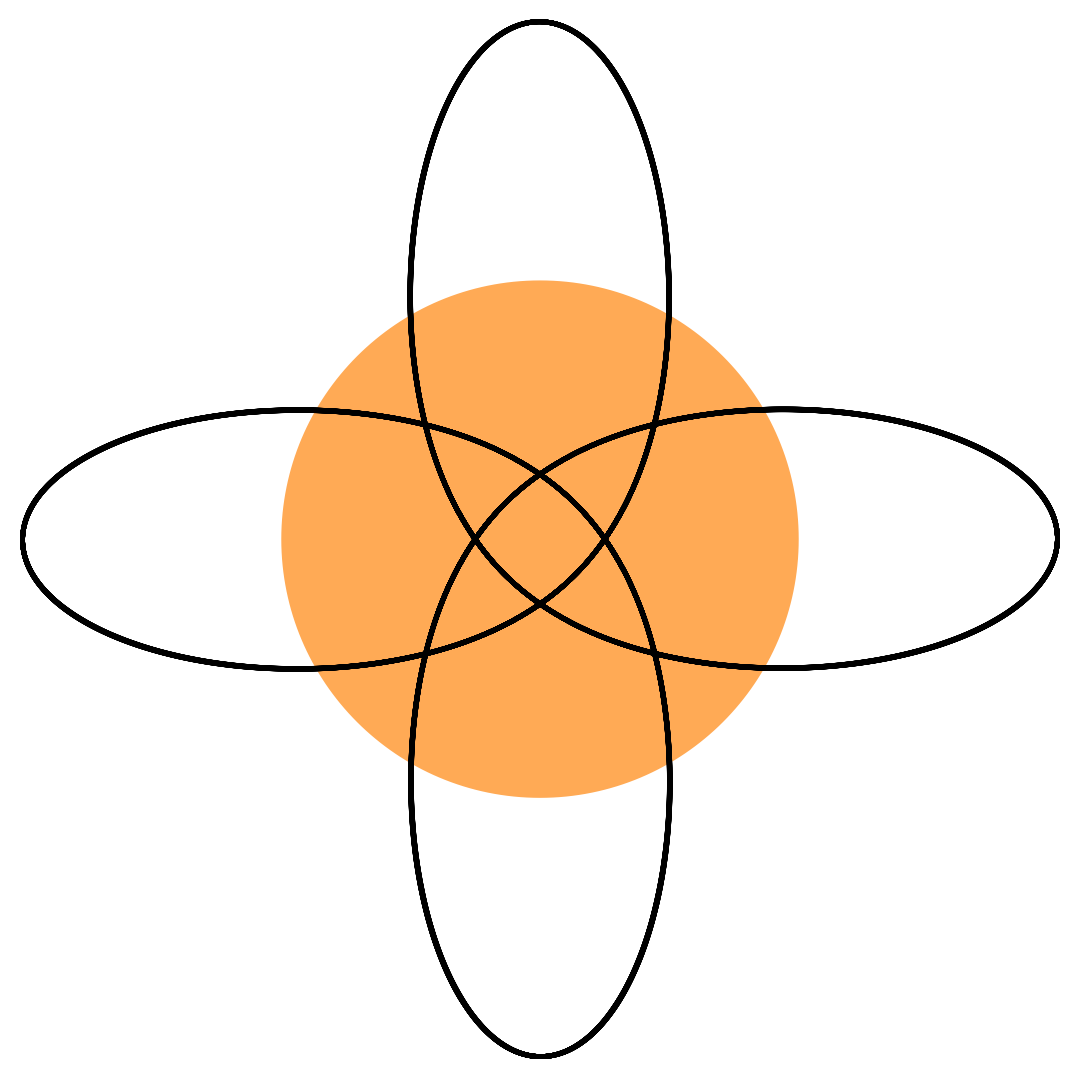} {\small $v(0)=105.95\, {\rm km\, s^{-1}}$} 
         & \includegraphics[width=\linewidth]{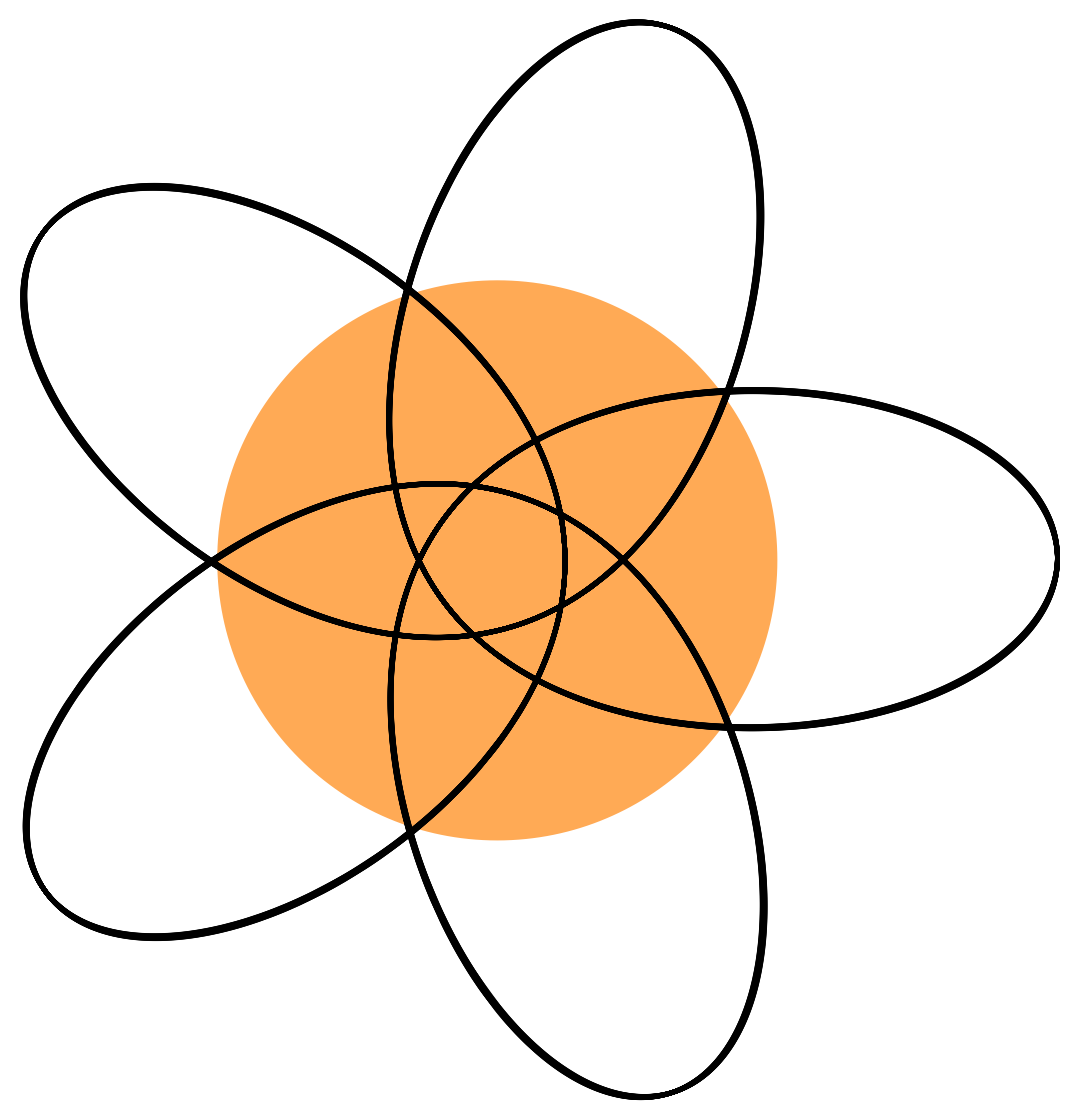} {\small $v(0)=125.89\, {\rm km\, s^{-1}}$} 
         & \includegraphics[width=\linewidth]{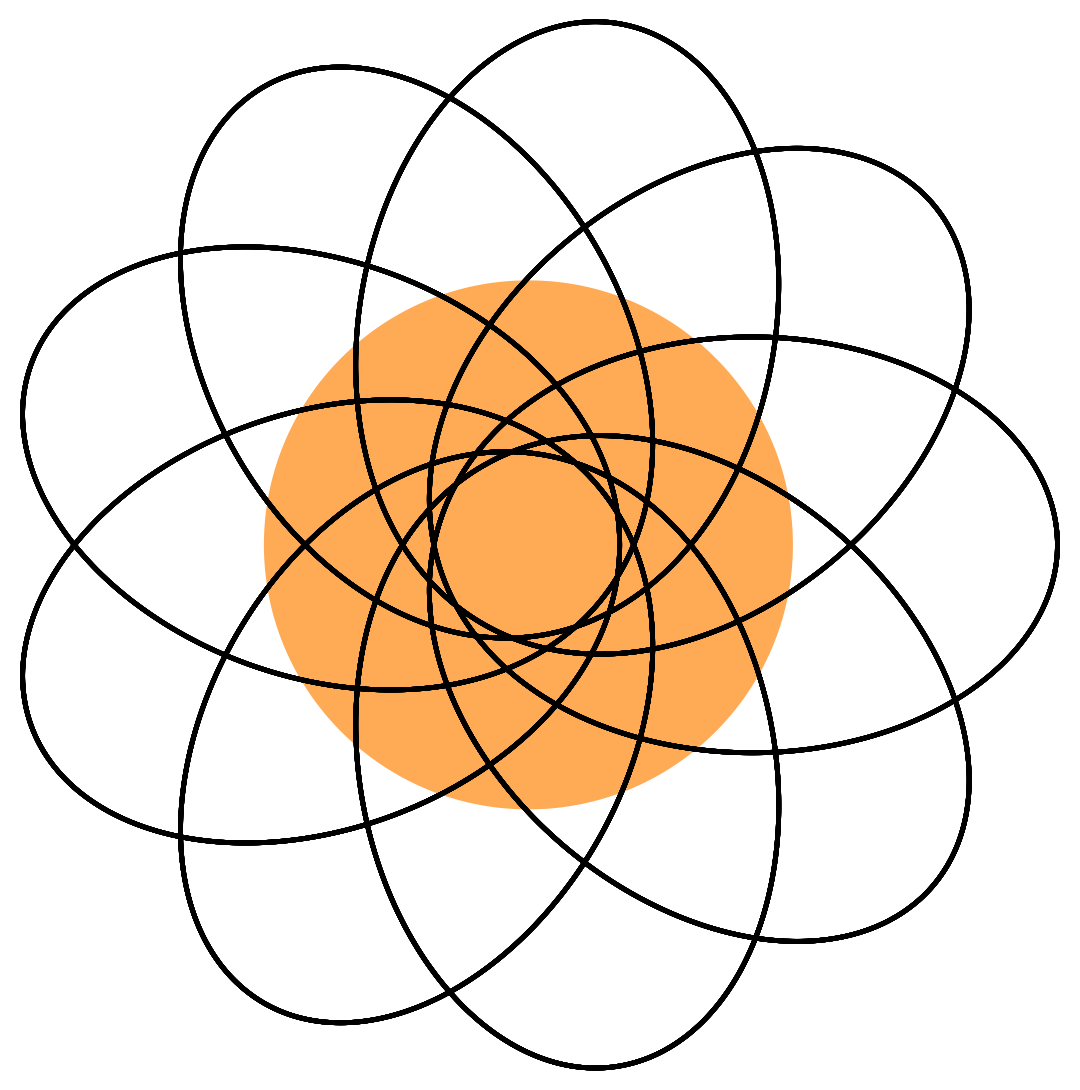}  {\small $v(0)=159.74\, {\rm km\, s^{-1}}$} 
         \\
        \hline
    \end{tabularx}
\caption{Closed orbit solutions for a PBH with an initial position outside the star ($r(0)=2R$). The interval of integration in all figures is $200\pi$. The orange disk represents the star.}
    \label{figs_out_closed}
        \end{minipage}

    \vspace{1cm} 

    \begin{minipage}{\textwidth}
    \begin{tabularx}{\textwidth}{>{\centering\arraybackslash}X|>{\centering\arraybackslash}X|>{\centering\arraybackslash}X|>{\centering\arraybackslash}X}
        \multicolumn{4}{c}{$r(0)=2R$} \\  
        \hline
         \includegraphics[width=\linewidth]{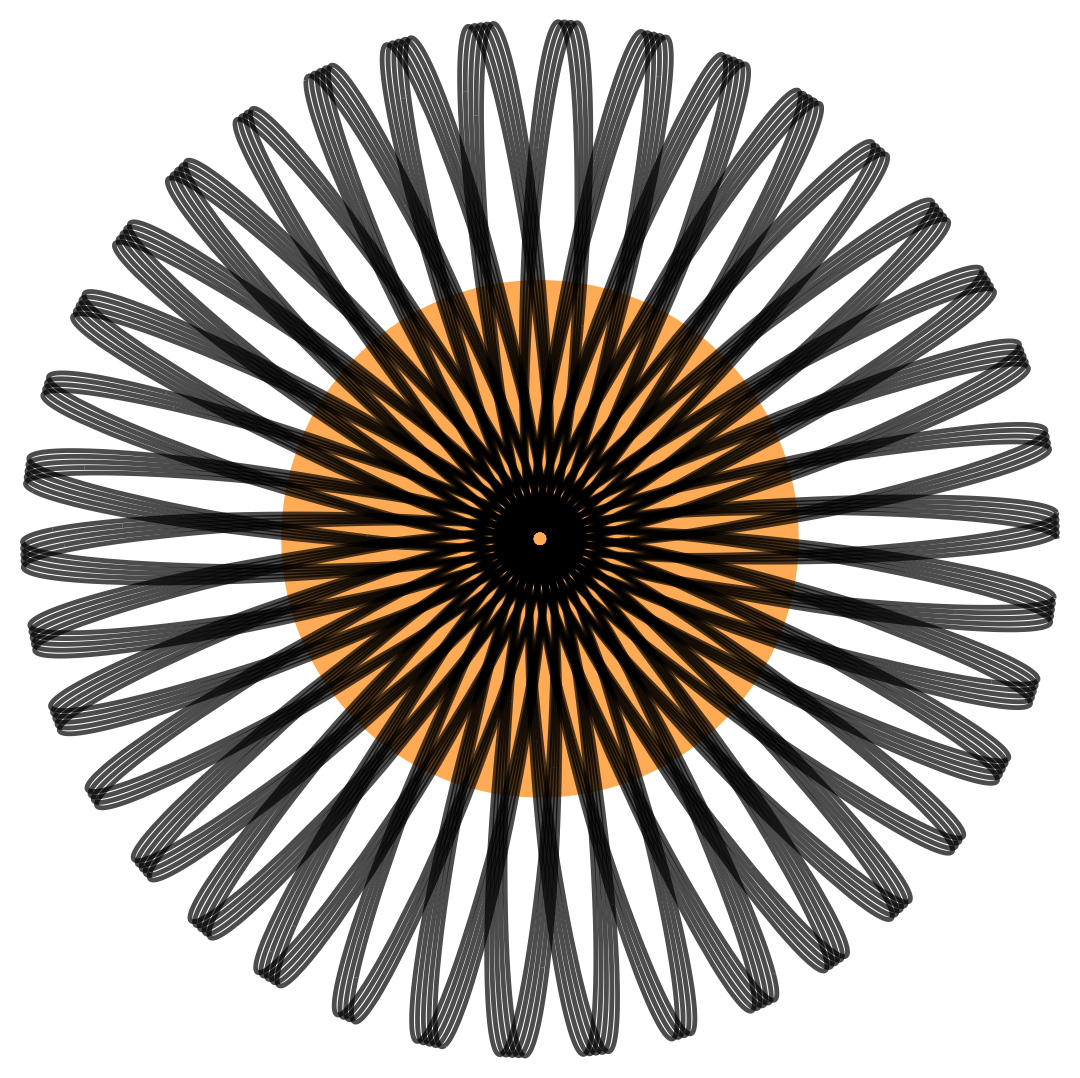} {\small $v(0)=22.196\, {\rm km\, s^{-1}}$}
         & \includegraphics[width=\linewidth]{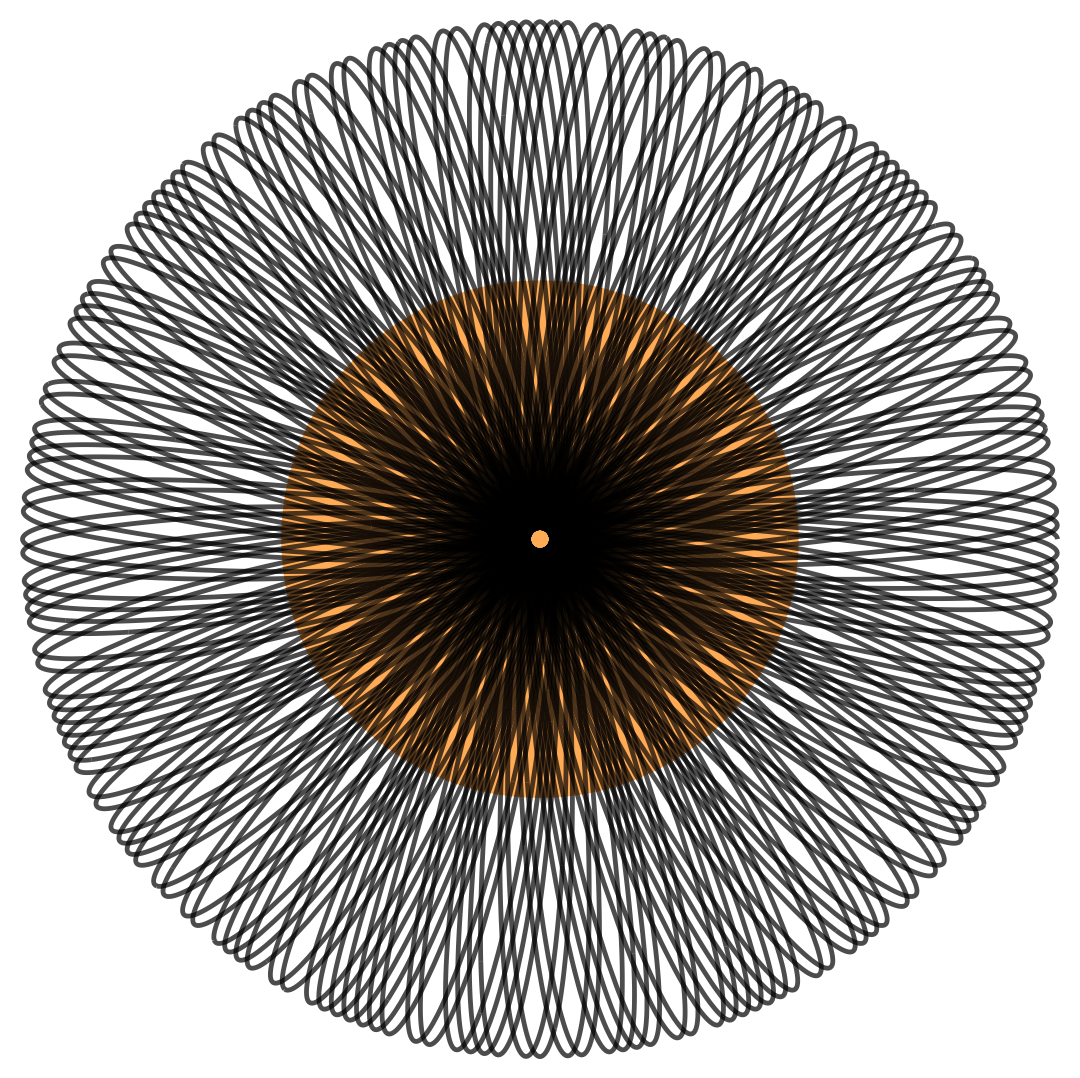} {\small $v(0)=27.3\, {\rm km\, s^{-1}}$}  
         & \includegraphics[width=\linewidth]{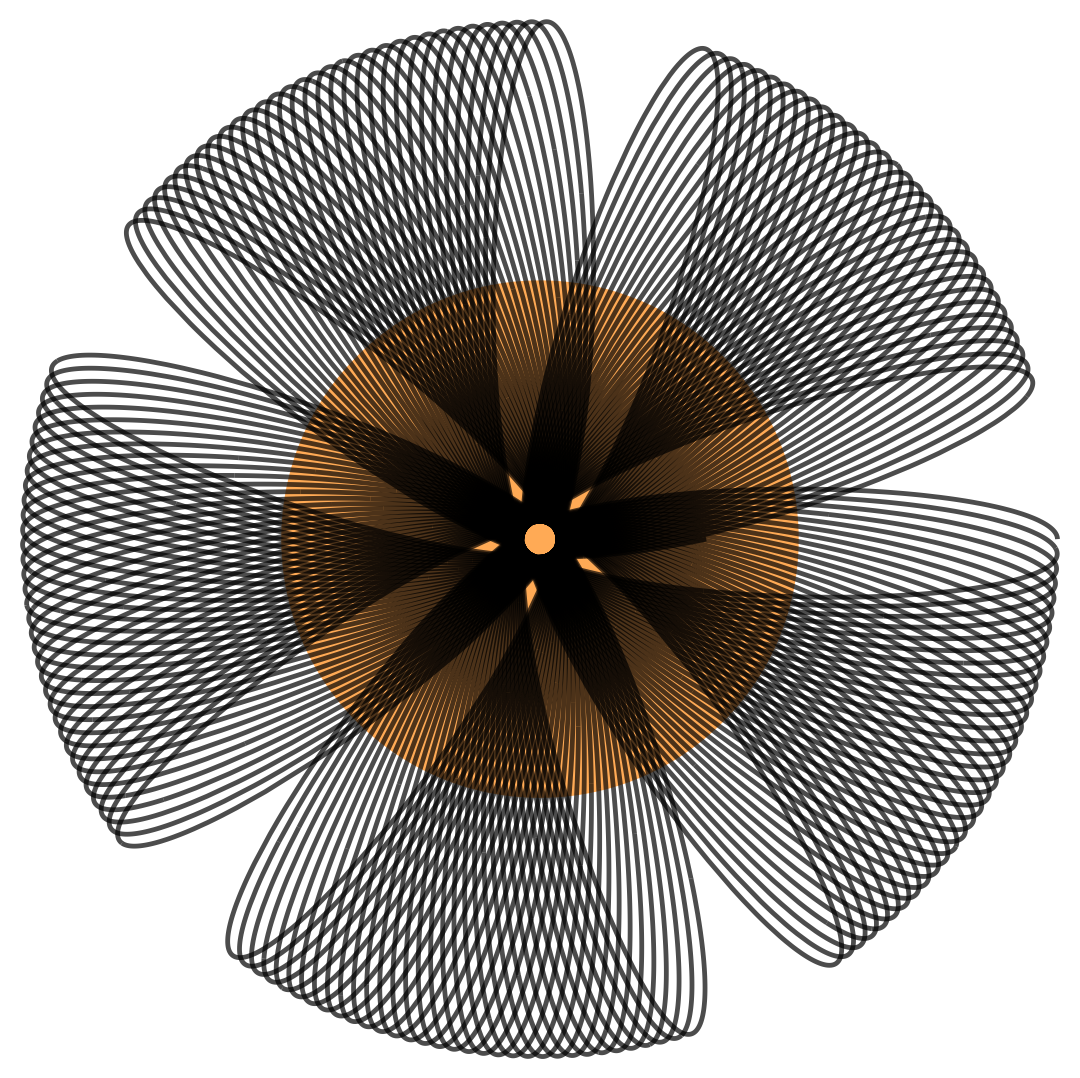} {\small $v(0)=42.065\, {\rm km\, s^{-1}}$} 
         & \includegraphics[width=\linewidth]{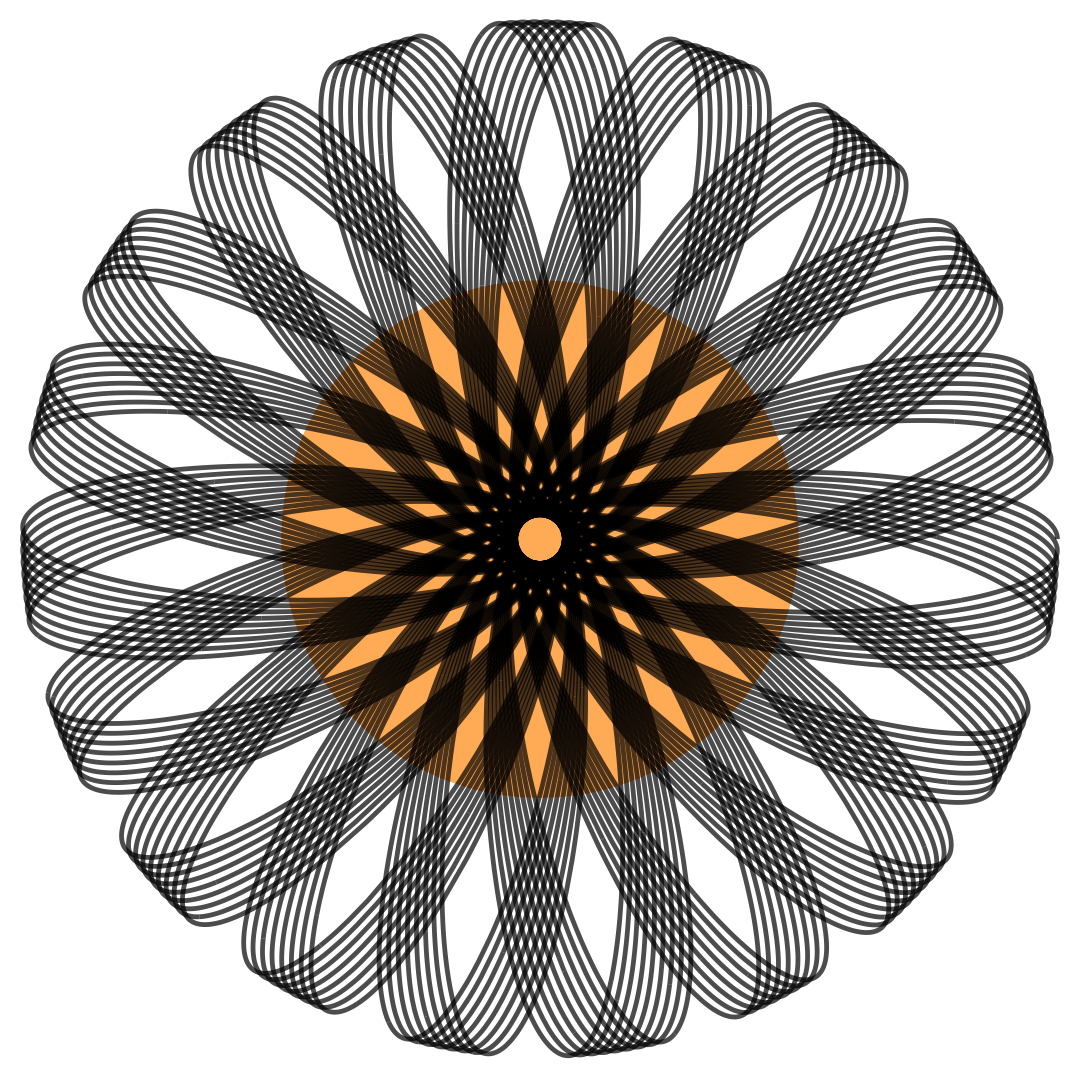} {\small $v(0)=55.98\, {\rm km\, s^{-1}}$} 
         \\ \hline
         \includegraphics[width=\linewidth]{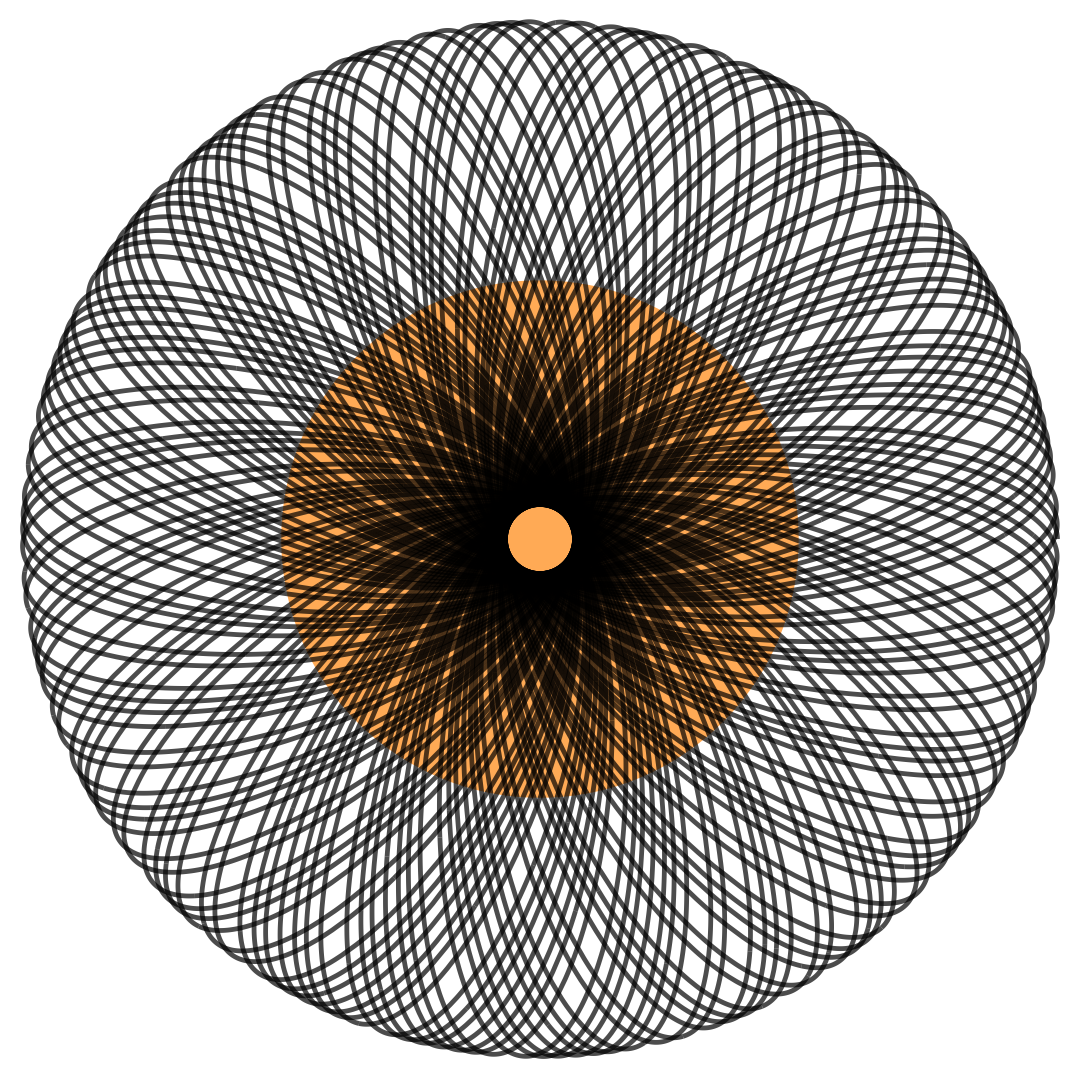} {\small $v(0)=77.25\, {\rm km\, s^{-1}}$}
         &\includegraphics[width=\linewidth]{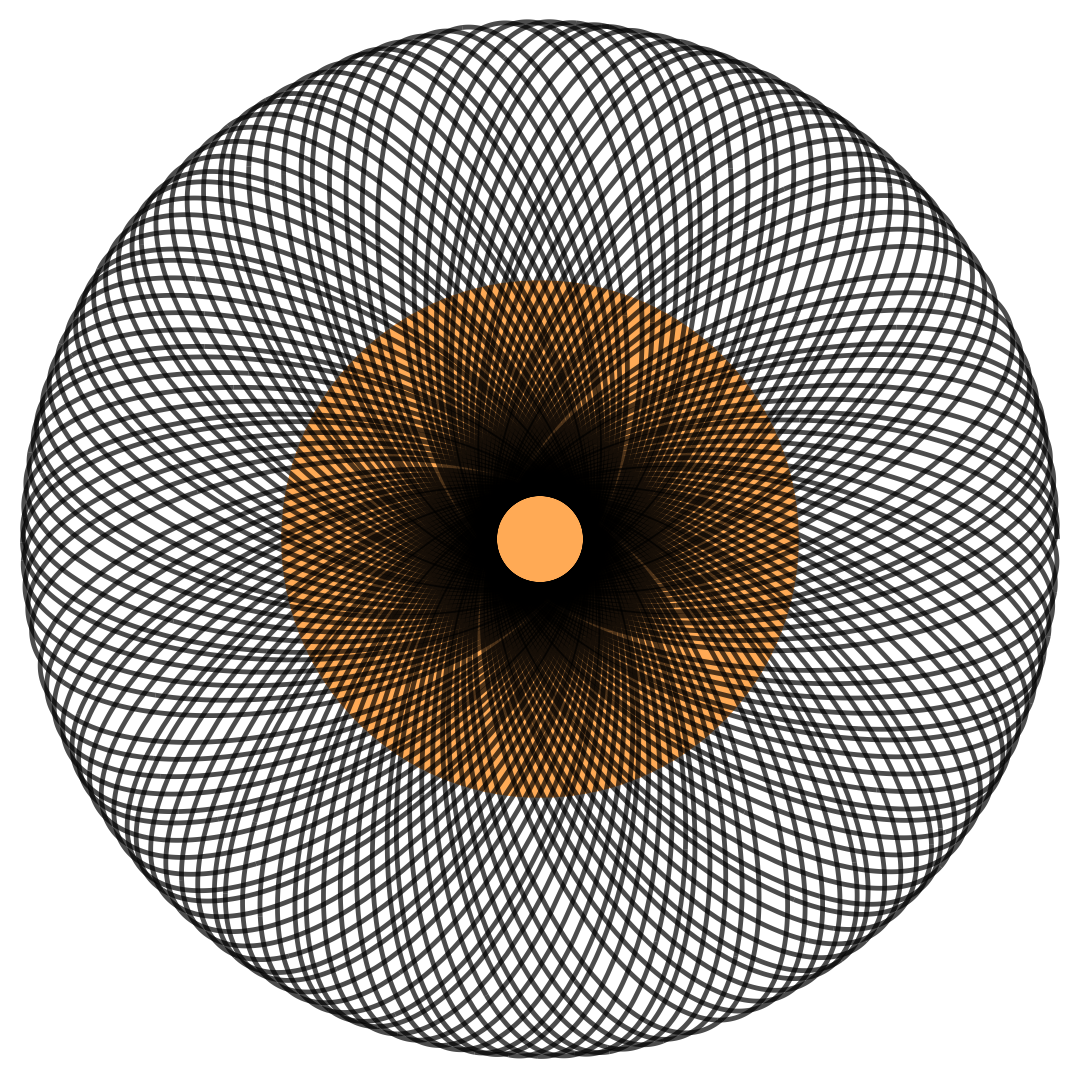} {\small $v(0)=97.95\, {\rm km\, s^{-1}}$} 
         & \includegraphics[width=\linewidth]{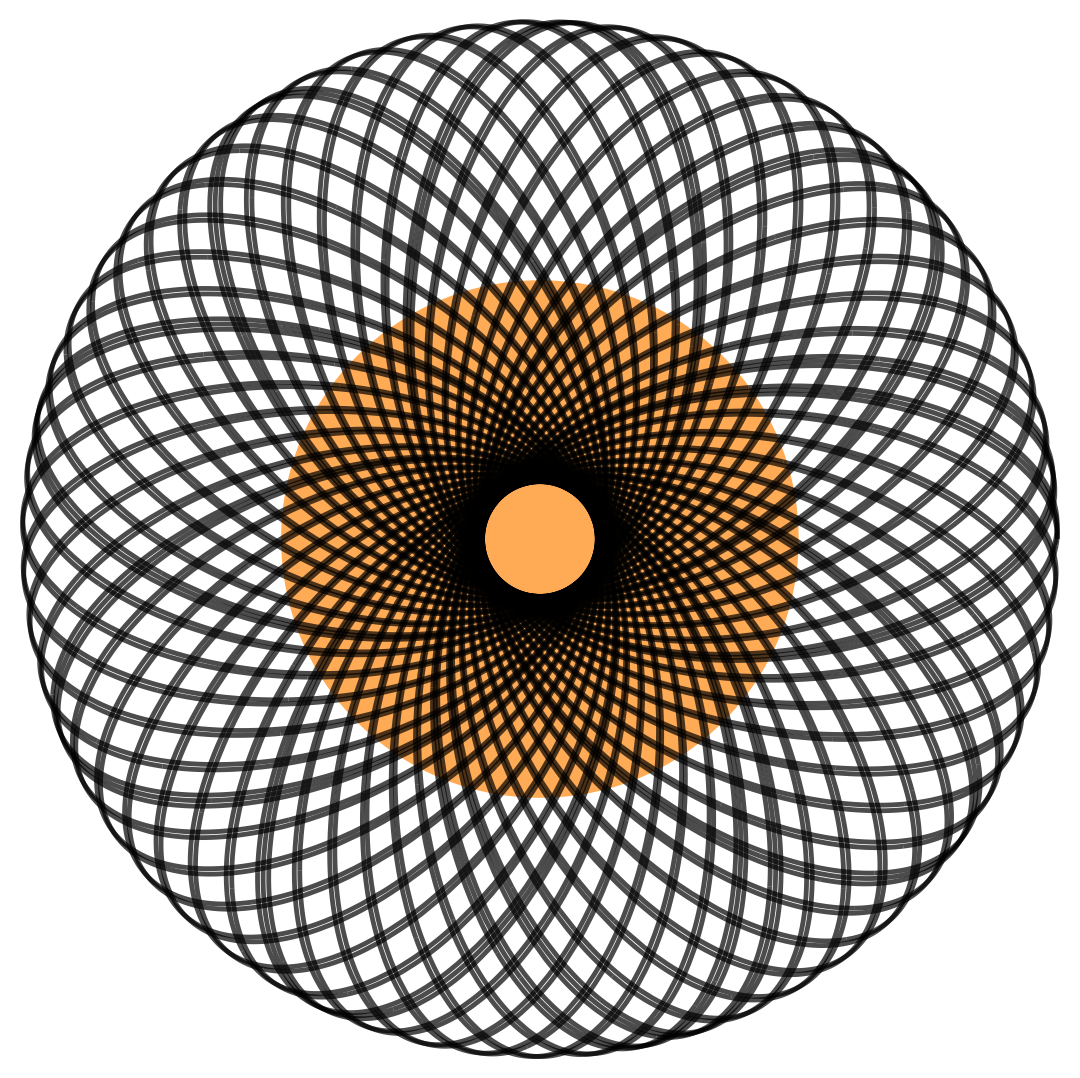} {\small $v(0)=117.89\, {\rm km\, s^{-1}}$} 
         & \includegraphics[width=\linewidth]{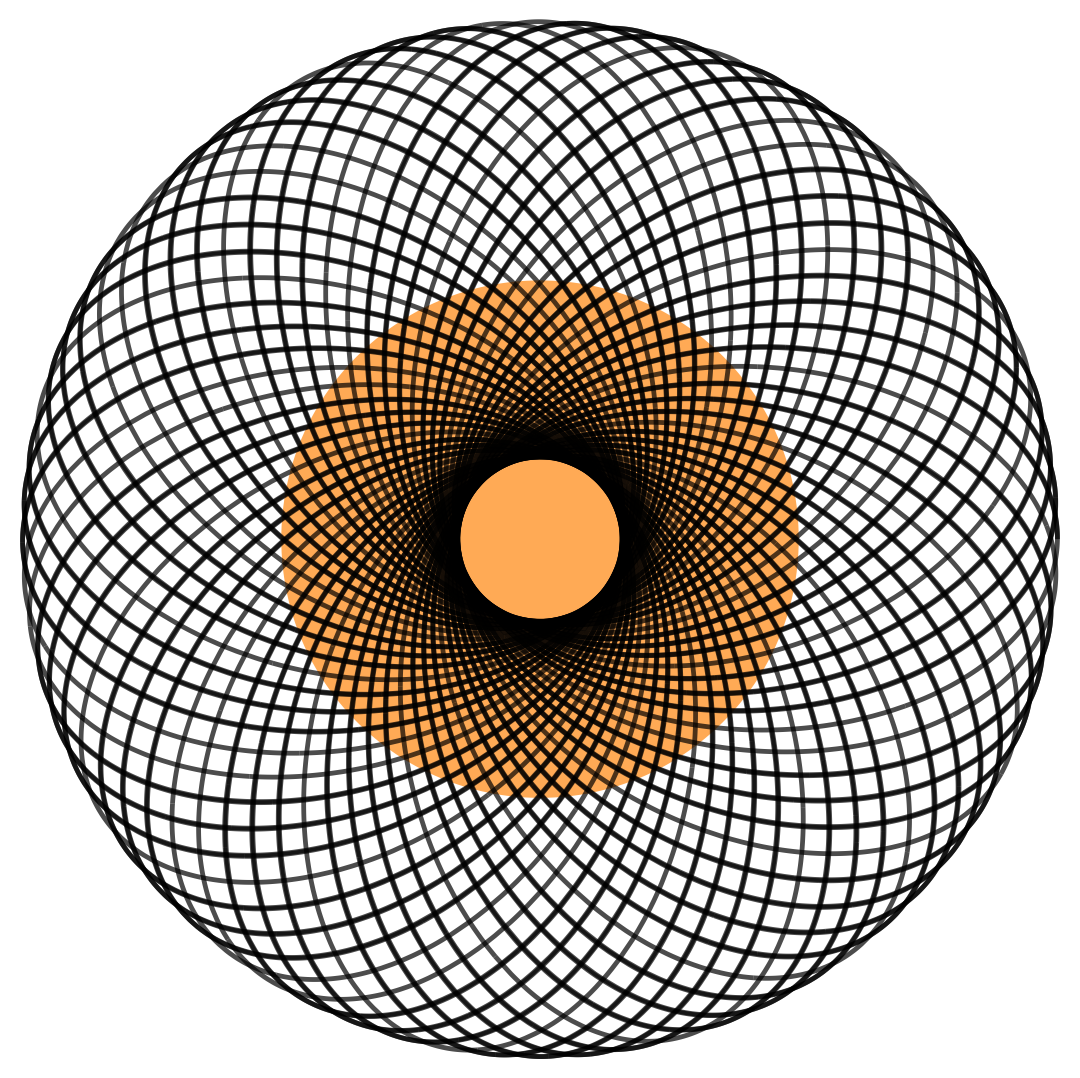} {\small $v(0)=151.74\, {\rm km\, s^{-1}}$} 
         \\
        \hline
    \end{tabularx}
   \caption{Here the initial velocity at $\varphi=0$ is perturbed compared to the initial velocities for the orbits shown in Fig.~\ref{figs_out_closed}. The resulting orbits are open. The plots illustrate an integration interval of $\Delta \varphi = 200\pi$.}
    \label{figs_out_open}
        \end{minipage}
\end{figure}

Similarly, Figs.~\ref{figs_out_closed} and \ref{figs_out_open} depict a set of solutions whose initial conditions are such that the orbit crosses the interior of the star in a certain sector. In Fig.~\ref{figs_out_closed} the solutions presenting closed trajectories are shown, while Fig.~\ref{figs_out_open} shows open trajectory solutions. 

\newpage

\bibliography{references}

\end{document}